\documentclass[a4paper,fleqn,usenatbib]{mnras}

\usepackage{mathptmx}

\usepackage[T1]{fontenc}
\usepackage{ae,aecompl}

\usepackage{graphicx}	
\usepackage{amsmath}	
\usepackage{amssymb}	
\usepackage{url}
\usepackage{morefloats}
\usepackage[flushleft]{threeparttable}
\usepackage{enumitem}
\setcitestyle{notesep={ }}

\title[Multicomponent H$_{2}$-DLA]{Multicomponent H$_2$ in DLA at $z_{abs}$ = 2.05: physical conditions through observations and numerical models\thanks{\scriptsize{Based on observations carried out with the Ultraviolet \& Visual Echelle Spectrograph at the Very Large Telescope, and with the High Resolution Echelle Spectrometer on the Keck I Telescope}}}

\author[]{Katherine Rawlins$^{1,2}$\thanks{\scriptsize{E-mail:
katherine.rawlins@gmail.com; Current affiliation: Department of Physics, University of Mumbai, Santa Cruz East, Mumbai 400 098, India}}, Raghunathan Srianand$^{3}$, Gargi Shaw$^{4}$, Hadi Rahmani$^{5,6}$,
\newauthor{Rajeshwari Dutta$^{3,7}$, Sajeev Chacko$^{1}$}
\\~\\
$^{1}$Department of Physics, University of Mumbai, Santa Cruz East, Mumbai 400 098, India\\
$^{2}$UM-DAE Centre for Excellence in Basic Sciences, University of Mumbai, Santa Cruz East, Mumbai 400 098, India\\
$^{3}$Inter-University Centre for Astronomy \& Astrophysics, Post Bag 4, Ganeshkhind, Pune University Campus, Pune 411 007, India\\
$^{4}$Tata Institute of Fundamental Research, Homi Bhabha Road, Mumbai 400 005, India\\
$^{5}$Aix Marseille Universite, CNRS, Laboratoire d'Astrophysique de Marseille UMR 7326, 13388, Marseille, France\\
$^{6}$GEPI, Observatoire de Paris, PSL Research University, CNRS, Place Jules Janssen, 92190 Meudon, France\\
$^{7}$European Southern Observatory, Karl-Schwarzschild-Str. 2, D-85748 Garching Near Munich, Germany\\
}

\date{Accepted ---. Received ---; in original form ---}

\pubyear{2018}

\begin{document}
\label{firstpage}
\pagerange{\pageref{firstpage}--\pageref{lastpage}}
\maketitle

\begin{abstract}
We perform detailed spectroscopic analysis and numerical modelling of an H$_{2}$-bearing damped Lyman-$\alpha$ absorber (DLA) at $z_{abs}$ = 2.05 towards the quasar FBQS J2340-0053. Metal absorption features arise from fourteen components spread over $\Delta v_{90}$ = 114 km s$^{-1}$, seven of which harbour H$_{2}$. Column densities of atomic and molecular species are derived through Voigt profile analysis of their absorption lines. We measure total \textit{N}(\ion{H}{i}), \textit{N}(H$_{2}$) and \textit{N}(HD) to be 20.35$\pm$0.05, 17.99$\pm$0.05 and 14.28$\pm$0.08 (log cm$^{-2}$) respectively. H$_{2}$ is detected in the lowest six rotational levels of the ground vibrational state. The DLA has metallicity, Z = 0.3 $Z_{\sun}$ ([S/H] = -0.52$\pm$0.06) and dust-to-gas ratio, $\kappa$ = 0.34$\pm$0.07. Numerical models of the H$_{2}$ components are constrained individually to understand the physical structure of the DLA. We conclude that the DLA is subjected to the metagalactic background radiation and cosmic ray ionization rate of $\sim$ 10$^{-15.37}$ s$^{-1}$. Dust grains in this DLA are smaller than grains in the Galactic interstellar medium. The inner molecular regions of the H$_{2}$ components have density, temperature and gas pressure in the range 30--120 cm$^{-3}$, 140--360 K and 7,000--23,000 cm$^{-3}$ K respectively. Micro-turbulent pressure is a significant constituent of the total pressure, and can play an important role in these innermost regions. Our H$_{2}$ component models enable us to constrain component-wise \textit{N}(\ion{H}{i}), and elemental abundances of sulphur, silicon, iron and carbon. We deduce the line-of-sight thickness of the H$_2$-bearing parts of the DLA to be 7.2 pc. 
\end{abstract}




\section{Introduction}
\label{sec:intro}

The spectra of luminous objects such as quasars and gamma ray bursts often indicate the presence of absorbing clouds along the line-of-sight. These absorbers are characterized by their content of neutral gas. Damped Lyman-$\alpha$ absorbers (DLAs) have the highest observed column densities of neutral hydrogen, \textit{N}(\ion{H}{i}) $\geq 10^{20.3}$ cm$^{-2}$ \citep* {Wolfe2005} and account for most of the neutral gas in the high-redshift Universe \citep {Noterdaeme2009, Noterdaeme2012}. Depending on the particular DLA sightline being probed, we may be able to observe the diffuse warm phase (n $\sim$ 0.6 cm$^{-3}$, T $\sim$ 5000 K), the dense cold phase (n $\sim$ 30 cm$^{-3}$ , T $\sim$ 100 K), or a combination of both \citep {Srianand2005b, Draine2011}. \par
\setlength{\parindent}{2ex}
The main sources of ionizing radiation in DLAs are the metagalactic background radiation from quasars and galaxies, and \textit{in situ} star formation. Studies indicate that the metagalactic background may be insufficient to account for the heating in DLAs \citep{Wolfe2003a, Wolfe2008, Srianand2005a, Dutta2014}. Emission lines have been detected in some high-redshift DLAs, indicating a local source of radiation; though the connection between DLAs and star formation remains open to further investigation \citep{Rahmani2010, Krogager2013, Fynbo2013, Fumagalli2015, Srianand2016}. \par
\setlength{\parindent}{2ex}
Various metal ions are observed in DLAs. The singly ionized state is usually the dominant form for most metals. Molecules are also detected in DLAs, but they can form only in the inner regions where there is sufficient H$_2$ self-shielding to protect them from being destroyed by incident ionizing radiation. In the context of this DLA, we find that dust shielding does not hold much significance. We thus, use the term `shielding' to refer to H$_2$ self-shielding throughout the paper, unless mentioned otherwise. H$_{2}$ is the most abundant molecule, and can be observed through the ultraviolet Lyman and Werner band transitions. These transitions occur when photons with energy 11.2--13.6 eV lead to excitation of the electronic states in the molecule. H$_{2}$ is detected in various rotational levels of the ground vibrational state. We use the notation H$_{2}$ (J), where J is the rotational level. As the first ionization potential of carbon is 11.2 eV, neutral carbon is also associated with the regions that harbour H$_{2}$. The ground state of carbon (\ion{C}{i}) has three fine structure levels $^3P_0$, $^3P_1$ and $^3P_2$. We denote these levels as \ion{C}{i*}, \ion{C}{i**} \& \ion{C}{i***}. Thus, observations of H$_{2}$ and \ion{C}{i} provide strong constraints for determining the physical state of cool gas in DLAs. Such studies to probe the physical environment prevalent in H$_{2}$-bearing DLAs (H$_{2}$-DLAs hereafter) have been attempted by \citet*{Ge2001}, \citet {Srianand2005a}, \citet{Jorgenson2010}, \citet{Noterdaeme2015a}, \citet{Klimenko2016}, \citet* {Shaw2016} and \citet{Noterdaeme2017}. \par
\setlength{\parindent}{2ex}
Most of the known H$_{2}$-DLAs are situated at $z_{abs} >$ 1.8, when the Lyman and Werner band transitions are redshifted into the optical region of the electromagnetic spectrum, and can be observed by ground-based telescopes. The atmospheric cut-off at 3000 {\AA} prevents observation of low-redshift H$_{2}$ from the ground. But recently, space-based missions have begun to be used to detect H$_{2}$ in low-redshift DLAs \citep {Crighton2013, Oliveira2014, Srianand2014, Muzahid2015}. It was earlier understood that 10--15 percent of DLAs at high redshift show the presence of H$_{2}$ absorption features \citep {Ledoux2003, Noterdaeme2008a}. However, recent surveys indicate that this number could be much lower, with the H$_2$ detection rate estimated to be less than 7 percent for \textit{N}(H$_2$) > 10$^{19}$ cm$^{-2}$ \citep {Balashev2014}, and less than 6 percent for \textit{N}(H$_2$) > 10$^{17.5}$ cm$^{-2}$ \citep {Jorgenson2014}. More recently, \citet* {Balashev2018} use composite absorption spectra to measure the H$_2$ detection rate to be 4 percent. So far, H$_{2}$ detections have been made in over 25 high-redshift DLAs \citep {Ledoux2003, Noterdaeme2008a, Bagdonaite2014, Balashev2015, Noterdaeme2015a, Krogager2016}. In addition to H$_2$, DLAs have been observed to harbour other molecules too. \citet {Varshalovich2001} reported the first high-redshift detection of HD. Subsequently, various other DLA systems have been found to contain HD molecules \citep {Noterdaeme2008b, Tumlinson2010, Ivanchik2010, Ivanchik2015, Balashev2010, Balashev2017, Albornoz2014, Klimenko2015a}. CO has also been observed along different high-redshift DLA sightlines \citep {Srianand2008, Noterdaeme2010, Noterdaeme2011, Noterdaeme2017, Noterdaeme2018}. \par
\setlength{\parindent}{2ex}	
Absorption in a DLA may arise from either a single clump of gas, or multiple associated clumps. In the case of multiple clumps, only a few may satisfy the high density and low temperature conditions necessary for the formation of molecules. It is rare for DLAs to show molecular absorption features in multiple components spread over large velocity intervals. Some examples are the DLAs at $z_{abs}$ = 2.6265 towards FBQS J0812+3208 which shows the presence of H$_{2}$ in three components \citep {Jorgenson2009, Tumlinson2010, Jorgenson2010}, at $z_{abs}$ = 1.973 towards Q 0013-004 with H$_{2}$ in 4 components \citep {Petitjean2002}, and at $z_{abs}$ = 2.418 towards the quasar SDSS J143912.04+111740.5 which has H$_{2}$ in 6 components \citep {Noterdaeme2008b, Srianand2008}. Such multicomponent absorbers with many observed species provide us an excellent opportunity to probe the variation of physical properties within the DLA, and hence, to understand the internal structure of the absorbing region.  \par
\setlength{\parindent}{2ex}
We present here spectroscopic analysis and detailed numerical modelling of a multicomponent H$_{2}$-DLA along the sightline to the QSO FBQS J2340-0053. There are two main absorption systems along this sightline -- an \ion{Mg}{ii} absorber at $z_{abs}$ = 1.36 \citep {Rahmani2012}, and a DLA at $z_{abs}$ = 2.05 \citep {Jorgenson2010}. \citet {Jorgenson2010} have detected H$_{2}$ absorption in the DLA and have extracted physical parameters through analysis of the spectrum obtained using the High Resolution Echelle Spectrometer (HIRES) on the Keck I Telescope. We study the DLA in greater detail in this paper using data obtained with the Ultraviolet and Visual Echelle Spectrograph (UVES) on the Very Large Telescope (VLT). The UVES spectrum has higher signal-to-noise ratio compared to the HIRES spectrum analysed by \citet {Jorgenson2010}. Voigt profile fitting of the H$_{2}$, \ion{C}{i} and metal absorption lines is performed to derive component-wise column densities. Numerical models are then constructed for each of the molecular components. By reproducing the observed column densities, we constrain the physical conditions in each molecular component. Such detailed modelling of a multicomponent H$_{2}$-absorber has as yet been unattempted. \par
\setlength{\parindent}{2ex}
This paper is organized as follows. In Section \ref{sec:reduction}, we mention details of the observations and data reduction techniques. The results of our Voigt profile fits can be found in Section \ref{sec:voigt}. In Section \ref{sec:phy_prop}, we obtain estimates of some physical properties of the DLA through the observed column densities of various species. Details of our numerical models are presented in Section \ref{sec:models}. In Section \ref{sec:result}, we combine the results of the observational analysis and the numerical models, to study the variation of different physical properties within the DLA. Section \ref{sec:conclude} provides a summary of our results.  

\section{Observations \& Data reduction}
\label{sec:reduction}

	The optical spectroscopic observations were carried out with the UVES mounted on the VLT, Chile [Programme ID: 082.A-0569]. The two arms of the instrument were operated with the beam splitter in the dichroic \#2 mode (390+580 setting). The wavelength coverage extends from 3284 to 4521 {\AA} on the blue CCD, and 4779 to 5759 {\AA} and 5837 to 6812 {\AA} on the two red CCDs, with a spectral resolution of $\sim$ 45,000 and FWHM $\sim$ 6.6 km s$^{-1}$. Data reduction was performed using the UVES Common Pipeline Library data reduction pipeline release 4.7.8\footnote{\url{http://www.eso.org/sci/facilities/paranal/instruments/uves/doc/}}, followed by wavelength calibration using ThAr emission lines. The spectrum was then corrected for the motion of the observatory around the barycentre of the Sun-Earth system. The air to vacuum wavelength conversion was carried out as per the formula given by \citet {Edlen1966}. Different exposures were then co-added to get a combined spectrum. \par
\setlength{\parindent}{2ex}
Here, we first try to reproduce the quasar continuum. We divide the entire wavelength range of the combined spectrum into $\sim$ 100 {\AA} intervals. We then manually identify spectral ranges unaffected by absorption lines, and fit these regions using cubic splines to get the unabsorbed quasar continuum. The continuum thus obtained is used to normalize the spectrum. Various metal and molecular absorption lines from the DLA are then identified. If required, we revisit the fit for the quasar continuum around the absorption lines of interest, and obtain a more accurate estimate of the local continuum level. Subsequently, we perform Voigt profile analysis, and derive column densities for the different absorbing species. We discuss the results of Voigt profile fitting in Section \ref{sec:voigt}. \par
	The Lyman-$\alpha$ emission peak of the QSO is located at redshift, $z_{em}$ $\sim$ 2.083. This corresponds to a velocity shift of $\sim$ 2710 km s$^{-1}$ from the nearest component of the DLA.\par
\setlength{\parindent}{2ex}
The wavelength coverage of the UVES spectrum prevents us from detecting transitions associated with the DLA having rest wavelengths lower than $\lambda$1075, causing us to miss out on quite a few H$_2$ transitions at the blue end of the spectrum. To compensate for this, we also include in the Voigt profile analysis, the clean low-wavelength H$_{2}$ transitions from the HIRES spectrum of this sightline (PI: Prochaska, August 2006) obtained through the Keck Observatory Database of Ionized Absorption toward Quasars (KODIAQ) \citep{Jorgenson2010, Lehner2014, OMeara2015, OMeara2017}. The HIRES spectrum has spectral resolution similar to the UVES spectrum, but its signal-to-noise ratio is lower. 

\section{Voigt profile analysis}
\label{sec:voigt}

The {\tiny VPFIT} package, version 10.0\footnote{\url{http://www.ast.cam.ac.uk/~rfc/vpfit.html}} is used to fit the absorption lines with multicomponent Voigt profiles. The profile fit to a given component depends on three parameters -- redshift ($z_{abs}$), Doppler parameter (\textit{b}) and column density (\textit{N}). The number of components is also ascertained during the profile fitting process. We assume that low ions such as \ion{Si}{ii}, \ion{S}{ii}, \ion{Ni}{ii}, \ion{Zn}{ii} and \ion{Cr}{ii} are associated with the same region of the gas cloud, and hence, link their values of $z_{abs}$ and \textit{b} for each component while performing the fit. {\tiny VPFIT} uses $\chi^2$ minimization to converge to the best fit to an absorption line. It fits each line with multiple components and measures the corresponding component-wise column densities. Similarly, we perform a different set of fits each for the \ion{C}{i}, H$_{2}$ and HD lines which arise from a cooler phase of gas. The Lyman-$\alpha$ line is also fitted separately to derive \textit{N}(\ion{H}{i}). Unlike metal and H$_{2}$ absorption, we get only total \textit{N}(\ion{H}{i}) from the Voigt profile fit to the Lyman-$\alpha$ line. \par
\setlength{\parindent}{2ex}
All the transitions of various ionic and molecular species used for profile fitting are listed in Table \ref{tab:transitions}. We discuss the profile fits of the \ion{H}{i}, metal, H$_{2}$, HD and \ion{C}{i} lines, individually in the following sub-sections.

\begin{table}
\setlength{\tabcolsep}{3pt}
\centering
  \caption{Transitions used for Voigt profile analysis}
  \label{tab:transitions}
\begin{threeparttable}
\begin{tabular}{ll}
\hline
Species & Transitions (Wavelengths in {\AA})\\
\hline
 \ion{H}{i} & $\lambda$1216 \\
 \ion{C}{i}$^\textit{a}$ & $\lambda$1155, $\lambda$1270, $\lambda$1277, $\lambda$1328, $\lambda$1656 \\
 \ion{C}{ii*} & $\lambda$1335.66, $\lambda$1335.71 \\
 \ion{Mg}{i} & $\lambda$1668, $\lambda$1683, $\lambda$1707, $\lambda$1747, $\lambda$1827, $\lambda$2026 \\
 \ion{Al}{iii} & $\lambda$1854, $\lambda$1862 \\ 
 \ion{Si}{ii} & $\lambda$1190, $\lambda$1304, $\lambda$1808 \\
 \ion{P}{ii} & $\lambda$1152 \\ 
 \ion{S}{ii} & $\lambda$1250, $\lambda$1253, $\lambda$1259 \\
 \ion{Cr}{ii} & $\lambda$2062, $\lambda$2066 \\
 \ion{Fe}{ii} & $\lambda$1106, $\lambda$1112, $\lambda$1127, $\lambda$1133, $\lambda$1143, $\lambda$1608, $\lambda$1611 \\
 \ion{Ni}{ii} & $\lambda$1317, $\lambda$1370, $\lambda$1454, $\lambda$1467.26, $\lambda$1467.76, $\lambda$1703, \\ & $\lambda$1709, $\lambda$1741, $\lambda$1751 \\
 \ion{Zn}{ii} & $\lambda$2026, $\lambda$2062 \\
 H$_{2}$ (0) & $\lambda$1092, $\lambda$1108; $\lambda$1049$^\textit{b}$ \\
 H$_{2}$ (1) & $\lambda$1077, $\lambda$1078, $\lambda$1092, $\lambda$1108, $\lambda$1110; $\lambda$1037$^\textit{b}$, $\lambda$1038$^\textit{b}$, \\ & $\lambda$1049$^\textit{b}$, $\lambda$1051$^\textit{b}$, $\lambda$1064$^\textit{b}$ \\
 H$_{2}$ (2) & $\lambda$1110, $\lambda$1112; $\lambda$1038$^\textit{b}$, $\lambda$1040$^\textit{b}$, $\lambda$1051$^\textit{b}$, $\lambda$1064$^\textit{b}$ \\
 H$_{2}$ (3) & $\lambda$1099, $\lambda$1112, $\lambda$1115; $\lambda$1019$^\textit{b}$, $\lambda$1028$^\textit{b}$, $\lambda$1041$^\textit{b}$, \\ & $\lambda$1043$^\textit{b}$, $\lambda$1056$^\textit{b}$, $\lambda$1067$^\textit{b}$, $\lambda$1070$^\textit{b}$ \\
 H$_{2}$ (4) & $\lambda$1085, $\lambda$1088, $\lambda$1100, $\lambda$1104, $\lambda$1116, $\lambda$1120; $\lambda$1035$^\textit{b}$, \\ & $\lambda$1044$^\textit{b}$, $\lambda$1060$^\textit{b}$, $\lambda$1074$^\textit{b}$ \\
 H$_{2}$ (5) & $\lambda$1109, $\lambda$1120; $\lambda$1048$^\textit{b}$, $\lambda$1061$^\textit{b}$, $\lambda$1075$^\textit{b}$ \\
 HD (0) & $\lambda$1021$^\textit{b}$, $\lambda$1042$^\textit{b}$, $\lambda$1054$^\textit{b}$, $\lambda$1066$^\textit{b}$\\
\hline
\end{tabular}
\begin{tablenotes}
\item $^\textit{a}$ \ion{C}{i*}, \ion{C}{i**} and \ion{C}{i***} transitions
\item $^\textit{b}$ Transitions covered only by the archival Keck HIRES spectrum, and included in Voigt profile analysis along with transitions covered by UVES
\end{tablenotes}
\end{threeparttable}
\end{table}

\subsection{Neutral hydrogen}
\label{ssec:hi}

\begin{figure}
\centering{\includegraphics[width=\columnwidth, height=17 cm, keepaspectratio]{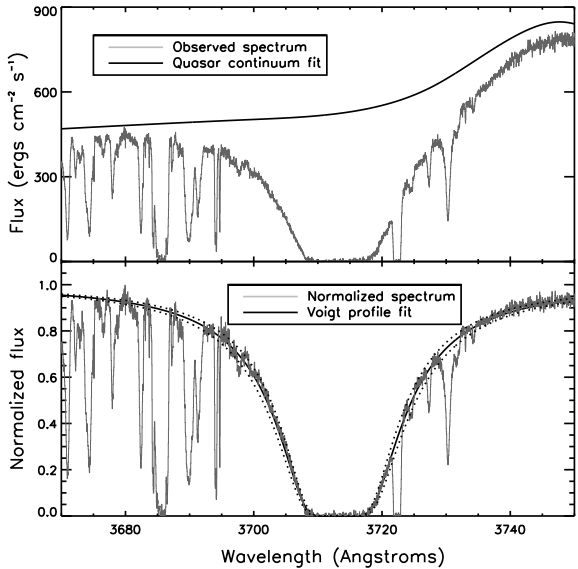}}
\caption{The upper panel shows the observed spectrum and the quasar continuum fit in the spectral range around the damped Lyman-$\alpha$ line. The quasar continuum is used to normalize the spectrum. The normalized spectrum is shown in the lower panel along with the Voigt profile fit to the damped Lyman-$\alpha$ line. We measure log[\textit{N}(\ion{H}{i})(cm$^{-2}$)] = 20.35$\pm$0.05.}
\label{fig:lyal}
\end{figure}

We observe \ion{H}{i} only through Lyman-$\alpha$ absorption. Other Lyman series transitions do not fall within the covered spectral range. As the  Lyman-$\alpha$ line is damped, we can only find the total content of neutral hydrogen in the DLA. It is impossible for Voigt profile analysis to determine the component-wise distribution of \ion{H}{i}. We consider a large value for \textit{b} $\sim$ 20--30 km s$^{-1}$, and perform the fit for different values of \textit{N}(\ion{H}{i}). The crucial factors in deciding the most appropriate fit are the damping wings and the turning points of the profile near the line core. Thus, in order to obtain the optimum Voigt profile fit, we try to normalize the region of the spectrum around the line profile by using different continua. Besides the fitting error from {\tiny VPFIT}, the uncertainty in the continuum level contributes significantly to the uncertainty in column density measurement. We consider various continua which trace the line profile appreciably well (as decided by eye). The two continua with the most deviation from the continuum producing the optimum Voigt profile fit are selected. Voigt profile analysis is then repeated using these continua. We compare these column density predictions with those from the optimum line profile fit, and estimate the uncertainty in continuum placement. We determine log[\textit{N}(\ion{H}{i})(cm$^{-2}$)] = 20.35$\pm$0.05. In comparison, \citet{Jorgenson2010} find log[\textit{N}(\ion{H}{i})(cm$^{-2}$)] = 20.35$\pm$0.15. The Lyman-$\alpha$ line profile, along with the continuum used for the profile fit, is shown in Fig. \ref{fig:lyal}. \par
\setlength{\parindent}{2ex}

\subsection{Metals}
\label{ssec:mets}

\begin{figure*}
\centering{\includegraphics[width=14.5cm, keepaspectratio]{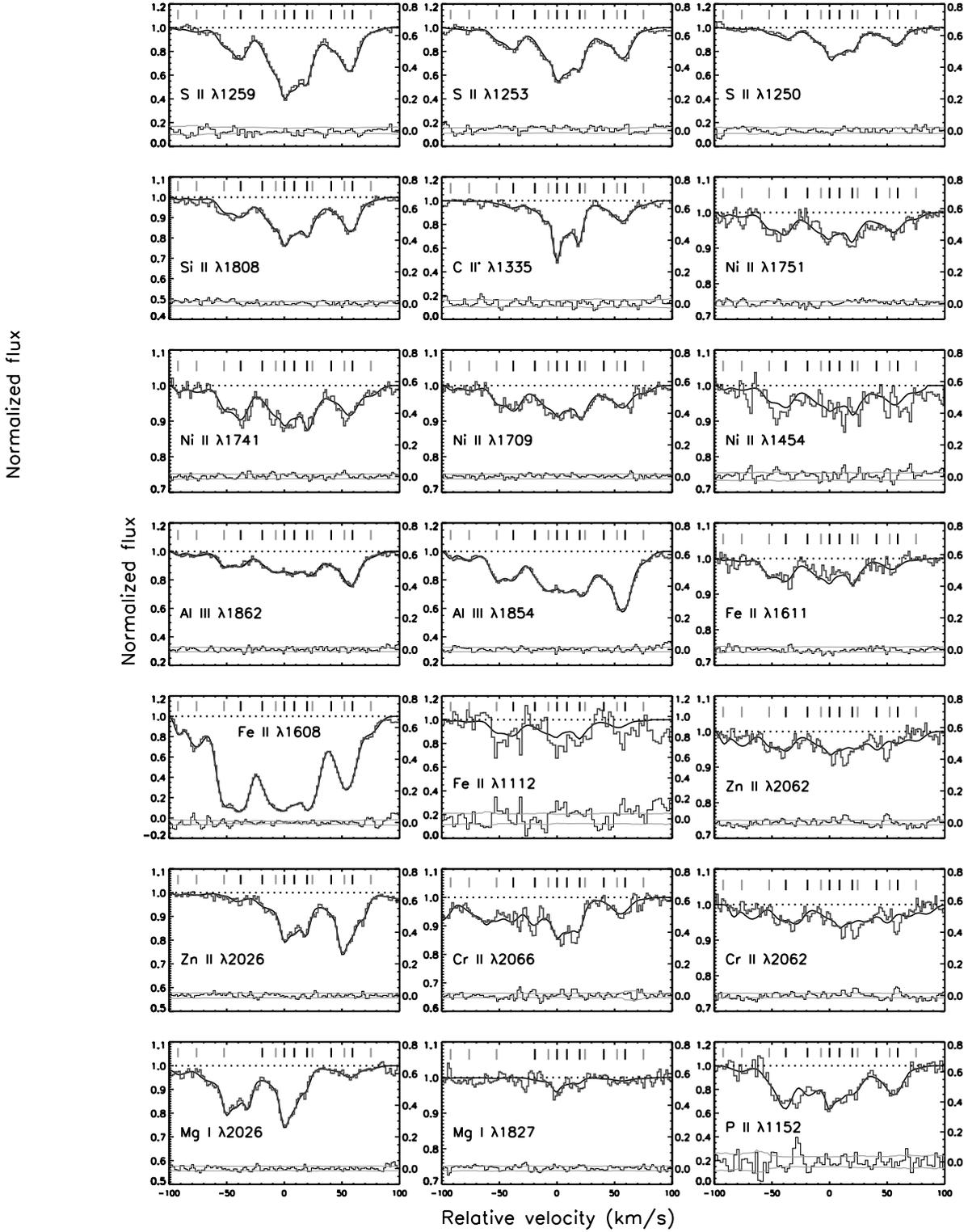}}
\caption{Multiple component Voigt profile fits to some of the metal absorption lines. The tick marks at the top indicate the positions of individual Voigt profile components in velocity space. Components associated with H$_2$ absorption are shown by black ticks, while the other components are indicated by grey ticks. Zero velocity is defined at the metal component with $z_{abs}$ = 2.054528. A residual indicating the difference between the observed line profile and its Voigt profile fit is shown at the bottom of each panel. The grey lines indicate the 1-$\sigma$ range in flux allowed by the normalized error spectrum. The scale of the residual plot is shown on the right-hand vertical axis.}
\label{fig:voigt_metals}
\end{figure*}

	Various absorption lines of the species \ion{C}{ii*}, \ion{Mg}{i}, \ion{Al}{iii}, \ion{Si}{ii}, \ion{P}{ii}, \ion{S}{ii}, \ion{Cr}{ii}, \ion{Fe}{ii}, \ion{Ni}{ii} \& \ion{Zn}{ii} are observed in the spectrum. These metal absorption features are seen to arise from fourteen components. Using the \ion{S}{ii} $\lambda$1259 transition, we determine the velocity width, $\Delta v_{90}$ to be 114 km s$^{-1}$ \citep*{Prochaska1997}. Table \ref{tab:redshifts} lists the redshift values of all the components, along with the Doppler parameter obtained from the Voigt profile fit. Some of the transitions along with the fitted profiles are shown in Fig. \ref{fig:voigt_metals}, while Table \ref{tab:coldens_metals} summarizes the column densities for all ionic species. Most of the \ion{Si}{ii} features are saturated. Saturated features of \ion{C}{ii}, \ion{N}{i}, \ion{O}{i} \& \ion{Al}{ii} are also seen in the spectrum. The \ion{Si}{iv} doublet lines associated with the DLA are also observed, but are not included in the Voigt profile fit as they trace a different phase of the gas. The presence of multiple absorbers along the sightline leads to many blended features. Some of the \ion{Fe}{ii} and \ion{Ni}{ii} transitions are affected by blends, and are excluded from Voigt profile fitting. We also discard the \ion{Cr}{ii} $\lambda$2056 transition from the fit. Of the three \ion{Cr}{ii} lines in the ultraviolet, this transition is the strongest. It is also free of any blends with lines from the DLA and the other known absorber. Yet, we are unable to obtain a good profile fit while including this line, and it also affects the profiles predicted for the other metal species. We conclude that there could be some undetected contamination, and leave out this line while performing the fit. \par
\setlength{\parindent}{2ex}
We determine an upper limit on the column density when the uncertainty in the Voigt profile fit is greater than 0.5 dex. This includes the column densities reported in Table \ref{tab:coldens_metals} for particular components of \ion{Mg}{i}, \ion{Al}{iii}, \ion{Cr}{ii}, \ion{Zn}{ii} and \ion{C}{ii*}. The method we follow to obtain these column density upper limits is briefly outlined here. The following relation from \citet {Hellsten1998} is used to calculate the rest equivalent width $W_r$ for an $N_\sigma$-sigma detection. 
\begin{equation}
    W_r(N_\sigma) = \frac{N_\sigma \sqrt{N_{pix}}\Delta \lambda_{pix}}{S/N (1+z_{abs})} \; \; \;.
	\label{eq:hellsten_limit}
\end{equation}
Here, $N_{pix}$, $\Delta \lambda_{pix}$ and S/N denote the number of pixels used for detection, wavelength per pixel and signal-to-noise ratio respectively. We consider $N_\sigma$ = 3. Further, the column density is computed from the equivalent width assuming the optically thin approximation. The following equation is used, where $\lambda_r$ is the rest wavelength  of the transition and $f_{osc}$ is the oscillator strength.
\begin{equation}
    N = 1.13 \times 10^{20}\frac{W_r(N_\sigma)}{\lambda_r^2 f_{osc}} \; \; \;.
	\label{eq:optically_thin}
\end{equation}
\par
\setlength{\parindent}{2ex}
\citet{Jorgenson2010} have performed analysis of the metal lines of the DLA through the apparent optical depth method. They have considered only the three major clumps of gas (super-components) within the DLA, and have derived column densities of \ion{C}{ii*}, \ion{Fe}{ii}, \ion{Ni}{ii}, \ion{S}{ii} and \ion{Ar}{i} for each super-component. We perform Voigt profile analysis for all these species except \ion{Ar}{i}, which is not covered in the UVES spectrum. The column densities that we derive agree closely with the values of \citet{Jorgenson2010}. Our detailed analysis enables us to probe the component-wise distribution of each metal ion. We later use these as observational constraints for our numerical models. Besides the ions included by \citet{Jorgenson2010} in their analysis, we also obtain column densities of other low ions such as \ion{Mg}{i}, \ion{Si}{ii}, \ion{P}{ii}, \ion{Zn}{ii} and \ion{Cr}{ii}. 

\begin{table}
 \centering
  \caption{Metal component redshifts \& Doppler parameter}
  \label{tab:redshifts}
  \begin{tabular}{@{}ccc@{}}
  \hline
   Component & Redshift & \textit{b} (km s$^{-1}$)  \\
 \hline
 1 & 2.053588$\pm$0.000002 & 1.3$\pm$0.2 \\
 2 & 2.053753$\pm$0.000002 & 9.8$\pm$0.4 \\
 3 & 2.053995$\pm$0.000003 & 6.7$\pm$0.3 \\
 4 & 2.054142$\pm$0.000003 & 8.6$\pm$0.5 \\
 5 & 2.054333$\pm$0.000011 & 6.9$\pm$1.1 \\
 6 & 2.054452$\pm$0.000008 & 8.7$\pm$1.3 \\ 
 7 & 2.054528$\pm$0.000002 & 1.4$\pm$0.3 \\ 
 8 & 2.054616$\pm$0.000006 & 8.5$\pm$1.4 \\ 
 9 & 2.054729$\pm$0.000002 & 2.1$\pm$0.4 \\
 10 & 2.054776$\pm$0.000008 & 7.5$\pm$0.6 \\
 11 & 2.054942$\pm$0.000010 & 6.9$\pm$1.4 \\
 12 & 2.055060$\pm$0.000006 & 7.3$\pm$1.0 \\
 13 & 2.055131$\pm$0.000014 & 7.7$\pm$1.2 \\
 14 & 2.055293$\pm$0.000010 & 7.7$\pm$1.1 \\
\hline
\end{tabular}
\end{table}
	
\begin{table*}
\setlength{\tabcolsep}{4pt}
 \centering
  \caption{Component-wise logarithmic column densities of different metal ions (in log cm$^{-2}$)}
  \label{tab:coldens_metals}
  \begin{tabular}{@{}ccccccccccc@{}}
  \hline
   Comp. & \ion{Mg}{i} & \ion{Al}{iii} & \ion{Si}{ii} & \ion{P}{ii} & \ion{S}{ii} & \ion{Cr}{ii} & \ion{Fe}{ii} & \ion{Ni}{ii} & \ion{Zn}{ii} & \ion{C}{ii*}\\
 \hline
 1 & $\leq$ 12.24 & 11.14$\pm$0.08 & 12.84$\pm$0.05 & - & 12.20$\pm$0.48 & 11.90$\pm$0.08 & 12.77$\pm$0.04 & 11.83$\pm$0.14 & 10.48$\pm$0.42 & - \\
 2 &  $\leq$ 11.58 & 11.63$\pm$0.04 & 13.37$\pm$0.01 & 12.11$\pm$0.24 & 13.11$\pm$0.10 & 12.45$\pm$0.04 & 13.44$\pm$0.01 & 12.24$\pm$0.09 & 10.87$\pm$0.26 & $\leq$ 12.00\\
 3 & $\leq$ 11.93 & 12.02$\pm$0.03 & 13.98$\pm$0.03 & 12.32$\pm$0.16 & 13.56$\pm$0.05 & 12.29$\pm$0.05 & 14.03$\pm$0.03 & 12.59$\pm$0.04 & 10.74$\pm$0.32 & 11.68$\pm$0.36\\
 4 & - & 12.13$\pm$0.03 & 14.23$\pm$0.02 & 13.00$\pm$0.03 & 13.97$\pm$0.02 & 12.51$\pm$0.04 & 14.27$\pm$0.02 & 12.88$\pm$0.03 & 11.28$\pm$0.18 & 12.38$\pm$0.09\\
 5 & 11.73$\pm$0.29 & 11.81$\pm$0.14 & 13.87$\pm$0.15 & 12.65$\pm$0.10 & 13.37$\pm$0.23 & 11.81$\pm$0.23 & 13.59$\pm$0.17 & 12.14$\pm$0.22 & 11.33$\pm$0.18 & 12.35$\pm$0.13\\
 6 & 12.19$\pm$0.15 & 12.25$\pm$0.08 & 14.42$\pm$0.08 & 12.61$\pm$0.15 & 14.13$\pm$0.08 & 12.38$\pm$0.10 & 14.18$\pm$0.08 & 12.81$\pm$0.08 & 11.59$\pm$0.15 & 12.40$\pm$0.25\\ 
 7 & 12.25$\pm$0.09 & $\leq$ 10.79 & 13.95$\pm$0.08 & 12.40$\pm$0.13 & 13.82$\pm$0.07 & 11.81$\pm$0.15 & 13.52$\pm$0.13 & 12.05$\pm$0.14 & 11.47$\pm$0.11 & 13.24$\pm$0.13\\ 
 8 & 12.45$\pm$0.11 & 12.28$\pm$0.08 & 14.50$\pm$0.08 & 12.86$\pm$0.08 & 14.34$\pm$0.07 & 12.48$\pm$0.09 & 14.11$\pm$0.08 & 12.83$\pm$0.08 & 12.06$\pm$0.07 & 13.05$\pm$0.07\\ 
 9 & 11.88$\pm$0.13 & $\leq$ 10.72 & 13.85$\pm$0.10 & 11.75$\pm$0.50 & 13.67$\pm$0.07 & $\leq$ 11.37 & 13.73$\pm$0.08 & 12.30$\pm$0.11 & 11.57$\pm$0.08 & 12.88$\pm$0.05\\
 10 & $\leq$ 12.10 & 12.26$\pm$0.06 & 14.22$\pm$0.07 & 12.61$\pm$0.10 & 13.82$\pm$0.08 & 12.37$\pm$0.09 & 14.00$\pm$0.06 & 12.73$\pm$0.07 & 11.39$\pm$0.18 & 12.34$\pm$0.19\\
 11 & 11.57$\pm$0.24 & 11.96$\pm$0.11 & 13.90$\pm$0.13 & 12.19$\pm$0.19 & 13.62$\pm$0.11 & 12.41$\pm$0.09 & 13.20$\pm$0.16 & 12.35$\pm$0.12 & 11.05$\pm$0.31 & 12.11$\pm$0.25\\
 12 & 11.63$\pm$0.48 & 12.05$\pm$0.35 & 14.06$\pm$0.30 & 12.43$\pm$0.29 & 13.73$\pm$0.29 & $\leq$ 12.20 & 13.77$\pm$0.13 & 12.47$\pm$0.27 & 11.61$\pm$0.20 & 12.42$\pm$0.34\\
 13 & 12.01$\pm$0.19 & 12.38$\pm$0.14 & 14.32$\pm$0.15 & 12.60$\pm$0.17 & 13.96$\pm$0.15 & 12.22$\pm$0.23 & 13.49$\pm$0.21 & 12.61$\pm$0.17 & 11.57$\pm$0.23 & 12.72$\pm$0.16\\
 14 & 11.59$\pm$0.21 & 11.62$\pm$0.09 & 13.44$\pm$0.19 & 11.66$\pm$0.49 & 12.98$\pm$0.16 & 12.15$\pm$0.09 & 13.04$\pm$0.07 & 12.19$\pm$0.10 & $\leq$ 10.80 & 12.14$\pm$0.16\\
\hline
 Total & $\leq$ 13.14 & 13.14$\pm$0.04 & 15.23$\pm$0.04 & 13.65$\pm$0.04 & 14.95$\pm$0.03 & 13.37$\pm$0.03 & 14.98$\pm$0.02 & 13.68$\pm$0.03 & 12.59$\pm$0.05 & 13.75$\pm$0.05 \\
\hline
\end{tabular}
\end{table*}

\subsection{Molecular hydrogen and deuterated molecular hydrogen}

Many Lyman band transitions of H$_{2}$ are detected in this DLA. These molecular absorption features arise from seven components. H$_2$ absorption is known to arise from a colder, compact region slightly offset from but associated with nearby warmer gas harbouring metals \citep{Ledoux2003, Srianand2005b, Noterdaeme2008a}. Comparing the redshift values listed in Tables \ref{tab:redshifts} and \ref{tab:coldens_h2}, the velocity separation between an H$_2$ component and its nearest metal component can be calculated. It is less than 5 km s$^{-1}$ for all seven H$_2$ components. The metal components in this DLA associated with H$_{2}$ are components 4, 5, 7, 8, 9, 11 \& 13.  \par
\begin{figure*}
\centering{\includegraphics[width=14.5cm, keepaspectratio]{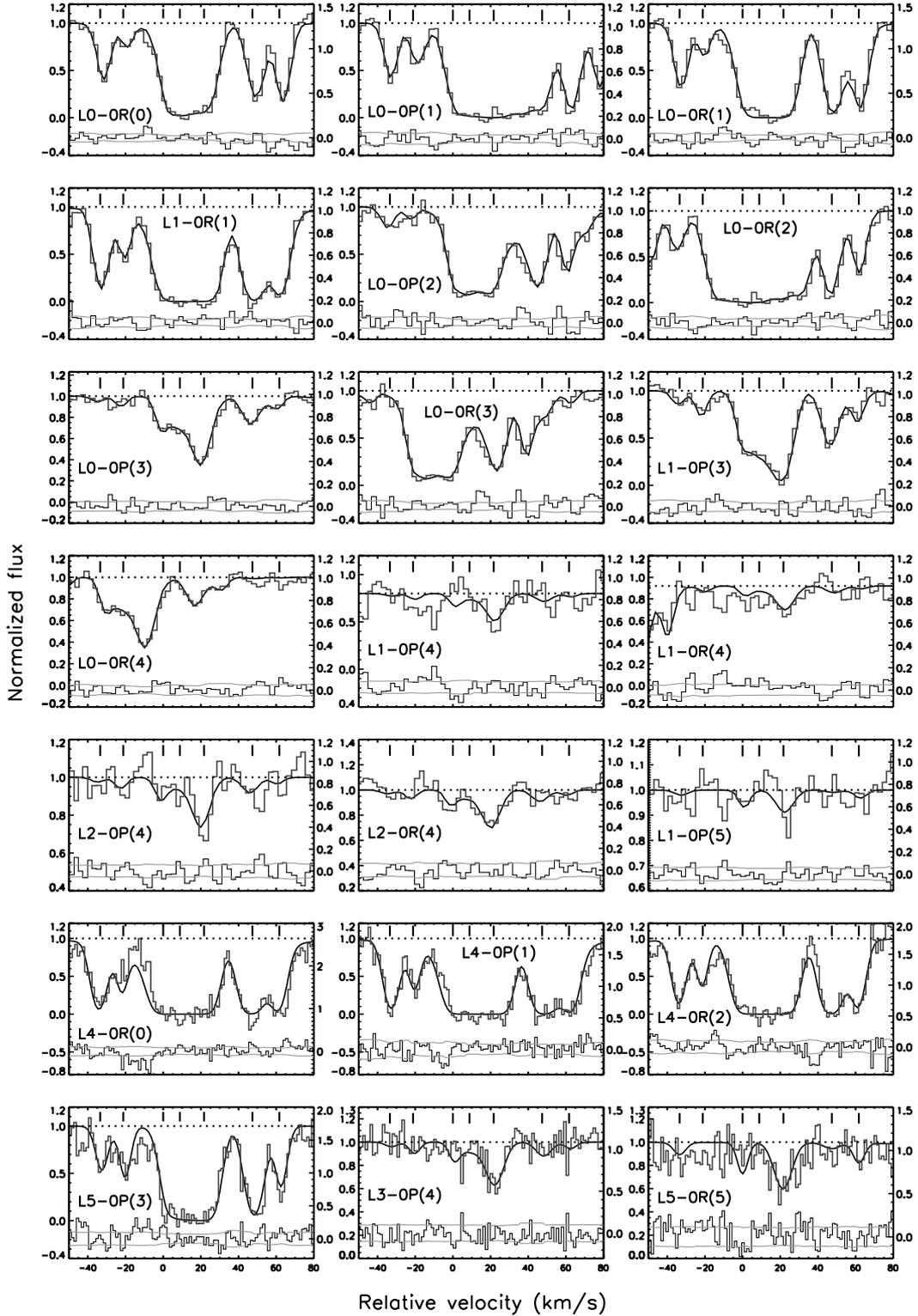}}
\caption{Multiple component Voigt profile fits to some of the H$_{2}$ absorption lines. The top five rows show transitions from the UVES spectrum, while the remaining two rows show the low-wavelength transitions covered only by the HIRES spectrum. The tick marks at the top indicate the positions of the Voigt profile components harbouring H$_2$ in velocity space. Zero velocity is defined at the H$_{2}$ component with $z_{abs}$ = 2.054509. A residual indicating the difference between the observed line profile and its Voigt profile fit is shown at the bottom of each panel. The grey lines indicate the 1-$\sigma$ range in flux allowed by the normalized error spectrum. The scale of the residual plot is shown on the right-hand vertical axis.}
\label{fig:voigt_h2}
\end{figure*}

\setlength{\parindent}{2ex}
Some of the H$_{2}$ absorption features suffer from saturation and some others are blended with either metal transitions or Lyman-$\alpha$ forest absorption. We perform Voigt profile fitting using the clean H$_{2}$ transitions covered by the UVES spectrum and additional lines from the lower signal-to-noise ratio HIRES spectrum. We constrain column densities from these lines, with $z_{abs}$ and \textit{b} linked for the lowest six rotational levels of the ground vibrational state. Upper limits on column density are determined by the method discussed in Section \ref{ssec:mets}, wherever Voigt profile analysis yields a measurement with high uncertainty. We determine log[\textit{N}(H$_{2}$)(cm$^{-2}$)] = 17.99$\pm$0.05 for the entire DLA. In comparison, \citet{Jorgenson2010} derived log[\textit{N}(H$_{2}$)(cm$^{-2}$)] = 18.20. As per their analysis, the H$_{2}$ absorption features comprise of six components. Our seven-component H$_{2}$ fit, however, is consistent with our analysis of the \ion{C}{i} absorption lines (discussed in Section \ref{ss:carb}). On the contrary, \citet{Jorgenson2010} require a nine-component fit for the \ion{C}{i} features, and these component positions do not agree as well with their H$_{2}$ component locations. Fig. \ref{fig:voigt_h2} shows the Voigt profile fits to some of the H$_{2}$ features, while Table \ref{tab:coldens_h2} lists the component-wise and level-wise column densities. \par
\begin{figure}
\centering{\includegraphics[width=\columnwidth, height=14 cm, keepaspectratio]{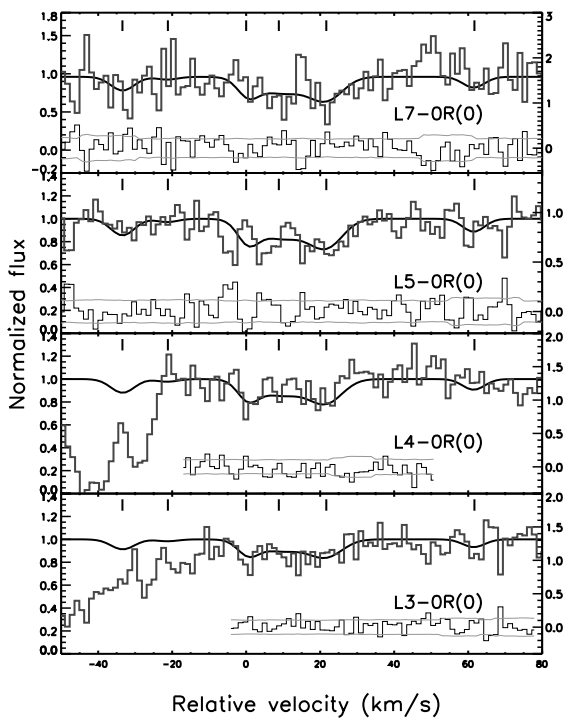}}
\caption{Multiple component Voigt profile fits to the HD absorption lines. The transitions used to perform this fit are covered only by the HIRES spectrum. Only the clean portions of the transitions in the lower two panels were used to perform profile fitting. The tick marks at the top indicate the positions of the Voigt profile components hosting HD in velocity space. Zero velocity is defined at the HD component with $z_{abs}$ = 2.054509. A residual indicating the difference between the observed line profile and its Voigt profile fit is shown at the bottom of each panel. The grey lines indicate the 1-$\sigma$ range in flux allowed by the normalized error spectrum. The scale of the residual plot is shown on the right-hand vertical axis.} 
\label{fig:voigt_hd}
\end{figure}

\setlength{\parindent}{2ex}
In addition to H$_{2}$, we also search for HD absorption. Most of the strong absorption features of HD lie blueward of the wavelength range covered by the UVES spectrum. We detect possible absorption features of HD at the expected locations of the ground rotational level J = 0. No absorption feature corresponding to the J = 1 rotational level is conspicuous. We use transitions covered by the HIRES spectrum to perform Voigt profile fitting and obtain component-wise column densities for HD (0). Possible HD absorption is seen to arise from 6 of the 7 H$_2$ components. The redshift and Doppler parameter are set to the values from the H$_2$ profile fit, while determining the component-wise column densities. We determine log[\textit{N}(HD (0))(cm$^{-2}$)] = 14.28$\pm$0.08. The HD transitions along with the Voigt profile fits are shown in Fig. \ref{fig:voigt_hd} and the column densities are listed in Table \ref{tab:coldens_hd}. \par
\setlength{\parindent}{2ex}
The ratio HD/2H$_{2}$ is used to constrain the D/H ratio in DLAs, as absorption features of \ion{D}{i} are usually undetected due to the stronger \ion{H}{i} absorption lines. The column densities of H$_{2}$ and HD in this DLA yield, log (HD/2H$_2$) -4.01$\pm$0.13, which is higher than the primordial D/H ratio of -4.59 (log-scale) determined through studies of the cosmic microwave background \citep{Komatsu2011}. High values of HD/2H$_{2}$ have been previously observed in high-redshift DLAs \citep{Ivanchik2010, Balashev2010, Ivanchik2015, Klimenko2015b}.

\begin{table*}
\setlength{\tabcolsep}{4pt}
 \centering
  \caption{Component-wise logarithmic column densities of H$_2$ in different rotational levels (in log cm$^{-2}$)}
  \label{tab:coldens_h2}
  \begin{tabular}{@{}cccccccccc@{}}
  \hline
   Component & Redshift & \textit{b} (km s$^{-1}$) & H$_2$ (0) & H$_2$ (1) & H$_2$ (2) & H$_2$ (3) & H$_2$ (4) & H$_2$ (5) & Total H$_2$ \\
 \hline
 4 & 2.054168$\pm$0.000001 & 2.5$\pm$0.1 & 15.35$\pm$0.05 & 15.74$\pm$0.05 & 15.03$\pm$0.06 & 14.30$\pm$0.07 & 13.17$\pm$0.36 & $\leq$ 13.67 & 15.96$\pm$0.03 \\
 5 & 2.054293$\pm$0.000002 & 1.5$\pm$0.2 & 14.77$\pm$0.10 & 15.15$\pm$0.08 & 14.79$\pm$0.12 & 14.62$\pm$0.08 & 13.55$\pm$0.18 & $\leq$ 13.90 & 15.49$\pm$0.05 \\
 7 & 2.054509$\pm$0.000004 & 0.9$\pm$0.2 & 17.38$\pm$0.11 & 17.57$\pm$0.11 & $\leq$ 15.90 & 15.24$\pm$0.21 & 13.77$\pm$0.15 & 13.92$\pm$0.22 & 17.79$\pm$0.08 \\ 
 8 & 2.054599$\pm$0.000007 & 7.5$\pm$0.5 & 16.06$\pm$0.07 & 16.54$\pm$0.12 & 16.29$\pm$0.04 & 15.36$\pm$0.04 & 13.78$\pm$0.15 & $\leq$ 13.91 & 16.83$\pm$0.06 \\ 
 9 & 2.054727$\pm$0.000003 & 5.0$\pm$0.2 & 15.76$\pm$0.06 & 16.60$\pm$0.09 & 16.07$\pm$0.06 & 15.74$\pm$0.03 & 14.37$\pm$0.04 & 14.14$\pm$0.09 & 16.80$\pm$0.06 \\
 11 & 2.054991$\pm$0.000001 & 4.2$\pm$0.1 & 15.57$\pm$0.03 & 16.18$\pm$0.04 & 15.66$\pm$0.06 & 15.13$\pm$0.02 & 13.77$\pm$0.12 & $\leq$ 13.88 & 16.39$\pm$0.03 \\
 13 & 2.055137$\pm$0.000001 & 1.9$\pm$0.2 & 16.43$\pm$0.17 & 17.20$\pm$0.11 & 16.03$\pm$0.11 & 14.69$\pm$0.06 & 13.29$\pm$0.30 & 13.55$\pm$0.30 & 17.29$\pm$0.09 \\
 \hline
 Total & - & - & 17.46$\pm$0.09 & 17.80$\pm$0.07 & $\leq$ 16.75 & 16.08$\pm$0.03 & 14.68$\pm$0.04 & $\leq$ 14.73 & 17.99$\pm$0.05 \\
\hline
\end{tabular}
\end{table*}

\begin{table}
 \centering
  \caption{Component-wise logarithmic column densities of HD (0) (in log cm$^{-2}$)}
  \label{tab:coldens_hd}
  \begin{tabular}{@{}ccc@{}}
  \hline
   Component & \textit{b} (km s$^{-1}$) & HD (0) \\
 \hline
 4 & 2.5 & 13.59$\pm$0.20 \\
 5 & 1.5 & $\leq$ 13.36 \\
 7 & 0.9 & 13.35$\pm$0.29 \\ 
 8 & 7.5 & 13.76$\pm$0.12 \\ 
 9 & 5.0 & 13.77$\pm$0.10  \\
 13 & 1.9 & 13.10$\pm$0.45 \\
 \hline
 Total & - & 14.28$\pm$0.08 \\
\hline
\end{tabular}
\end{table}

\subsection{Neutral carbon \& neutral chlorine}
\label{ss:carb}

\ion{C}{i}  absorption is seen to arise from the seven H$_{2}$ components, in agreement with our understanding that \ion{C}{i} is usually associated with regions harbouring H$_{2}$. Our analysis shows the \ion{C}{i} absorption to mostly coincide with the presence of H$_{2}$. Only in the case of metal components 11 and 12, it is seen that the H$_{2}$ absorption occurs in a region nearer to component 11, while the \ion{C}{i} features arise from a region near component 12. For all calculations throughout this paper, we assume components 11 and 12 to constitute a single cloud of gas comprising of both H$_{2}$ and \ion{C}{i} absorption features. In comparison, \citet{Jorgenson2010} reproduce the observed \ion{C}{i} lines with a nine-component fit. Some of the \ion{C}{i} components are thus, not associated with molecular absorption.\par
\begin{figure}
\centering{\includegraphics[width=\columnwidth, height=14 cm, keepaspectratio]{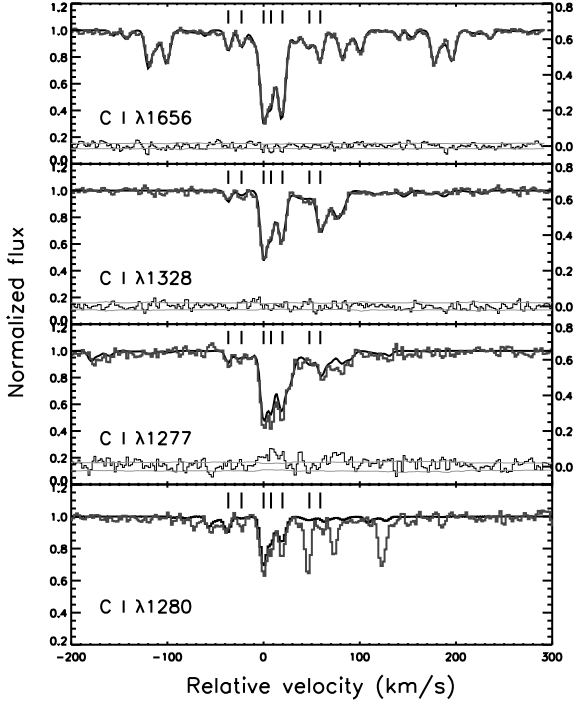}}
\caption{Multiple component Voigt profile fits to the \ion{C}{i} transitions at 1656 {\AA}, 1328 {\AA} \& 1277 {\AA} are shown here. In the lowest panel, the fit is laid over the observed profile of the 1280 {\AA} transition, which is a blended feature. The tick marks at the top indicate the positions of the Voigt profile components harbouring \ion{C}{i} (with reference to the strongest \ion{C}{i*} transition) in velocity space. Zero velocity is defined at the \ion{C}{i} component with $z_{abs}$ = 2.054529. A residual indicating the difference between the observed line profile and its Voigt profile fit is shown at the bottom of each panel. The grey lines indicate the 1-$\sigma$ range in flux allowed by the normalized error spectrum. The scale of the residual plot is shown on the right-hand vertical axis.} 
\label{fig:voigt_ci}
\end{figure}

\setlength{\parindent}{2ex}
We fit the \ion{C}{i} features at $\lambda$1656, $\lambda$1328, $\lambda$1277, $\lambda$1270 and $\lambda$1155. We determine column densities of \ion{C}{i*}, \ion{C}{i**} and \ion{C}{i***} by fitting their line profiles simultaneously with $z_{abs}$ and \textit{b} linked together. The feature at $\lambda$1560 is not covered by the spectrum, while the $\lambda$1280 feature is blended with the \ion{C}{i} $\lambda$1656 transition of the \ion{Mg}{ii} absorber. The other weaker lines too, are blended and are excluded from the fit. \citet{Jorgenson2010} observe the $\lambda$1560 transition too, but the $\lambda$1155 line is not included in their analysis. We show our Voigt profile fits in Fig. \ref{fig:voigt_ci}, and list the column densities in Table \ref{tab:coldens_ci}. \par
\begin{figure}
\centering{\includegraphics[width=\columnwidth, keepaspectratio]{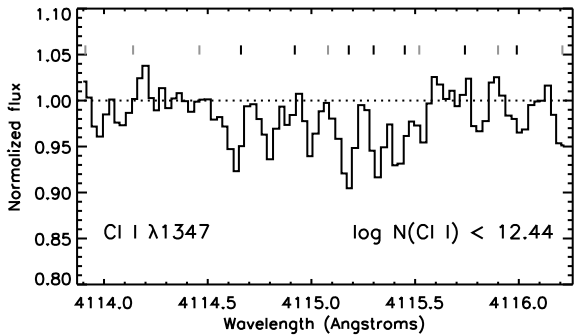}}
\caption{The \ion{Cl}{i} $\lambda$1347 absorption feature is shown here. log[\textit{N}(\ion{Cl}{i})(cm$^{-2}$)] is found to be $\leq$ 12.44. The tick marks at the top indicate the positions of the components in velocity space. H$_{2}$ and non-H$_{2}$ components are shown by black and grey ticks respectively.}
\label{fig:cl1}
\end{figure}

\setlength{\parindent}{2ex}
\ion{Cl}{i} is known to arise in cool gas harbouring H$_{2}$ \citep {Noterdaeme2007a, Noterdaeme2015a, Balashev2015}. We detect a very weak $\lambda$1347 transition of \ion{Cl}{i}. Over the expected wavelength range for \ion{Cl}{i} absorption corresponding to the 14 components of the DLA, we derive an upper limit on the column density of neutral chlorine using the method outlined in Section \ref{ssec:mets}. We find log[\textit{N}(\ion{Cl}{i})(cm$^{-2}$)] $\leq$ 12.44. This is in close agreement with the relation, \textit{N}(\ion{Cl}{i}) $\sim$ $1.5\times10^{-6}$ \textit{N}(H$_{2}$), deduced by \citet{Balashev2015} from a study of 18 high-redshift H$_{2}$-DLAs. Fig. \ref{fig:cl1} shows the expected spectral range for the \ion{Cl}{i} $\lambda$1347 transition.

\begin{table}
\setlength{\tabcolsep}{4pt}
 \centering
  \caption{Component-wise logarithmic column densities of the \ion{C}{i} fine structure levels (in log cm$^{-2}$)}
  \label{tab:coldens_ci}
  \begin{tabular}{@{}ccccc@{}}
  \hline
   Component & \textit{b} (km s$^{-1}$) & \ion{C}{i*} & \ion{C}{i**} & \ion{C}{i***} \\
 \hline
 4 & 0.7$\pm$0.3 & 12.42$\pm$0.09 & 12.10$\pm$0.08 & 11.92$\pm$0.14 \\
 5 & 3.7$\pm$0.6 & 12.13$\pm$0.07 & 12.40$\pm$0.04 & 12.29$\pm$0.04 \\
 7 & 1.9$\pm$0.1 & 13.53$\pm$0.03 & 13.06$\pm$0.01 & 11.87$\pm$0.14 \\ 
 8 & 1.9$\pm$0.1 & 13.16$\pm$0.03 & 12.64$\pm$0.03 & 11.80$\pm$0.19 \\ 
 9 & 3.4$\pm$0.1 & 13.25$\pm$0.01 & 13.03$\pm$0.01 & 12.27$\pm$0.04 \\
 12 & 6.5$\pm$1.1 & 12.42$\pm$0.04 & 11.93$\pm$0.14 & 11.75$\pm$0.23 \\
 13 & 0.4$\pm$0.1 & 12.75$\pm$0.18 & 12.06$\pm$0.08 & 11.88$\pm$0.11 \\
 \hline
 Total & - & 13.89$\pm$0.02 & 13.51$\pm$0.01 & 12.86$\pm$0.04\\
\hline
\end{tabular}
\end{table}

\section{Estimation of physical properties of the DLA}
\label{sec:phy_prop}
Some of the physical properties of the DLA can be estimated by performing simple calculations involving the observed column densities of various species. In the following sub-sections, we obtain constraints on the metallicity, dust content, temperature, density and radiation field of this DLA.

\subsection{Metallicity \& dust content}
\label{ssec:met_kap}

We calculate the average metallicity of the DLA using the column densities of \ion{S}{ii} and \ion{H}{i} in the following relation, \par
\begin{equation}
    Z = [X/H] = log(\frac{N(X)}{N(H I)})-log(\frac{N(X)}{N(H I)})_\odot \; \; \;.
	\label{eq:z}
\end{equation}
In DLAs, sulphur is almost entirely present in the singly ionized state. It is also known to occur chiefly in the gas phase as it is largely unaffected by depletion. This permits us to use X = \ion{S}{ii} in equation (\ref{eq:z}). We refer to \citet {Asplund2009} for the solar abundances. It is impossible to compute the metallicity of individual components as the component-wise distribution of \ion{H}{i} is unknown. We find [S/H] = -0.52$\pm$0.06 for the entire DLA, implying an overall metallicity of 0.3 $Z_\odot$. \par
\setlength{\parindent}{2ex}
The presence of H$_{2}$ in the DLA points to the likelihood that grain surface processes play a crucial role \citep*{Cazaux2002, Cazaux2004a, Cazaux2004b}. The dust content of a system is expressed as a fraction of the dust content of the Milky Way, and is known as the dust-to-gas ratio $\kappa$. The following relation from \citet* {Wolfe2003a} is used to calculate $\kappa$,
\begin{equation}
    \kappa = 10^{[X/H]}(1-10^{[Y/X]}) \; \; \;.
	\label{eq:kappa}
\end{equation}
Here, X is a volatile species which is undepleted on grain surface, and Y is a refractory species. In our calculations, we use X = Zn and Y = Fe. We first determine the dust content of each component of the DLA by using the component-wise values of [Fe/Zn]. As [Zn/H] can only be calculated for the entire DLA, its mean value of -0.32 is used while estimating $\kappa$ for each individual component. $\kappa$ lies between 0.20 and 0.50 for most of the components. As the ratio Fe/Zn is super-solar in components 3 and 4, negative $\kappa$ values of -0.61 and -0.06 are obtained for these respective components. The mean dust-to-gas ratio over the entire DLA is 0.34$\pm$0.07, compatible with the metal enrichment of the system.

\subsection{Temperature \& density in the molecular region}
\label{ssec:temp_den}
\begin{figure*}
\centering{\includegraphics[width=\textwidth, keepaspectratio]{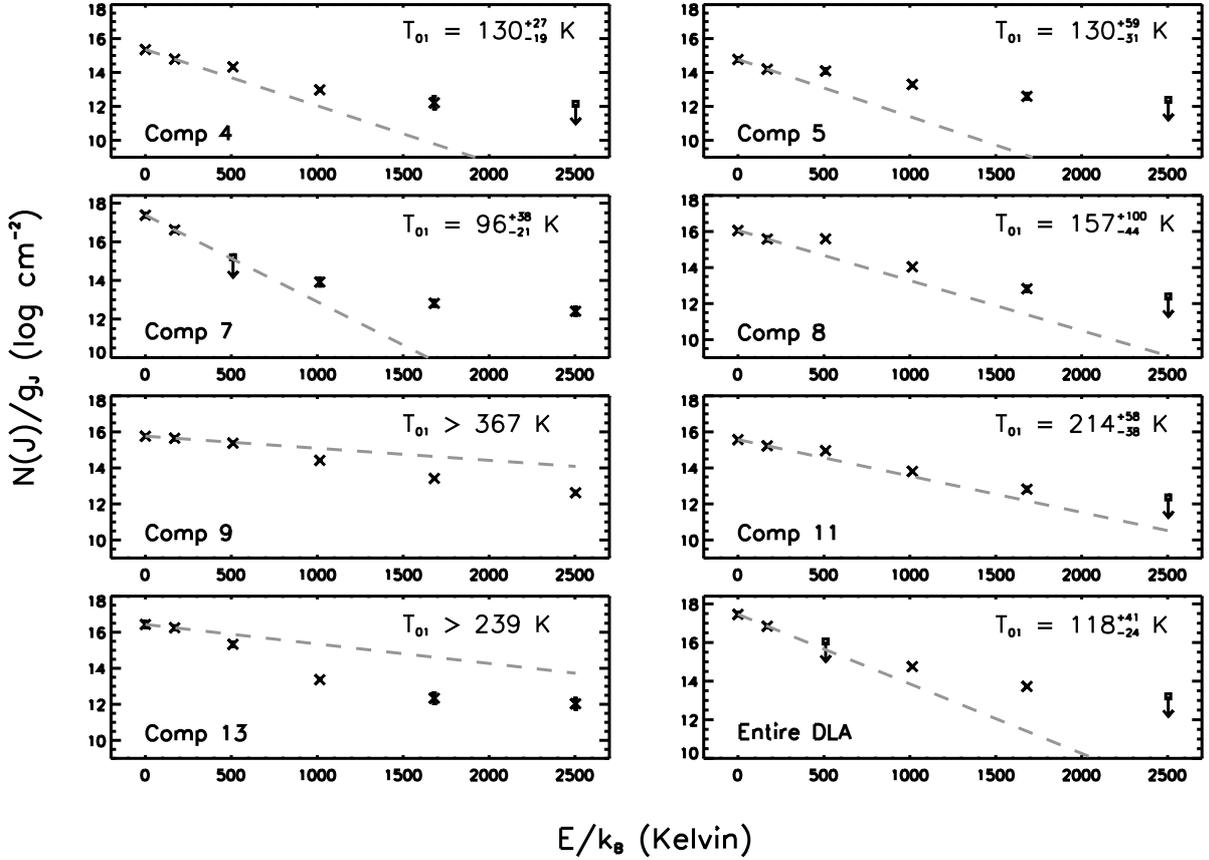}}
\caption{H$_{2}$ excitation diagrams with the estimated excitation temperature $T_{01}$ for the individual H$_{2}$ components. The bottom right panel shows the overall excitation diagram for the DLA. The grey dashed lines correspond to the excitation inferred from the H$_{2}$ (0) \& H$_{2}$ (1) rotational levels. $T_{01}$ is calculated from the observed column densities of these two levels. Barring components 8 and 11, $T_{01}$ does not explain the population of the higher rotational levels. This indicates the involvement of other excitation processes.}
\label{fig:h2_exc}
\end{figure*}

The excitation temperature corresponding to different rotational levels of H$_{2}$ can be calculated using
\begin{equation}
    \frac{N_j}{N_i} = {\frac{g_j}{g_i}}\cdot \exp{(\frac{-E_{ij}}{k_B\cdot T_{ij}})}
	\label{eq:boltzmann}
\end{equation}
where $N_i$ and $N_j$ are the column densities of the rotational levels \textit{i} and \textit{j}, while $g_i$ and $g_j$ are the respective statistical weights. $E_{ij}$ is the energy difference between the levels, $T_{ij}$ is the corresponding temperature difference and $k_B$ is the Boltzmann constant. We use this equation to determine $T_{01}$, the excitation temperature for the rotational levels J = 0 and 1. If these levels are thermalized and the gas is well-shielded, $T_{01} \sim T_{kin}$, where $T_{kin}$ is the kinetic temperature of the gas \citep*{Roy2006}. In Fig. \ref{fig:h2_exc}, we plot the excitation diagrams for each of the H$_{2}$ components. The bottom right panel shows the excitation diagram for the entire DLA.\par
\setlength{\parindent}{2ex}
We determine $T_{01}$ for five H$_{2}$ components (components 4, 5, 7, 8 and 11) to be in the range 75--272 K. We obtain lower limits on $T_{01}$ for the other two H$_{2}$ components. Components 9 and 13 have $T_{01}$ higher than 367 K and 239 K respectively. The coolest H$_{2}$ component is component 7, with $T_{01}$ = $96^{+38}_{-21}$ K; while components 9, 11 and 13 are the warmest, having $T_{01}$ in excess of 200 K. In comparison, the median $T_{01}$ in high-redshift H$_{2}$-DLAs is 143$\pm$19 K \citep[and references therein]{Muzahid2015}. We also calculate $T_{01}$ for the entire DLA by adding the level-wise column densities over the various components. This yields $T_{01}$ = $118^{+41}_{-24}$ K. Such a calculation is unable to trace the small clumps of gas within the DLA, which exist at temperatures significantly different from the mean value.\par
\citet{Muzahid2015} study the dependence of $T_{01}$ on the total H$_{2}$ column density in the Milky Way, and in DLAs at low and high redshift. $T_{01}$ is almost always less than 100 K, for log[\textit{N}(H$_{2}$)(cm$^{-2}$)] > 19. At column densities lower than this, $T_{01}$ varies over a wide range of temperatures between 25 and 600 K, and does not show a particular trend. The H$_{2}$ column densities in the molecular components of the DLA under present study, lie in the range 15 < log[\textit{N}(H$_{2}$)(cm$^{-2}$)] < 18. Most of the total H$_{2}$ is present in components 7 and 13, both of which exhibit log[\textit{N}(H$_{2}$)(cm$^{-2}$)] > 17. Components 4 and 5 have the least H$_{2}$, with log[\textit{N}(H$_{2}$)(cm$^{-2}$)] < 16. $T_{01}$ for all the components lies in the range 75--400 K, which is consistent with literature. We note that $T_{01}$ is mostly higher for the components with 16 < log[\textit{N}(H$_{2}$)(cm$^{-2}$)] < 17.4. For most components, the higher rotational levels of H$_{2}$ (J $\geq$ 2) follow a different temperature distribution from the lower levels. Due to insufficient shielding, the two weakest H$_{2}$ components are not likely to be thermalized, and hence $T_{01}$ cannot be used to trace the kinetic temperature of the gas.

\begin{figure*}
\centering{\includegraphics[width=\textwidth, height=17 cm, keepaspectratio]{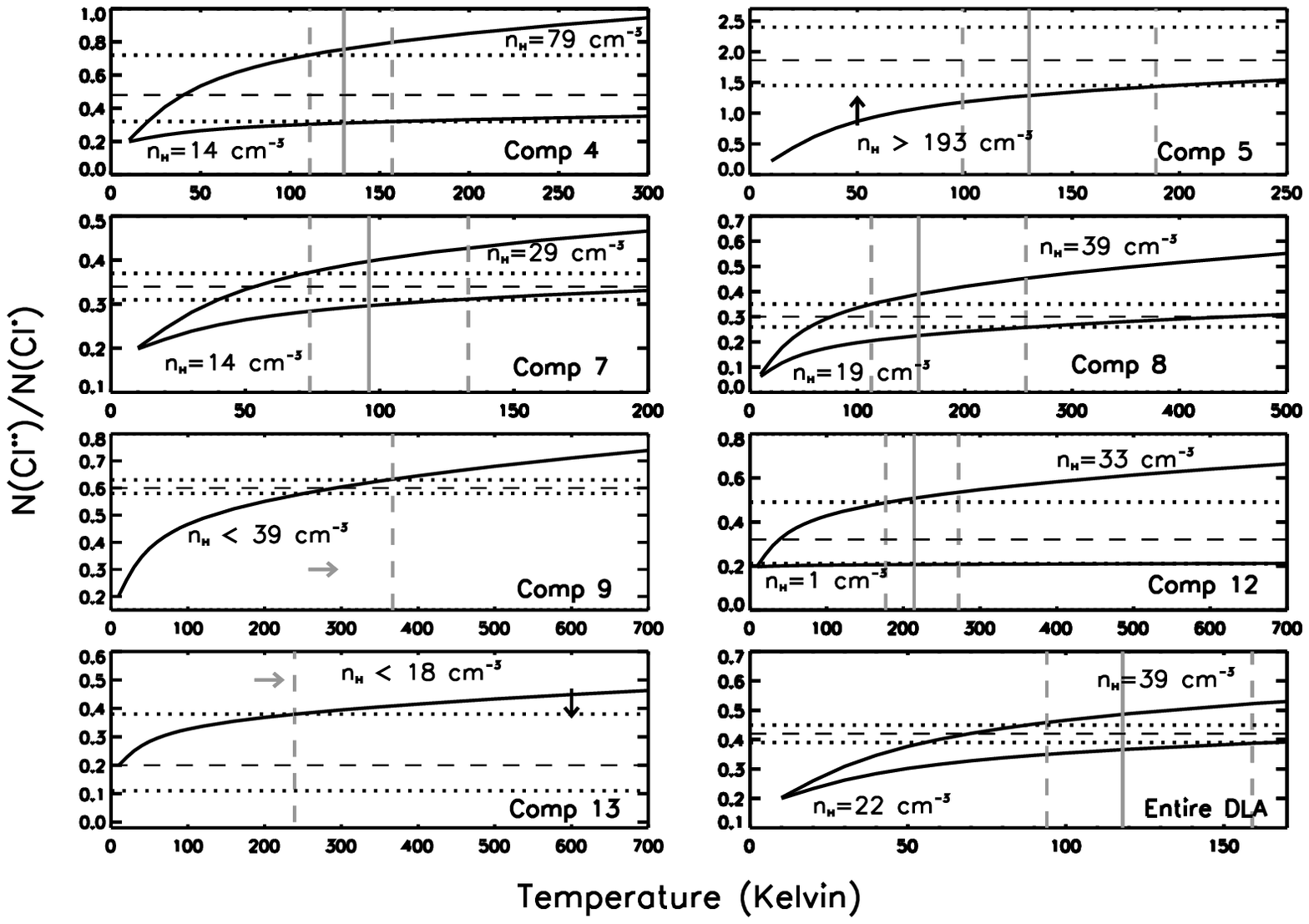}}
\caption{The observed \textit{N}(\ion{C}{i**})/\textit{N}(\ion{C}{i*}) is used to constrain the atomic hydrogen density, $n_\mathrm{H}$ in each of the \ion{C}{i} (also H$_{2}$) components. We solve the rate equations and estimate $n_\mathrm{H}$ in the molecular regions, assuming the temperature to be equal to $T_{01}$, as determined from the corresponding observations of the H$_{2}$ (0) \& H$_{2}$ (1) rotational levels. The horizontal black dashed lines indicate the observed \textit{N}(\ion{C}{i**})/\textit{N}(\ion{C}{i*}), with the dotted lines representing the 1-$\sigma$ error range. The vertical grey lines indicate $T_{01}$, with the dashed lines representing the error range. In the bottom right panel, $n_\mathrm{H}$ is constrained assuming the entire DLA to be a single cloud.}
\label{fig:ci_den}
\end{figure*}

\begin{figure*}
\centering{\includegraphics[width=\textwidth, height=17 cm, keepaspectratio]{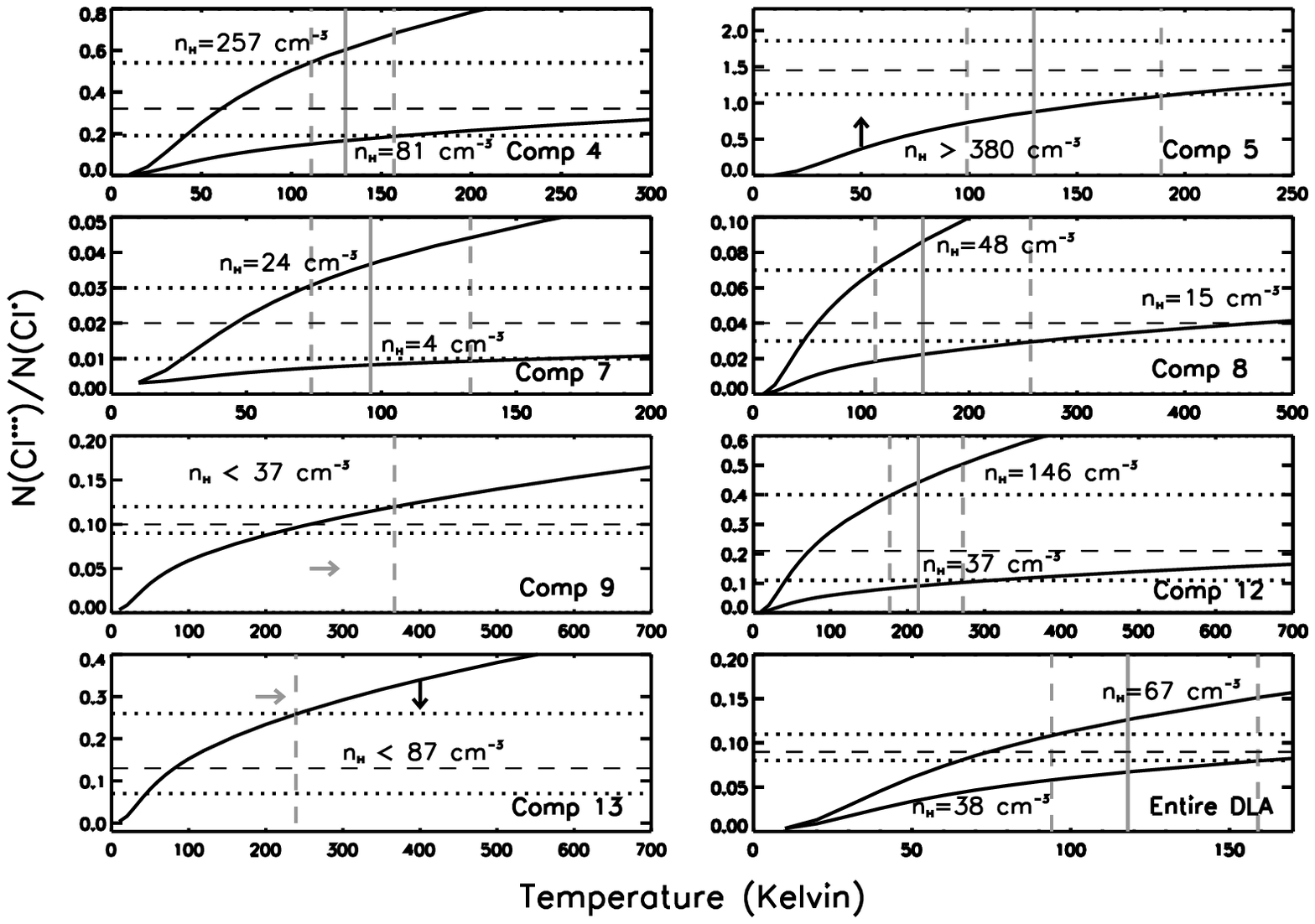}}
\caption{The observed \textit{N}(\ion{C}{i***})/\textit{N}(\ion{C}{i*}) is used to constrain the atomic hydrogen density, $n_\mathrm{H}$ in each of the \ion{C}{i} (also H$_{2}$) components. We solve the rate equations and estimate $n_\mathrm{H}$ in the molecular regions, assuming the temperature to be equal to $T_{01}$, as determined from the corresponding observations of the H$_{2}$ (0) \& H$_{2}$ (1) rotational levels. The horizontal black dashed lines indicate the observed \textit{N}(\ion{C}{i***})/\textit{N}(\ion{C}{i*}), with the dotted lines representing the 1-$\sigma$ error range. The vertical grey lines indicate $T_{01}$, with the dashed lines representing the error range. In the bottom right panel, $n_\mathrm{H}$ is constrained assuming the entire DLA to be a single cloud.}
\label{fig:cistar_den}
\end{figure*}

We also get an estimate of the hydrogen density in each of the seven components by using the rate equations corresponding to the \ion{C}{i} fine structure levels. Applying the physics of the three-level atom, we can write the following relations,
\begin{equation}
    \frac{\textit{N}(\ion{C}{i**})}{\textit{N}(\ion{C}{i*})} = \frac{R_{01}R_{20}+R_{01}R_{21}+R_{21}R_{02}}{R_{10}R_{20}+R_{10}R_{21}+R_{12}R_{20}}
	\label{eq:ci_rate}
\end{equation}
and
\begin{equation}
    \frac{\textit{N}(\ion{C}{i***})}{\textit{N}(\ion{C}{i*})} = \frac{R_{02}R_{10}+R_{02}R_{12}+R_{12}R_{01}}{R_{10}R_{20}+R_{10}R_{21}+R_{12}R_{20}} \; \; \;.
	\label{eq:ci_rate}
\end{equation}
The subscripts 0, 1 and 2 indicate the states \ion{C}{i*}, \ion{C}{i**} and \ion{C}{i***} respectively. The rates $R_{ij}$ refer to upward transitions when \textit{i} < \textit{j}, and downward transitions when \textit{i} > \textit{j}. These reaction rates constitute contributions from radiative processes -- UV pumping, excitation by CMB photons, spontaneous emission -- as well as collisional excitation and de-excitation \citep[refer to][for the rate equations]{Draine2011}. We consider all these processes in our calculations. Of the various species present in neutral gas, atomic hydrogen is the dominant collision partner for carbon atoms. As the gas becomes molecular, H$_2$ becomes a major collision partner. As a simplification, we assume that atomic hydrogen is the only collision partner and use the corresponding rate coefficients from \citet* {Launay1977}. We lay constraints on the gas temperature according to $T_{01}$ derived from the observed H$_{2}$ (0) and H$_{2}$ (1) levels; and on \textit{N}(\ion{C}{i**})/\textit{N}(\ion{C}{i*}) (or \textit{N}(\ion{C}{i***})/\textit{N}(\ion{C}{i*})), by the ratio of the observed column densities of the two concerned levels. The 1-$\sigma$ uncertainties in the observed column densities, as obtained through Voigt profile fitting, are also taken into consideration here. This helps us to get an estimate of the atomic hydrogen density of the gas $n_\mathrm{H}$, which we plot in Figs. \ref{fig:ci_den} and \ref{fig:cistar_den}. We consider the \ion{C}{i} in component 12 to be associated with the H$_{2}$ in component 11, and use the $T_{01}$ constraints from this H$_{2}$ component. \par
The value of $n_\mathrm{H}$ constrained for the various components through \textit{N}(\ion{C}{i**})/\textit{N}(\ion{C}{i*}) lies within the range of 1--80 cm$^{-3}$, while the density predictions from \textit{N}(\ion{C}{i***})/\textit{N}(\ion{C}{i*}) stretch over 4--260 cm$^{-3}$. However, the corresponding component-wise estimates from both ratios are mostly consistent. For component 5, the estimated $n_\mathrm{H}$ is higher than the density predicted for the other components. \textit{N}(\ion{C}{i**})/\textit{N}(\ion{C}{i*}) and \textit{N}(\ion{C}{i***})/\textit{N}(\ion{C}{i*}) yield estimates of $n_\mathrm{H}$ > 193 cm$^{-3}$ and > 380 cm$^{-3}$ respectively. In case of components 4 and 12, the density predictions from \textit{N}(\ion{C}{i**})/\textit{N}(\ion{C}{i*}) are lower than those from \textit{N}(\ion{C}{i***})/\textit{N}(\ion{C}{i*}). They lie in the ranges 14--79 cm$^{-3}$ and 81--257 cm$^{-3}$ respectively for component 4; and in the ranges 1--33 cm$^{-3}$ and 37--146 cm$^{-3}$ respectively for component 12. However, we note that the density estimates from both ratios are consistent at the 2-$\sigma$ level. 

\subsection{Radiation field}
\label{ssec:rad_fld_obs}
Singly ionized carbon has two fine structure levels. \ion{C}{ii*} represents the higher of these two energy levels. The observed column density of \ion{C}{ii*} can be used to estimate the strength of the radiation field incident on the DLA. Grain photoelectric emission and [\ion{C}{ii}] 158 $\umu$m emission are assumed to be the main sources of heating and cooling respectively in neutral gas. The cooling rate can thus, be calculated by considering the system to be in thermal equilibrium. Equation(\ref{eq:cool_rate}), described by \citet* {Pottasch1979}, follows from this and can be used to determine the cooling rate per hydrogen atom of gas in the DLA,
\begin{equation}
    l_c = \frac{h \nu \cdot \textit{N}(\ion{C}{ii*}) \cdot A_{21}}{\textit{N}(\ion{H}{i})} \ \mathrm{erg \ s^{-1} \ H^{-1}} \; \; \;.
	\label{eq:cool_rate}
\end{equation}
The [\ion{C}{ii}] 158 $\umu$m emission occurs due to the $^2P_{3/2} \rightarrow ^2P_{1/2}$ radiative transition, which has decay rate $A_{21}\sim 2.4 \times 10^{-6}$ s$^{-1}$ \citep{Pottasch1979}. From Voigt profile analysis, we have measured \textit{N}(\ion{H}{i}) and \textit{N}(\ion{C}{ii*}) to be 20.35 and 13.75 (log cm$^{-2}$) respectively for this DLA. This yields a cooling rate of $10^{-26.1}$ erg s$^{-1}$ H$^{-1}$. The typical cooling rate in the Galactic disc is $l^{MW}_{c}$ $\sim 10^{-25.2}$ erg s$^{-1}$ H$^{-1}$ \citep{Wolfe2003a}.  \par
\setlength{\parindent}{2ex}
\citet{Wolfe2008} studied the cooling rate in DLAs and reported bimodal distribution around the critical value, $l_c$ = $10^{-27.0}$ erg s$^{-1}$ H$^{-1}$. They found that ``low cool" DLAs (with cooling rate $\leq$ $l_c$) were strongly connected with \textit{in situ} star formation, while ``high cool" DLAs (with cooling rate > $l_c$) were not. ``High cool" DLAs are likely to be located in the peripheral regions of star-forming galaxies and are heated predominantly by the metagalactic background. We calculate here $l_c$ = $10^{-26.1}$ erg s$^{-1}$ H$^{-1}$, which puts this DLA in the ``high cool" population. \par
\setlength{\parindent}{2ex}
Following \citet {Noterdaeme2007b} and \citet {Albornoz2014} for further calculations, we consider the temperature, density and dust grain properties in the DLA to be similar to the Galaxy. The flux of ultraviolet radiation incident on the DLA ($F_{UV}$) can then be estimated relative to the ultraviolet flux in the Milky Way ($F^{MW}_{UV}$), through the following approximation.
\begin{equation}
    \frac{l_c}{l^{MW}_{c}} \sim \kappa \frac{F_{UV}}{F^{MW}_{UV}}
	\label{eq:uv_flux}
\end{equation}
Using $\kappa$ = 0.34, as derived from the observed [Fe/Zn], we find $F_{UV} \sim 0.35 \times F^{MW}_{UV}$.

\setlength{\parindent}{2ex}
We now proceed to construct individual numerical models of the H$_{2}$ components, around the range of parameter values obtained with simplified assumptions. These models help us gain more insight into the true physical environment of this DLA.

\section{Numerical models}
\label{sec:models}

We use the spectral synthesis code {\tiny CLOUDY} version c13.04\footnote{\url{http://www.nublado.org/}} \citep {Ferland1998, Ferland2013} to model the H$_{2}$ components, treating  the observed column densities of various atomic \& molecular species as constraints. \citet {Srianand2005a} and \citet {Shaw2016} have previously studied the physical conditions in DLAs through detailed numerical models. These models assume the DLA to have the plane parallel geometry of a photodissociation region \citep* {TH1985}. However, individual modelling of multiple components of a DLA has not yet been explored. This kind of study will enable us to understand the variation of physical properties within the DLA. \par
\setlength{\parindent}{2ex}
{\tiny CLOUDY} requires as input the geometry of the system, spectral shape and intensity of the ionizing radiation, and chemical composition and total hydrogen density of the gas. Here, total hydrogen density includes hydrogen in all its forms -- ionic, atomic and molecular. With this information for a given astrophysical system, various radiative and collisional processes are considered by {\tiny CLOUDY} to self-consistently determine the thermal, ionization and chemical balance at each point within the cloud, and predict respectively the intensities or column densities of the resultant emission or absorption lines. {\tiny CLOUDY} includes an extensive chemistry network and a comprehensive treatment of grain physics \citep {vanHoof2004}. Our models use the detailed micro-physics of the H$_{2}$ molecule incorporated in {\tiny CLOUDY} and discussed in \citet {Shaw2005}.    

\subsection{Modelling approach}
\label{ssec:approach}

Each H$_{2}$ component is modelled as a constant pressure cloud, similar to the approach discussed in \citet {Shaw2016}. The total pressure at each point in the cloud remains constant, though the contribution of individual pressure constituents to the total pressure may vary at different points in the cloud. Both faces of the cloud are illuminated with radiation. There are two sources of ionizing radiation in our models -- the metagalactic background which we adopt from \citet {Khaire2015a}, and the cosmic microwave background. This combined radiation field first passes through a slab of intervening gas corresponding to log[\textit{N}(\ion{H}{i})(cm$^{-2}$)] = 19, before it is incident on the cloud. As a consequence, the continuum is partially attenuated due to absorption by atomic hydrogen and helium. It is necessary to account for this attenuation as each of the H$_{2}$ components that we model is surrounded by other components of the DLA. We present a detailed discussion on the nature of the radiation field in Section \ref{ssec:rad}. \par
\setlength{\parindent}{2ex}
As an initial guess for each of our component models, we use the hydrogen density estimated from the rate equations of the \ion{C}{i} fine structure levels, and the dust-to-gas ratio calculated from the observed [Fe/Zn]. We start with a metallicity of 0.3 $Z_\odot$, as deduced from [S/H] for the entire DLA. The solar abundances used in {\tiny CLOUDY} are from \citet{Grevesse2010}. We consider cosmic ray ionization in our models, initially equivalent to a rate of $2\times10^{-16}$ s$^{-1}$ (-15.7 in log-scale), as found for particular Galactic sightlines by \citet {Indriolo2007}. The study of cosmic ray ionization rates is an active area of research. A discussion on this, in the context of Galactic and extragalactic environments, is provided in Section \ref{sssec:tdpcr}. Our calculations stop at a suitable value of \textit{N}(H$_{2}$) for each of the respective components. We constrain our models to match the observed rotational level population of H$_{2}$ and fine structure levels of \ion{C}{i}. The physical parameters that are varied include the total hydrogen density, metallicity, dust-to-gas ratio, dust grain sizes and cosmic ray ionization rate. However, we are unable to reproduce the observed relative strengths of the \ion{C}{i} fine structure levels and \ion{C}{ii*} without varying the carbon abundance. This also influences the H$_{2}$ rotational level population. Besides this, we note that the abundances of Si, Fe and C are closely related to various heating and cooling processes, and therefore affect the column densities predicted by the model. The abundances of other metals though, have minimal effect on the column density predictions. Thus, we vary the abundances of Si, Fe and C to produce the optimal model. We try to match the H$_{2}$, \ion{C}{i}, \ion{C}{ii*}, \ion{S}{ii}, \ion{Si}{ii} and \ion{Fe}{ii} column densities to within 0.3 dex of the corresponding observed values. \par
\setlength{\parindent}{2ex}
An additional constraint on our models is imposed by the observed \textit{N}(\ion{H}{i}). As Voigt profile analysis only allows measurement of the total \textit{N}(\ion{H}{i}) in the DLA, component-wise distribution of \ion{H}{i} can be understood only through the results of our {\tiny CLOUDY} models. We know that the DLA has seven metal components without H$_{2}$, which also contribute to the total \ion{H}{i} budget of the DLA. But our models can only account for the \ion{H}{i} associated with H$_{2}$. The sum of the predicted \textit{N}(\ion{H}{i}) from the individual models must therefore, be lower than the observed \textit{N}(\ion{H}{i}). \par
\setlength{\parindent}{2ex}
The constraint on total \textit{N}(\ion{H}{i}) enables us to determine the dust grain sizes in the DLA. Models with smaller grains, similar to those described in \citet {Shaw2016}, yield \textit{N}(\ion{H}{i}) consistent with observation. These grains have radii in the range 0.0025-0.125 $\mu$m, which are half the sizes of grains in the local interstellar medium (ISM), and follow the same power law as the MRN size distribution \citep* {MRN1977}. An alternate possible scenario was presented by \citet{Shaw2016}. For the 3 DLA sightlines modelled using smaller dust grains, it was found that porous ISM-sized dust grains composed of silicate and amorphous carbon with porosity (vacuum content by volume) 0.55, predict similar column densities. This behaviour is exhibited by our component models for this DLA too. All observed column densities agree within 0.05 dex when the models with smaller compact grains are compared with the corresponding ones having ISM-sized porous grains. For the sake of further discussion in this paper, we refer only to the models with the smaller-sized compact dust grains.

\subsection{Model for component 8}
\label{ssec:8_model}

Among all the components, the highest H$_{2}$ content is exhibited by component 7 (log \textit{N}(H$_{2}$) = 17.79 cm$^{-2}$), while component 5 has the least H$_{2}$ (log \textit{N}(H$_{2}$) = 15.49 cm$^{-2}$). Component 8 has an intermediate H$_{2}$ column density, log[\textit{N}(H$_{2}$)(cm$^{-2}$)] = 16.83$\pm$0.06. The model for component 8 is discussed here in detail. The other components have been modelled along similar lines. \par
\setlength{\parindent}{2ex}
The total hydrogen density of the gas at each illuminated face of the cloud is $\sim$ 31.6 cm$^{-3}$ (1.50 in log-scale). The {\tiny CLOUDY} calculations predict both the total and atomic hydrogen density at different depths into the cloud. In the inner molecular region, we find that the density increases to 40 cm$^{-3}$. Of this, the atomic hydrogen density contributes 38 cm$^{-3}$. This is in good agreement with the $n_\mathrm{H}$ range of 15--48 cm$^{-3}$ obtained through the \ion{C}{i} rate equations. The metallicity for this component ([S/H]) is found to be 0.45 $Z_\odot$, higher than the overall metallicity of the DLA. We also scale the elemental abundances of Si, Fe and C to match the observed column densities of \ion{Si}{ii}, \ion{Fe}{ii}, \ion{C}{i*}, \ion{C}{i**}, \ion{C}{i***} and \ion{C}{ii*}. The dust-to-gas ratio for this component is 0.50, and lies close to the value of 0.42$\pm$0.07 obtained using the observed [Fe/Zn]. We require cosmic ray ionization rate of 10$^{-15.30}$ s$^{-1}$ to produce the required excitation for the higher rotational levels of H$_{2}$. This is 2.5 times the value of 10$^{-15.7}$ s$^{-1}$ found by \citet {Indriolo2007} for particular sightlines within the Galaxy. To include the effects of micro-turbulence, we use a Doppler parameter of 6.5 km s$^{-1}$, slightly lower than the value 7.5$\pm$0.5 km s$^{-1}$ derived from Voigt profile analysis. This is the highest value for the Doppler parameter seen among the H$_{2}$ components. All the input parameters of our {\tiny CLOUDY} model are listed in Table \ref{tab:paras_8}. We compare the observed and predicted column densities in Table \ref{tab:model_8}.\par
\begin{figure*}
\centering{\includegraphics[width=\textwidth, height=17 cm, keepaspectratio]{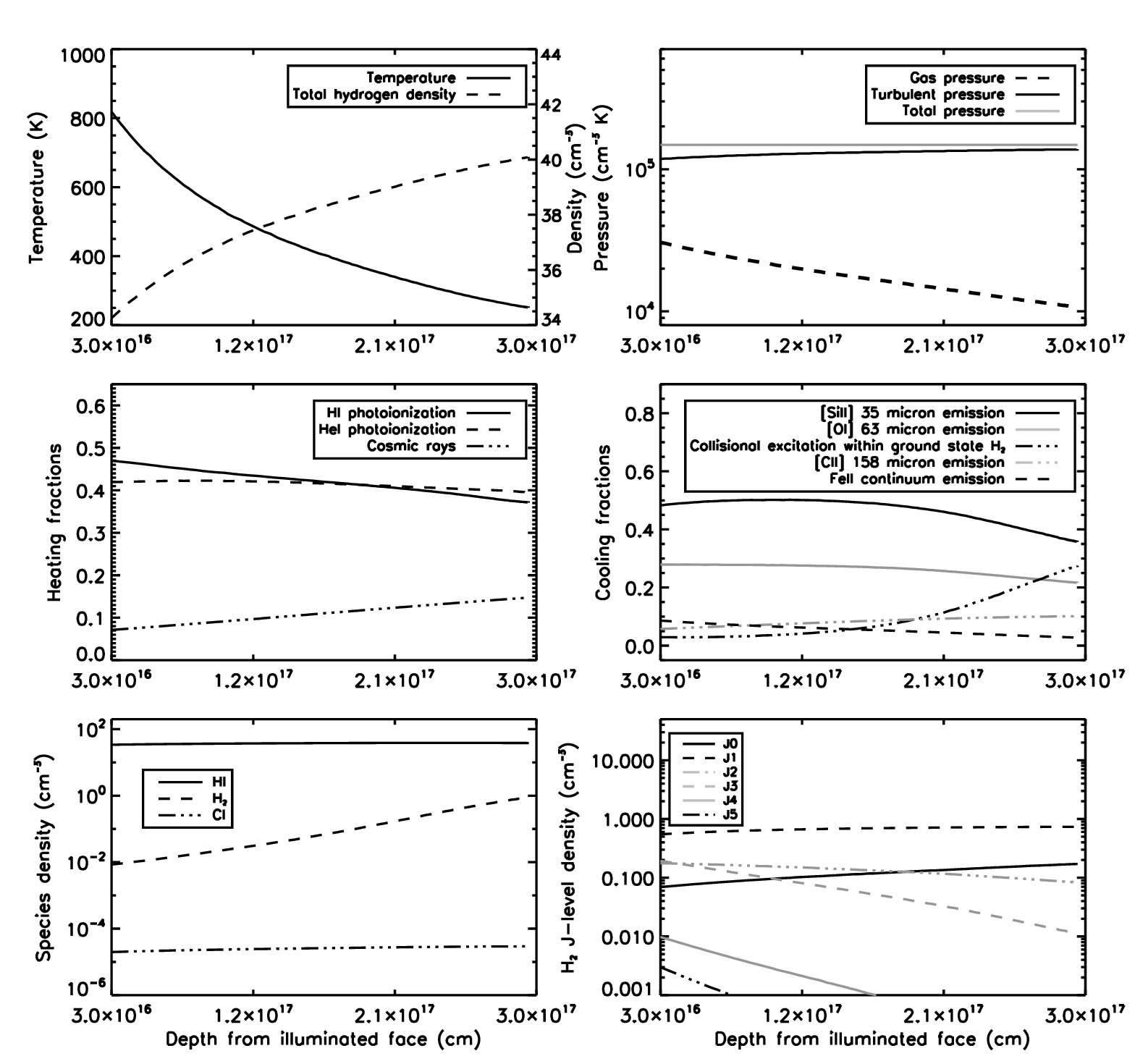}}
\caption{The physical conditions within H$_{2}$ component 8 are summarized in this figure. Each panel shows the variation of different physical properties with depth from one of the illuminated faces of the cloud. The plots here represent one half of the cloud, with the other half having a symmetrical profile. The properties include gas temperature and total hydrogen density (top row, first panel), total pressure and its major constituents (top row, second panel), heating and cooling fractions due to various physical processes (middle row), density of \ion{H}{i}, H$_2$ and \ion{C}{i} (bottom row, first panel), and the densities of various rotational levels of H$_{2}$ (bottom row, second panel).}
\label{fig:phycon_8}
\end{figure*}

\setlength{\parindent}{2ex}
	The electron temperature in the innermost region of the cloud is 251 K. This lies within the excitation temperature range, $T_{01}$ = $157^{+100}_{-44}$ K. The pressure within each DLA component has many constituents including gas pressure, turbulent pressure and radiation pressure. Gas pressure is typically the dominant contributor. We find in our models, that the other major contribution to the total pressure of the system arises from micro-turbulence. As an exception, the high micro-turbulence in this component implies that turbulent pressure actually dominates thermal pressure. But our models for the other components indicate much lower contributions from turbulent pressure. While the total pressure within the cloud remains constant, gas pressure $P/k$ drops down from 39,900 to 10,700 cm$^{-3}$ K, while going from the hotter to the colder phase. The gas pressure in the shielded region lies towards the higher end of the pressure range found by \citet{Srianand2005b} for high-redshift H$_{2}$-DLAs. The molecular regions of the other H$_{2}$ components show even higher gas pressure, but are compatible with the results of \citet{Srianand2005b}. From all the component models, we note that turbulent pressure acquires more significance in the molecular region of the cloud where the thermal pressure is lower than in the hotter atomic phase. In this component however, it is higher than the gas pressure throughout the extent of the cloud. It becomes even more dominant, by about an order of magnitude, in the H$_{2}$ region. From the model predictions, the line-of-sight size of H$_{2}$ component 8 is 0.19 pc. We show the variation of density, temperature and pressure constituents with depth in the top two panels of Fig. \ref{fig:phycon_8}. All panels in this figure show variations in the respective quantities for only one half of the component cloud, starting at one of the two illuminated faces. The other half of the cloud is understood to have a symmetrical profile. \par 
\setlength{\parindent}{2ex}	
Further, we study the physical processes occurring in the cloud, which determine the thermal balance of the gas. In the middle row of plots in Fig. \ref{fig:phycon_8}, we show the heating and cooling fractions associated with the  most significant physical processes. The important heating processes are \ion{H}{i} and \ion{He}{i} photoionization, and heating due to cosmic rays. Near the illuminated face of the cloud, the photoionization processes dominate. As we move deeper into the cloud, the significance of photoionization heating drops due to shielding of incident radiation. Though cosmic ray heating rises in importance in these regions, the photoionization processes still remain the major contributors to overall heating. Grain photoelectric emission accounts for less than 6 percent of the total heating. The total heating in the various components of the DLA lie in the range of $10^{-25}-10^{-23}$ erg cm$^{-2}$ s$^{-1}$. Among the major cooling processes are [\ion{Si}{ii}] and [\ion{O}{i}] line cooling. [\ion{C}{ii}] 158 $\umu$m emission becomes significant only in the well-shielded regions. There is insufficient shielding in most of the individual component environments for [\ion{C}{ii}] 158 $\umu$m emission to dominate the cooling. \ion{Fe}{ii} continuum emission is another important cooling process, especially at shallower depths in the cloud. The relevance of these cooling processes reiterates the necessity of proper elemental abundances in the models. While we use the available information on Fe, Si and C, there are no observational constraints on the abundance of oxygen. Our models assume [O/H] = [S/H]. In molecular gas, H$_2$ plays an important role in gas cooling. H$_{2}$ molecules undergo collisions with various species such as electrons, protons, atomic hydrogen, helium, and other H$_{2}$ molecules \citep {LeBourlot1999, Glover2008}, leading to collisional excitation within the ground electronic state of H$_{2}$ which is then followed by emission. This results in the loss of kinetic energy of the colliding species and cools the gas. \par
\setlength{\parindent}{2ex}
The first panel in the lowest row of Fig. \ref{fig:phycon_8} shows the density of \ion{H}{i}, H$_2$ and \ion{C}{i} at different depths. The species densities of \ion{H}{i} and \ion{C}{i} are almost constant at all depths. However, there is a gradual increase in the H$_{2}$ density till it becomes abundant in the inner, shielded region of the cloud. The plots in the last panel of Fig. \ref{fig:phycon_8} offer further insight into the H$_{2}$ level population and show the density of the observed rotational levels at different depths in the cloud. The densities of H$_2$ levels J = 4 and 5 are highest in the unshielded region, due to UV pumping in the hotter regions of the cloud. At the shallow depths where these levels are most abundant, the H$_2$ density is still very low, $\sim$ 10$^{-2}$ cm$^{-3}$. The densities of levels J = 4 and 5 eventually drop off at depths of $\sim 10^{17}$ cm. Subsequently, the rotational levels J = 0 and 1 become the dominant form of H$_{2}$ in the denser and cooler shielded regions. This is also the region where the H$_2$ density rises and attains its maximum value in the cloud, $\sim$ 0.9 cm$^{-3}$.\par
\setlength{\parindent}{2ex}
The other H$_{2}$ components are also modelled in a similar way, as discussed here for component 8. We are able to successfully match more than 91 percent of all the column density constraints to within a factor of 2. As mentioned before, we consider metal components 11 and 12 to constitute a single cloud with both H$_{2}$ and \ion{C}{i} absorption features, and construct a combined model accordingly. The details of the individual models are available for reference in the Appendix (Tables \ref{tab:paras_4}--\ref{tab:model_13}). We also make plots depicting the physical conditions in these components (Figs. \ref{fig:phycon_4}--\ref{fig:phycon_13}), along the lines of Fig. \ref{fig:phycon_8}. 

\begin{table}
 \centering
  \caption{Physical parameters for {\tiny CLOUDY} model of component 8}
  \label{tab:paras_8}
  \begin{tabular}{@{}ll@{}}
   \hline
   Physical parameter     &     Model value\\
 \hline
 Radiation field & KS15 background$^{\textit{a}}$ \\
 Total hydrogen density & 31.6 cm$^{-3}$ (1.50 in log-scale) \\
 (at illuminated face) & \\
 Metallicity (log-scale) & -0.35 \\
 $[C/H]$ & -1.50 \\
 $[Si/H]$ & -0.57 \\
 $[Fe/H]$ & -0.87 \\
 Dust-to-gas ratio & 0.50 (0.42$\pm$0.07)$^{\textit{b}}$ \\
 Size range of dust grains & 0.0025-0.125 $\umu$m \\
 Micro-turbulence & 6.5 (7.5$\pm$0.5)$^{\textit{b}}$ km s$^{-1}$ \\
 Cosmic ray ionization rate & 10$^{-15.30}$ s$^{-1}$ \\
 \hline
\end{tabular}
\newline
\raggedright $^{\textit{a}}$ Khaire-Srianand background \citep{Khaire2015b, Khaire2015a}\\
\raggedright $^{\textit{b}}$ Numbers within the brackets indicate observed values
\end{table}

\begin{table*}
 \centering
  \caption{Observed and {\tiny CLOUDY} model column densities for component 8}
  \label{tab:model_8}
  \begin{threeparttable}
  \begin{tabular}{@{}lccc@{}}
  \hline
   Species     &     Observed log(\textit{N}) (cm$^{-2}$) & Best-fitting model log(\textit{N}) (cm$^{-2}$) & ISM model log(\textit{N}) (cm$^{-2}$)$^\textit{a}$ \\
 \hline
 \ion{H}{i} & - & 19.34 & 19.56 \\
 H$_2$ & 16.83$\pm$0.06 & 16.98 & 16.98 \\
 H$_2$ (0) & 16.06$\pm$0.07 & 16.16 & 16.21 \\
 H$_2$ (1) & 16.54$\pm$0.12 & 16.84 & 16.84 \\
 H$_2$ (2) & 16.29$\pm$0.04 & 15.99 & 15.91 \\
 H$_2$ (3) & 15.36$\pm$0.04 & 15.40 & 15.18 \\ 
 H$_2$ (4) & 13.78$\pm$0.15 & 13.64 & 13.38 \\ 
 H$_2$ (5) & $\leq$ 13.91 & 12.91 & 12.70 \\ 
 \ion{C}{i*} & 13.16$\pm$0.03 & 12.90 & 13.17 \\
 \ion{C}{i**} & 12.64$\pm$0.03 & 12.73 & 12.98 \\
 \ion{C}{i***} & 11.80$\pm$0.19 & 12.09 & 12.32 \\
 \ion{C}{ii*} & 13.05$\pm$0.07 & 12.83 & 13.02 \\
 \ion{Si}{ii} & 14.50$\pm$0.08 & 14.28 & 14.50 \\
 \ion{S}{ii} & 14.34$\pm$0.07 & 14.11 & 14.33 \\
 \ion{Fe}{ii} & 14.11$\pm$0.08 & 13.97 & 14.19 \\
 \hline
\end{tabular}
\begin{tablenotes}
\item $^\textit{a}$ The ISM model has exactly the same parameters as the best-fitting model, except for dust grain size. It employs ISM-sized grains, as compared to the smaller grains used for the best-fitting model. 
\end{tablenotes}
\end{threeparttable}
\end{table*}

\subsection{Model uncertainties}
\label{ssec:lim}

Further in this section, we use the best-fitting model as our point of reference. We incorporate smaller grains in our models in accordance with the results discussed in \citet{Shaw2016}. As described earlier, these grains range from 0.0025-0.125 $\umu$m in size, and follow the standard MRN size distribution. We compare the best-fitting model for component 8, with the column densities predicted by the same model when the smaller grains are replaced with ISM-sized grains. The column densities are reported in Table \ref{tab:model_8}. It is clear that the model with ISM-sized grains is not feasible as it produces \ion{H}{i} and metal species column densities greater by about 0.2 dex, and a less pronounced effect is seen on the H$_2$ column densities. The larger prediction of \textit{N}(\ion{H}{i}) when using ISM-sized grains is unacceptable as the models are constrained by the total observed \ion{H}{i} budget of the DLA. However, as pointed out in Section \ref{ssec:approach}, there is a possibility that the grains may be porous and ISM-sized \citep[see discussion in][]{Shaw2016}. \par
\setlength{\parindent}{2ex}
For a given set of constraints, {\tiny CLOUDY} is capable of optimizing the variable input parameters through $\chi^2$-minimization using the inbuilt {\tiny PHYMIR} algorithm \citep {vanHoof1997, Ferland2013}. In this approach, it is possible that some observables are constrained very well, while others are not. The best-fitting model is decided on the basis of the overall $\chi^2$ value. Thus, we fine-tune our models manually in order to construct one which reproduces to the best extent all the species column densities which are to be constrained. Carrying this out for seven components associated with the same DLA is a challenging task. Each component is initially modelled separately. Later, we revisit the models to fine-tune parameters for overall consistency. At this stage, we ensure that the sum of predicted \ion{H}{i} column densities from the modelled H$_2$ components agrees with the overall observed \ion{H}{i}, as well as the remaining \ion{H}{i} which arises from the non-H$_2$ components. We also try to attain maximum possible agreement between the elemental abundances in the various components, as they are associated with the same DLA. Even as we arrive at a single value for each parameter for each modelled component, it is clear that the model may still hold good for some variation in the value of each parameter. Here, we study the effects of varying each parameter by a small specified step around the value used in the best-fitting model for component 8. The results are represented in Table \ref{tab:uncert}.\par
\setlength{\parindent}{2ex} 
In varying the total hydrogen density by a factor of two (0.3 in log-scale), the H$_2$ rotational levels J = 4 and 5, along with the \ion{H}{i} and metal species column densities are seen to change by more than 0.1 dex. Lower hydrogen density at the illuminated faces of the cloud produces higher column densities of these levels and species. On the other hand, lower values of cosmic ray ionization rate decrease the population of the higher H$_2$ rotational levels. A change of 0.3 dex introduced in the cosmic ray ionization rate causes a change of 0.1 dex in the column densities of H$_2$ rotational levels J = 4 and 5. The impact on \ion{H}{i} and metal species column densities is less pronounced. Varying metallicity and dust-to-gas ratio by 0.1 dex and 0.1 respectively influences the higher H$_2$ rotational levels and metal species column densities. Less dust produces lower population of the higher H$_2$ rotational levels, along with more \ion{H}{i} and metals. Meanwhile, lower metallicity increases the column densities of the higher H$_2$ rotational levels, and reduces the column densities of metal species. Column densities of H$_2$ levels J = 3, 4 and 5 change by more than 0.1 dex due to the variation in metallicity, but the \ion{H}{i} column density barely shows any difference. In the best-fitting model, [Si/H], [Fe/H] and [C/H] were scaled separately from the overall metallicity, which set the abundance of other metal species. We re-run the best-fitting model thrice, each time setting [Si/H], [Fe/H] and [C/H] respectively to the overall metallicity of -0.35 for the component. The abundances of these species affect the higher rotational levels of H$_2$ (J = 3, 4 and 5), and thus, they must be carefully ascertained. Allowing for a greater amount of these metal ions to be present in the gas, reduces the column densities of the higher H$_2$ rotational levels. It is the abundance of carbon that produces the strongest impact on the species column densities. It affects almost all H$_2$ levels, with column densities of levels J = 3, 4 and 5 changing by 0.6 dex or more. The increase in the amount of available carbon also causes a spike in the column densities of the \ion{C}{i} and \ion{C}{ii*} levels, all of which increase by more than 0.8 dex. Finally, we vary the micro-turbulence in the model by 1 km s$^{-1}$, but find only minor variations in the predicted column densities.  

\begin{table*}
 \centering
  \caption{Maximum column density variation (in dex) in the {\tiny CLOUDY} results for component 8, for variation in different input parameters}
  \label{tab:uncert}
  \begin{threeparttable}
  \begin{tabular}{@{}lcccccccc@{}}
  \hline
   Species & $\Delta$n$_{tot} ^{\textit{a}}$ & $\Delta$(CR rate) & $\Delta$Z & [Si/H]$^{\textit{b}}$ & [Fe/H]$^{\textit{b}}$ & [C/H]$^{\textit{b}}$ & $\Delta \kappa$ & $\Delta$\textit{b}\\
 & = 10$^{0.3}$ cm$^{-3}$ & = 10$^{0.3}$ s$^{-1}$ & = $\pm$0.1 dex & = -0.35 & = -0.35 & = -0.35 & = $\pm$0.1 & = $\pm$1 km s$^{-1}$ \\
 \hline
 \ion{H}{i} & 0.28 & 0.02 & 0.01 & 0.02 & 0.00 & 0.13 & 0.07 & 0.06\\
 $H_2$ & 0.00 & 0.00 & 0.00 & 0.00 & 0.00 & 0.00 & 0.00 & 0.00 \\
 $H_2$ (0) & 0.01 & 0.02 & 0.02 & 0.03 & 0.01 & 0.19 & 0.01 & 0.00 \\
 $H_2$ (1) & 0.00 & 0.00 & 0.01 & 0.00 & 0.00 & 0.00 & 0.00 & 0.00 \\
 $H_2$ (2) & 0.02 & 0.02 & 0.03 & 0.04 & 0.01 & 0.37 & 0.02 & 0.00 \\
 $H_2$ (3) & 0.04 & 0.08 & 0.10 & 0.16 & 0.05 & 0.90 & 0.06 & 0.01 \\ 
 $H_2$ (4) & 0.15 & 0.11 & 0.22 & 0.26 & 0.10 & 0.85 & 0.08 & 0.00 \\ 
 $H_2$ (5) & 0.37 & 0.11 & 0.34 & 0.31 & 0.16 & 0.60 & 0.07 & 0.02 \\ 
 \ion{C}{i*} & 0.15 & 0.04 & 0.14 & 0.04 & 0.03 & 1.24 & 0.09 & 0.05 \\
 \ion{C}{i**} & 0.03 & 0.05 & 0.11 & 0.01 & 0.00 & 1.12 & 0.08 & 0.04 \\
 \ion{C}{i***} & 0.07 & 0.06 & 0.07 & 0.05 & 0.02 & 0.99 & 0.08 & 0.03 \\
 \ion{C}{ii*} & 0.14 & 0.04 & 0.06 & 0.07 & 0.03 & 0.87 & 0.06 & 0.04  \\
 \ion{Si}{ii} & 0.28 & 0.02 & 0.10 & 0.20 & 0.00 & 0.13 & 0.07 & 0.06  \\
 \ion{S}{ii} & 0.28 & 0.02 & 0.10 & 0.02 & 0.00 & 0.13 & 0.07 & 0.06  \\
 \ion{Fe}{ii} & 0.27 & 0.03 & 0.09 & 0.02 & 0.53 & 0.13 & 0.08 & 0.06  \\
 \hline
\end{tabular}
\begin{tablenotes}
\item $^{\textit{a}}$ Change in total hydrogen density at the illuminated face of the cloud
\item $^{\textit{b}}$ The abundances of Si, Fe and C are not varied over a range, but set to the metallicity deduced from the best-fitting model
\end{tablenotes}
\end{threeparttable}
\end{table*}

\section{Discussion}
\label{sec:result}
\subsection{Comparative study of physical properties of the components}

The physical properties we deduce from the H$_{2}$ models vary from one component to another. Here, we study this variation by combining the results of observational analysis and numerical modelling.

\subsubsection{Temperature, density, pressure and cosmic ray ionization}
\label{sssec:tdpcr}

As discussed in Section \ref{ssec:temp_den}, the observed population of the \ion{C}{i} fine structure lines was used to derive the atomic hydrogen density $n_\mathrm{H}$ in the shielded, molecular regions of the cloud. The calculations assumed a radiation field similar to the Galaxy and gas temperature similar to $T_{01}$ determined from the H$_{2}$ rotational levels. However, $T_{01}$ is a good approximation for the gas temperature only when there is sufficient shielding, which is clearly not the case for some of the components. Besides, we only consider collisions with atomic hydrogen. In the real DLA environment, there would be many atomic and molecular species which can collisionally excite the neutral carbon fine structure levels. While atomic hydrogen is the dominant collider, other species including electrons, protons, neutral helium and H$_{2}$ also play a significant role \citep{Silva2002}. Collisions with all these species are considered in {\tiny CLOUDY}. Our models simulate the physical environment of the DLA more accurately than the simplified calculations outlined in Section \ref{sec:phy_prop}. We find that this DLA is irradiated mainly by the metagalactic background. The radiation field is thus, different from the interstellar radiation field of the Milky Way, which we assumed for the preliminary calculations outlined in Section \ref{ssec:temp_den}. Yet, the hydrogen density $n_\mathrm{H}$ from the models agrees well with the calculations for component 8. The model predictions vary from the simplified calculation, in case of the other components. The value of $n_\mathrm{H}$ constrained from each component model is the atomic hydrogen density attained in the innermost region of the particular component, as discussed for component 8 in Section \ref{ssec:8_model}. The values of $n_\mathrm{H}$ constrained through the {\tiny CLOUDY} models of components 4 and (11+12) agree well with the respective calculations using the \textit{N}(\ion{C}{i**})/\textit{N}(\ion{C}{i*}) ratio, but not with the higher density values predicted by the \textit{N}(\ion{C}{i***})/\textit{N}(\ion{C}{i*}) ratio calculations. In the case of component 13, the ratio \textit{N}(\ion{C}{i**})/\textit{N}(\ion{C}{i*}) implies that $n_\mathrm{H}$ < 18 cm$^{-3}$, while the ratio \textit{N}(\ion{C}{i***})/\textit{N}(\ion{C}{i*}) provides a much higher upper limit of $n_\mathrm{H}$ < 87 cm$^{-3}$. Our {\tiny CLOUDY} model for this component predicts $n_\mathrm{H}$ = 36 cm$^{-3}$, which agrees with the calculation based on \textit{N}(\ion{C}{i***})/\textit{N}(\ion{C}{i*}). For component 7, $n_\mathrm{H}$ determined through the ratios \textit{N}(\ion{C}{i**})/\textit{N}(\ion{C}{i*}) and \textit{N}(\ion{C}{i***})/\textit{N}(\ion{C}{i*}) lies within the combined range 4--29 cm$^{-3}$, while our {\tiny CLOUDY} model predicts a higher density of 120 cm$^{-3}$. In the inner regions of component 9, $n_H$ reaches a value of 52 cm$^{-3}$. However, both the \ion{C}{i} calculations together imply that this value should be < 39 cm$^{-3}$. In case of component 5, the calculations indicate $n_\mathrm{H}$ > 193 cm$^{-3}$, while the value from our model is much lower, $n_\mathrm{H}$ = 62 cm$^{-3}$. We do note that our model for component 5 is ineffective in reproducing the observed \textit{N}(\ion{C}{i**})/\textit{N}(\ion{C}{i*}) very well, though both column densities individually agree within 0.3 dex of the respective observed quantities. Besides, from our analysis of all other properties, it appears unlikely that this component should have density drastically different from the others. \par
\setlength{\parindent}{2ex}
The temperature in the molecular regions of all the H$_{2}$ components lies in the range of 140--360 K. Components 4 and 5 have very low column densities of H$_{2}$, log[\textit{N}(H$_{2}$)(cm$^{-2}$)] $\lesssim$ 16. As a result, the gas is not sufficiently shielded, for the assumption $T_{01} \sim T_{kin}$ to hold good. There is much discrepancy here, between $T_{01}$ and the actual gas temperature, with the gas actually tracing a warmer phase than predicted by $T_{01}$. There is better agreement between $T_{01}$ and $T_{kin}$ in case of the other components, all of which have log[\textit{N}(H$_{2}$)(cm$^{-2}$)] > 16. \par
\setlength{\parindent}{2ex}
\citet {Srianand2005b} analysed various physical properties of 33 high-redshift DLAs, some of which contain H$_{2}$. Assuming a radiation field similar to the Galaxy, they found that the gas pressure in H$_{2}$ regions lies in the range 824--30,000 cm$^{-3}$ K. 20 percent of the DLAs were seen to have pressure exceeding 5,000 cm$^{-3}$ K, and only 8 percent had pressure higher than 10,000 cm$^{-3}$ K. In comparison, the cold neutral medium in the Galaxy has a lognormal distribution of thermal pressure, with mean at $P/k$ = 3800 cm$^{-3}$ K \citep{Jenkins2011}. The pressure in the molecular regions of all the H$_{2}$ components in this DLA lie towards the higher end of the pressure range known for high-redshift H$_{2}$-DLAs. \par
\setlength{\parindent}{2ex}
The cosmic ray ionization rate of neutral hydrogen shows slight variation between the components. The mean cosmic ray ionization rate in the DLA is more than twice the Galactic rate of $2 \times 10^{-16}$ s$^{-1}$ deduced by \citet {Indriolo2007} through observations of H$_3^+$ in diffuse interstellar clouds. The highest value attained by the cosmic ray ionization rate in any of the modelled H$_2$ components is four times this Galactic rate. The Galactic cosmic ray ionization rate has been subsequently revisited by \citet* {Indriolo2012} and \citet {Indriolo2015}. \citet {Indriolo2012} also observed H$_3^+$ along diffuse Galactic sightlines, but with a larger survey sample. They inferred cosmic ray ionization rates  in the range $(1.7\pm1.3)\times10^{-16} - (10.6\pm8.2)\times 10^{-16}$ s$^{-1}$, with a mean of $3.5^{+5.3}_{-3.0}\times10^{-16}$ s$^{-1}$. \citet {Indriolo2015} then probed the cosmic ray ionization rate using the ions OH$^+$, H$_2$O$^+$ and H$_3$O$^+$. They found the mean rate to be $1.78\times10^{-16}$ s$^{-1}$. Thus, the study of cosmic ray ionization is an ongoing endeavour even in the local Galactic environment. Much less is known about high-redshift sightlines. We do know that cosmic rays are produced and accelerated due to supernovae and stellar winds of massive stars. Thus, enhanced cosmic ray ionization is typically linked with high star formation rates and magnetic fields \citep{Dalgarno2006, Veritas2009, Ceccarelli2011}. It is thus, unusual to associate a region of low star formation like this DLA, with high cosmic ray ionization. \citet{Dutta2014} studied low-metallicity DLAs at high redshift, and inferred enhanced cosmic ray ionization rates for the ``high cool" DLAs therein. They argued that the need for higher cosmic ray heating could simply be a manifestation of additional heating sources such as hydrodynamical heating, which are not included in \tiny{CLOUDY} \normalsize calculations. This highlights the necessity for both more detailed calculations, as well as an improved understanding of astrophysical environments beyond our Galaxy and in the high-redshift Universe. Fig. \ref{fig:props} provides a comparison between the density, temperature, pressure and cosmic ray ionization rate in the different H$_{2}$ components. 

\subsubsection{Component-wise neutral hydrogen and extent}

\begin{figure*}
\centering{\includegraphics[width=\textwidth, height=17 cm, keepaspectratio]{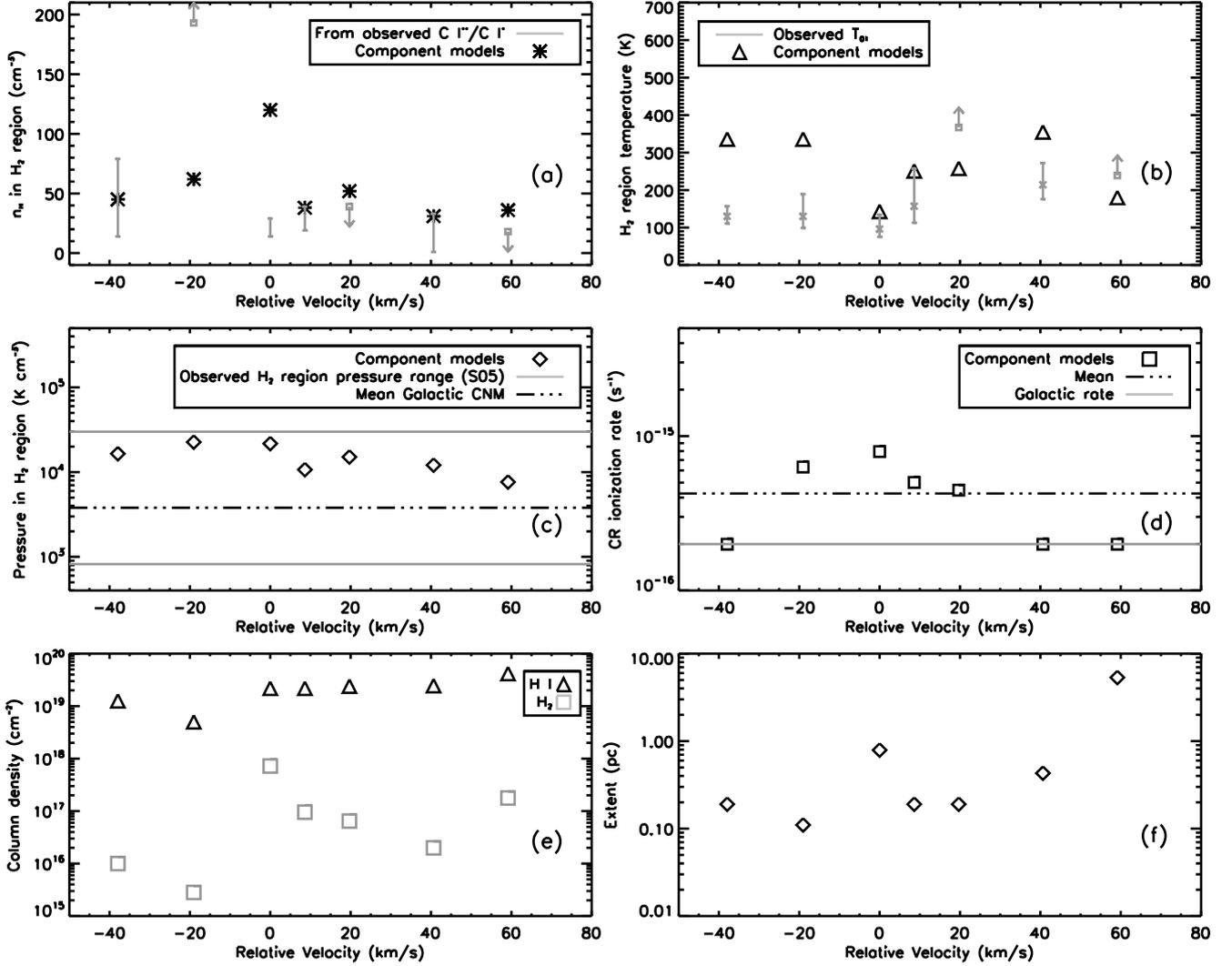}}
\caption{Physical properties constrained from the {\tiny CLOUDY} models for the individual H$_{2}$ components are plotted here. Zero velocity is defined at the H$_{2}$ component with $z_{abs}$ = 2.054509. The properties in panels (a), (b) and (c) are the atomic hydrogen density $n_\mathrm{H}$, gas temperature \& gas pressure in the molecular region. The observed pressure range for H$_{2}$ regions in DLAs is from the study of \citet{Srianand2005b}. The mean gas pressure in the Galactic CNM is taken from \citet{Jenkins2011}. Panel (d) shows the cosmic ray ionization rate of neutral hydrogen. The Galactic rate is from \citet{Indriolo2007}. Panels (e) and (f) represent \textit{N}(\ion{H}{i}) and \textit{N}(H$_2$); and the sizes of the H$_{2}$ components respectively, as constrained from the {\tiny CLOUDY} models.}
\label{fig:props}
\end{figure*}

As we already mention in Section \ref{ssec:hi}, component-wise distribution of \ion{H}{i} cannot be inferred from the spectrum through Voigt profile analysis. So, we can only obtain component-wise information on \textit{N}(\ion{H}{i}) from our numerical models. In Table \ref{tab:h1}, we list the model  predictions of \textit{N}(\ion{H}{i}) for the H$_{2}$ components. The sum of the predicted \textit{N}(\ion{H}{i}) over all the H$_{2}$ components is 0.17 dex lower than the observed \textit{N}(\ion{H}{i}) for the whole DLA. The remaining \ion{H}{i} naturally arises from the components without H$_{2}$, with log $\textit{N}(\ion{H}{i})_{\mathrm{non-H_{2}}}$ = 19.86 cm$^{-2}$. This accounts for $\sim$ 32 percent of the total \ion{H}{i} in the DLA. In contrast, previous studies have shown that most of the \ion{H}{i} absorption in H$_{2}$-DLAs arises from regions which do not produce H$_{2}$ \citep{Noterdaeme2015b, Srianand2012}. \citet {Srianand2012} used the lack of 21-cm absorption from regions directly associated with H$_{2}$ to infer that $\leq$ 10 percent of the total \ion{H}{i} in a DLA arises from H$_{2}$-rich regions. Thus, the system that we analyse here appears to be special compared to other known H$_{2}$-DLAs. However, we note that the individual H$_{2}$ components each host between 2 and 19 percent of the total \textit{N}(\ion{H}{i}) of the DLA. Only one of the seven components has an \ion{H}{i} contribution of more than 11 percent of the total observed \textit{N}(\ion{H}{i}). The uniqueness of the system then is in the presence of multiple H$_{2}$ components, which causes a greater percentage of \ion{H}{i} to be associated with the total H$_{2}$ content. From their study, \citet {Srianand2012} also concluded that the regions producing H$_{2}$ absorption are usually very small and compact, likely to be $\leq$ 15 pc across. The H$_{2}$ components of this DLA each have an extent of less than a few parsecs along the line-of-sight, and even their total size sums up to 7.2 pc, which agrees with these findings. Panel (e) of Fig. \ref{fig:props} shows the component-wise distribution of \textit{N}(\ion{H}{i}) and H$_2$, while panel (f) plots the extent of each H$_{2}$ component. \par

\begin{table*}
 \centering
  \caption{Component-wise \ion{H}{i} column density, with other important quantities based on the {\tiny CLOUDY} models}
  \label{tab:h1}
  \begin{tabular}{@{}ccccc@{}}
  \hline
   Component & \textit{N}(\ion{H}{i}) (log cm$^{-2}$) & \textit{N}(H$_2$) (log cm$^{-2}$) & Molecular fraction, \textit{f} & Metallicity, $Z_{\odot}$\\
 \hline
 4 & 19.10 & 16.00 & -2.80 & -0.46\\
 5 & 18.70 & 15.45 & -2.95 & -0.46\\
 7 & 19.34 & 17.86 & -1.21 & -0.60\\
 8 & 19.34 & 16.98 & -2.06 & -0.35\\
 9 & 19.38 & 16.81 & -2.27 & -0.82\\
 11+12 & 19.39 & 16.30 & -2.79 & -0.40\\
 13 & 19.62 & 17.25 & -2.07 & -0.52\\
 \hline
 Total (H$_{2}$ components) & 20.18 & 18.04 & -1.90 (Mean) & -0.49 (Mean)\\
 \hline
 Observed value for the DLA & 20.35$\pm$0.05 & 17.99$\pm$0.05 & -2.06$\pm$0.10 & -0.52$\pm$0.06\\
\hline
\end{tabular}
\end{table*}

\subsubsection{Elemental abundances and dust content}

\begin{figure*}
\centering{\includegraphics[width=\textwidth, height=17 cm, keepaspectratio]{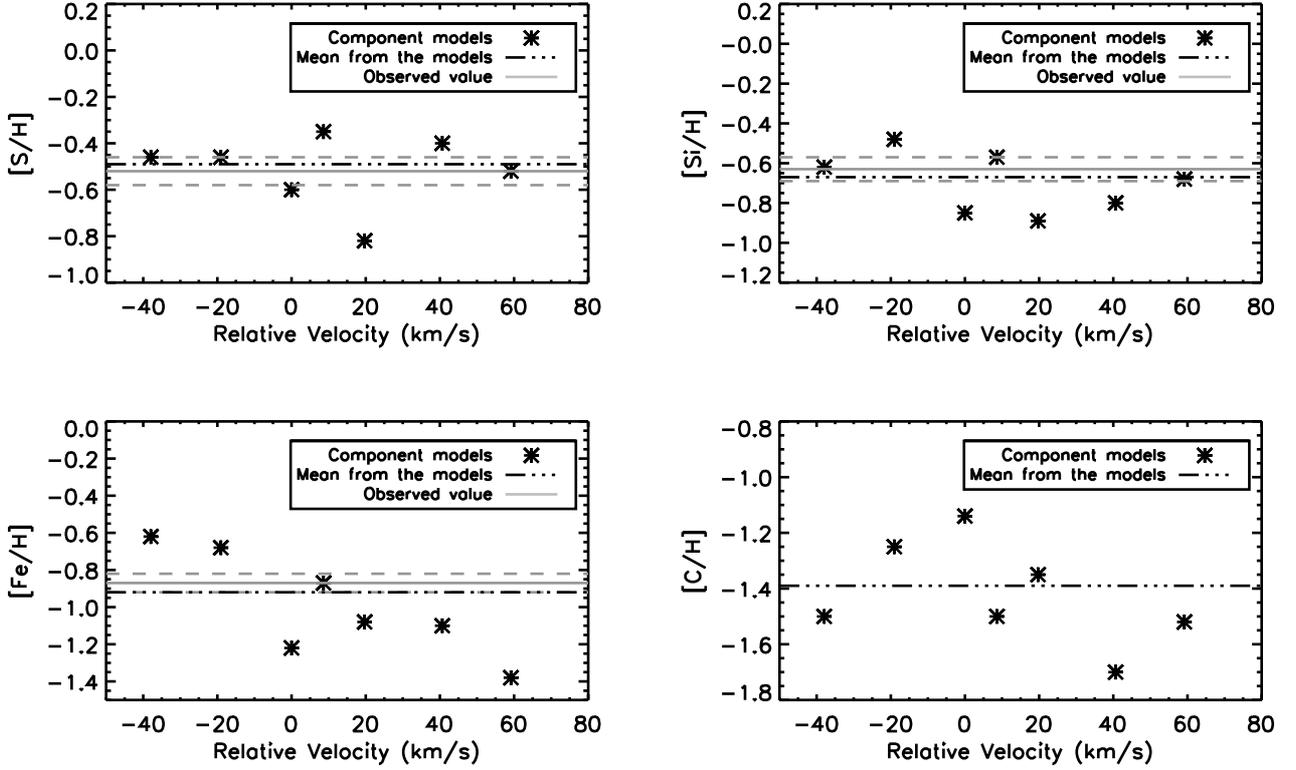}}
\caption{The abundances of S, Si, Fe \& C, as constrained from the H$_{2}$ component models, are plotted here. Zero velocity is defined at the H$_{2}$ component with $z_{abs}$ = 2.054509. The mean [X/H] from the models is compared with the observed [X/H] for the entire DLA, for X = S, Si and Fe. Observational constraints are not available for the carbon abundance. }
\label{fig:abund}
\end{figure*}

We compute the metallicity of the DLA from the observed [S/H]. Hence, the metallicity of the component models are constrained through the observed \textit{N}(\ion{S}{ii}). Similarly, we also constrain the abundances of Si, Fe and C using the observed column densities of the respective species. We note that the elemental abundances vary between the components. However, the mean abundance values calculated over the seven H$_{2}$ components, agree closely with the observationally deduced values. Fig. \ref{fig:abund} shows the component-wise variation in [S/H] (or, metallicity), [Si/H], [Fe/H] and [C/H] in comparison with the observed values. Further, knowing log $\textit{N}(\ion{H}{i})_{\mathrm{non-H_{2}}}$ = 19.86 cm$^{-2}$ and log $\textit{N}(\ion{S}{ii})_{\mathrm{non-H_{2}}}$ = 14.44 cm$^{-2}$, we also compute the mean metallicity of the non-H$_{2}$ components. We find  $[S/H]_{\mathrm{non-H_{2}}}$ = -0.54, which is slightly lower than $[S/H]_{\mathrm{H_{2}}}$ = -0.49, but still similar to the [S/H] derived for the entire DLA. \par
\setlength{\parindent}{2ex}
Through our models, we are able to constrain the carbon abundance in this DLA. The carbon species column densities are reasonably well-produced across the seven H$_{2}$ component models. A major part of the carbon is likely to be present in the form of \ion{C}{ii}, but we are unable to observationally constrain the \ion{C}{ii} column density on account of saturation. However, our {\tiny CLOUDY} models predict the component-wise values of log \textit{N}(\ion{C}{ii}) to be between 13.85--14.62 cm$^{-2}$ for the seven H$_2$ components. This points to the likelihood of a saturated \ion{C}{ii} transition, as is observed for this DLA. While this may not be an accurate mode of comparison, the prediction does indicate consistency with observation. Besides, the [S/H], [Si/H] and [Fe/H] values show similar scatter as [C/H] among the component models, and their mean values agree well with the corresponding observationally constrained values. All this together builds a logically sound case in favour of the [C/H] constrained from the models.\par
\setlength{\parindent}{2ex}
Determining [C/H] through observations is usually difficult, except at very low metallicities. As much of the carbon exists in the form of \ion{C}{ii}, transitions of \ion{C}{ii} are mostly saturated at higher metallicities. Observational studies of carbon abundance in DLAs have thus, been limited to the low-metallicity end \citep{Pettini2008, Dutta2014, Cooke2011, Cooke2015, Cooke2017}. \citet{Pettini2008} found good agreement between the observed values of [C/O] for low-metallicity DLAs and metal-poor Galactic halo stars, thus expecting similar trends at higher metallicities too. Accordingly, an increase in [C/O] is expected with increase in metallicity. For the DLA at $z_{abs}$ = 4.2 towards J0953-0504, \citet{Dutta2014} constrained [C/O] = -0.50$\pm$0.03. This is the lowest observed value of [C/O] in low-metallicity DLAs. Due to the higher metallicity of the DLA of our interest, we expect higher [C/O] for our system. But, the mean [C/H] from our component models is -1.39, and the resultant [C/O] is -0.87, assuming [O/H] = [S/H]. We are limited by the lack of observational constraints on the oxygen abundance both for this system in particular, and for high-metallicity DLAs in general. Despite the lack of agreement with the expected [C/O], the [C/Fe] ratio is similar to values constrained for Galactic dwarfs. Using [C/H] = -1.39 and [Fe/H] = -0.92, as constrained by our \tiny{CLOUDY} \normalsize component models, yields [C/Fe] = -0.47 for this DLA. \citet*{Tomkin1986} studied halo dwarfs at metallicities in the range -2.5 $\leq$ [Fe/H] $\leq$ +0.2, and observed [C/Fe] comparable with this DLA at a similar metallicity. Further such studies on dwarfs were performed by \citet{Carbon1987} and \citet{Andersson1994}. In their study of elemental abundances, \citet{Prochaska2002} find low values of [C/Fe] in DLAs too. For instance, they establish for the DLA at $z_{abs}$ = 4.224 towards PSS 1443+27, [Fe/H] = -1.096 and [C/Fe] > -0.642 \citep*{Prochaska2001,Howk2005}. We note here, that the DLAs in their database have [Fe/H] $\leq$ -1, and the values obtained for [C/Fe] are only lower limits.  

\begin{figure}
\centering{\includegraphics[width=\columnwidth, height=17 cm, keepaspectratio]{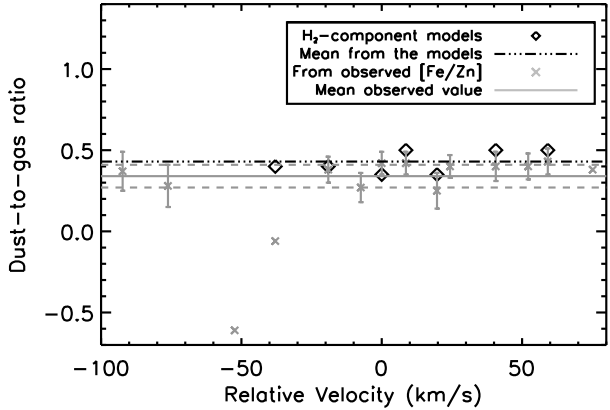}}
\caption{The values of dust-to-gas ratio $\kappa$, as constrained from the H$_{2}$ component models, are plotted here. Zero velocity is defined at the H$_{2}$ component with $z_{abs}$ = 2.054509. Observed $\kappa$ for all the H$_{2}$ and non-H$_{2}$ components are also shown. The mean $\kappa$ from the models is compared with the mean observed $\kappa$ for the DLA.}
\label{fig:dtg}
\end{figure}

We estimate the dust-to-gas ratio, $\kappa$ for the entire DLA and for the individual components, using the depletion from the observed values of [Fe/Zn]. We study the variation in the component-wise values and provide a comparison with the value of $\kappa$ constrained from the H$_{2}$ models. The mean $\kappa$ from these models agrees with the observational estimate of 0.34$\pm$0.07 for the entire DLA. For most of the components, the observed value of $\kappa$ varies closely around the mean value over the entire DLA. But, as already mentioned in Section \ref{ssec:met_kap}, the observed [Fe/Zn] is super-solar in components 3 and 4. Hence, the calculation yields a negative dust-to-gas ratio for these components, which is physically unreasonable. This could merely imply that these regions of the DLA have very low $\kappa$ compared to the other regions. However, as the individual component clouds are of parsec-scale or smaller, it is unlikely that there is significant difference in their dust content. In our model for H$_{2}$ component 4, we assume it to have dust-to-gas ratio similar to the other regions of the DLA, and are able to reproduce the observed column densities reasonably well. Fig. \ref{fig:dtg} summarizes all the information on the dust-to-gas ratio. 

\subsection{Radiation field}
\label{ssec:rad}

In Section \ref{ssec:rad_fld_obs}, we estimate the intensity of ultraviolet radiation in the vicinity of the DLA through simplified calculations involving the observed column densities of \ion{C}{ii*} and \ion{H}{i}. Here, we compare those inferences with the predictions of the numerical models which were explained in Section \ref{sec:models}. Our discussion is focussed on the two sources of ultraviolet radiation -- local star formation within the DLA, and the metagalactic background. First, we address the nature of the metagalactic background, and then discuss the intensity of the interstellar radiation field. 

\subsubsection{Metagalactic background}
\label{sssec:bkgd}

\begin{figure}
\centering{\includegraphics[width=\columnwidth, height=17 cm, keepaspectratio]{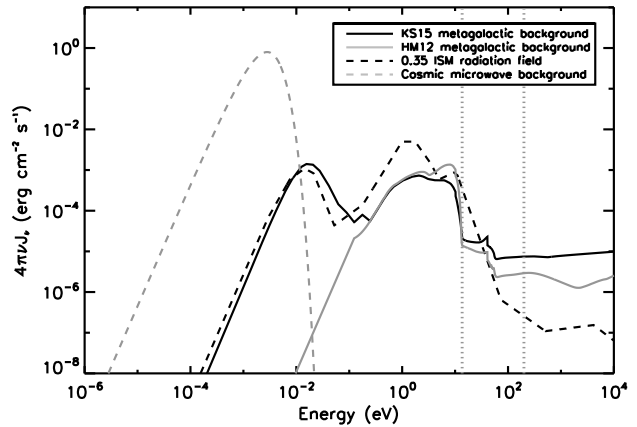}}
\caption{The incident radiation field in our models constitutes the KS15 \citep{Khaire2015a} metagalactic background and the cosmic microwave background, both of which we plot here. The continuum is generated and attenuated to account for neutral gas absorption, before it is incident on the illuminated face of the cloud. The region bordered by the dotted grey lines corresponds to the energy interval affected by attenuation. We also show here the HM12 \citep{HM2012} metagalactic background, and a radiation field with the spectral shape of the ISM radiation field but with intensity 0.35 times the ISM field. The observed \ion{C}{ii*} column density indicates that the ultraviolet flux in the vicinity of the DLA is 0.35 times that seen in the Milky Way.}
\label{fig:rad}
\end{figure}

The metagalactic background which constitutes the ultraviolet and X-ray radiation from background quasars and galaxies at any given epoch, is an area of active study \citep{HM2012, Kollmeier2014, Shull2015, Khaire2015b, Khaire2015a}. Two of the existing models for the metagalactic background have been formulated by \citet{HM2012} and \citet{Khaire2015a}. In Fig. \ref{fig:rad}, the spectral energy distributions of both these backgrounds are shown for the redshift of this DLA. We use the Khaire-Srianand background in our models. Here, we try to understand how the choice of this metagalactic background influences our {\tiny CLOUDY} model predictions. We consider component 9 for the purpose of this discussion, and recompute its original {\tiny CLOUDY} model by replacing the Khaire-Srianand background with the Haardt-Madau background. All other physical parameters are the same as in the original model and are listed in Table \ref{tab:paras_7}. We refer to the original model and the new model with the Haardt-Madau background as KS15 and HM12 respectively. The column densities predicted by both models are compared in Table \ref{tab:hm_sf}. \par
\setlength{\parindent}{2ex}
As can be seen in Fig. \ref{fig:rad}, the Haardt-Madau background is more intense in the energy interval 1-13.6 eV, and thus causes stronger ionization of the absorbing gas. The effect is clearly evident in case of the carbon species for which we observe four quantities -- \ion{C}{i*}, \ion{C}{i**}, \ion{C}{i***} and \ion{C}{ii*} -- corresponding to two different ionization stages of the atom. The HM12 model produces more \ion{C}{ii*} and less \ion{C}{i} than observed. We consider the possibility that the HM12 model could produce a closer match to the observed quantities if some of the other physical parameters are varied. As the KS15 model has enhanced cosmic ray ionization, which we retained in the HM12 model, we first attempt lowering the cosmic ray ionization rate. Though this affects the H$_{2}$ level population, it does not significantly alter the trend of higher \ion{C}{ii*} and lower \ion{C}{i}. Indeed, this is difficult to change unless the radiation field itself is changed. In comparison, the KS15 model reproduces most of the observed column densities satisfactorily for this as well as the other six H$_{2}$ components of this DLA. Recently, \citet{Hussain2017} have also used the Khaire-Srianand background to model \ion{Ne}{viii} absorbers at low redshift. Our study here, highlights the significant effect on the resultant predicted spectrum, caused merely by choosing a particular model of the metagalactic background. This reinforces the importance of a more refined understanding of the background radiation. 

\subsubsection{Local star formation}

As the calculations from observed \textit{N}(\ion{C}{ii*}) indicate, the DLA should be irradiated by an ultraviolet radiation field 0.35 times the intensity of the interstellar radiation field in the Milky Way. These calculations involve a few assumptions. It is understood that the main heating mechanism in the neutral gas is grain photoelectric emission, while cooling chiefly occurs through [\ion{C}{ii}] 158 $\umu$m emission. The DLA is considered to have similar temperature and density conditions as the Milky Way, and dust grains are assumed to have similar properties as grains in the ISM. On the contrary, our models indicate that there are other important heating and cooling processes acting within this DLA than those assumed for the calculation. Also, the DLA harbours grains which are smaller in size than the ISM grains. Our models are solely irradiated by the metagalactic background, without the requirement of additional ultraviolet photons from local star formation. However, as DLAs are understood to be associated with star formation, we study the effect of introducing an interstellar radiation field along with the metagalactic background. \par
\setlength{\parindent}{2ex}
We recompute the KS15 model with added \textit{in situ} star formation and observe the effect on the column densities of various species. We approximate the strength of the interstellar radiation field to the intensity calculated using the \ion{C}{ii*} cooling rate. Accordingly, we use a radiation field similar in shape to the ISM radiation field, but with an intensity 0.35 times that of the ISM field. All other physical parameters are the same as in the KS15 model. We refer to the model with star formation as the SF model. It produces excess \textit{N}(\ion{H}{i}) and \textit{N}(\ion{C}{ii*}). However, the high cosmic ray ionization rate required in the KS15 model and which we initially retain in the SF model, would produce additional ionization in the gas. Hence, we also lower the cosmic ray ionization in the SF model to the value 10$^{-15.7}$ s$^{-1}$ found along certain Galactic sightlines by \citet {Indriolo2007}. The radiation field in this model is equivalent to $\sim$ 0.5 $G_0$, while the Khaire-Srianand metagalactic background itself accounts for $\sim$ 0.2 $G_0$. Here, $G_0$ refers to the intensity of the interstellar ultraviolet radiation field, as defined by \citet{TH1985}. This model underpredicts the higher rotational levels of H$_{2}$, though it produces more \textit{N}(\ion{H}{i}) and \textit{N}(\ion{C}{ii*}) than the KS15 model. Besides, due to the increase in ionizing radiation, we see an increase in the predicted \ion{C}{ii*}, while the \ion{C}{i} level population decreases. This is similar to the effect seen in the HM12 model discussed in Section \ref{sssec:bkgd}. \par
Further, if we reduce the intensity of the interstellar radiation field (such that the intensity of the overall radiation field is $\sim$ 0.3 $G_0$) to improve the column densities of the carbon species, we are again unable to reproduce the observed high-J excitation of H$_{2}$. While the higher H$_{2}$ rotational levels can be populated with increased cosmic ray excitation, this also produces excess \ion{H}{i}, which is undesirable. Increase in either intensity of ultraviolet radiation or cosmic ray ionization both result in the production of more \ion{H}{i}. The effect of both these parameters on the column densities of other species must be consulted in order to decide the most apt solution. The KS15 model has been arrived at, after much deliberation in this manner. The column density predictions of the SF model are compared with the KS15 model in Table \ref{tab:hm_sf}. \par
\setlength{\parindent}{2ex}
We have already deduced from our calculations in Section \ref{ssec:rad_fld_obs}, that this DLA belongs to the ``high cool" population, and is likely to be linked with low star formation \citep{Wolfe2008}. Further, our numerical models here, indicate that UV photons from the metagalactic background are sufficient to simulate the DLA environment. This implies a weak interstellar radiation field, and supports our previous conclusion about the radiation incident on the DLA.  

\begin{table*}
 \centering
  \caption{Observed and {\tiny CLOUDY} model column densities for component 9 when different types of radiation illuminate the cloud}
  \label{tab:hm_sf}
  \begin{threeparttable}
  \begin{tabular}{@{}lcccc@{}}
 \hline
   Species     &     Observed log(\textit{N}) (cm$^{-2}$) & KS15$^\textit{a}$ log(\textit{N}) (cm$^{-2}$) & HM12$^\textit{a}$ log(\textit{N}) (cm$^{-2}$) & SF$^\textit{a}$ log(\textit{N}) (cm$^{-2}$) \\
 \hline
 \ion{H}{i} & - & 19.38 & 19.49 & 19.90\\
 H$_2$ & 16.80$\pm$0.06 & 16.81 & 16.81 & 16.81\\
 H$_2$ (0) & 15.76$\pm$0.06 & 15.94 & 16.06 & 16.16\\
 H$_2$ (1) & 16.60$\pm$0.09 & 16.65 & 16.67 & 16.67\\
 H$_2$ (2) & 16.07$\pm$0.06 & 15.88 & 15.71 & 15.50\\
 H$_2$ (3) & 15.74$\pm$0.03 & 15.52 & 14.91 & 14.77\\ 
 H$_2$ (4) & 14.37$\pm$0.04 & 14.17 & 13.26 & 13.63\\ 
 H$_2$ (5) & 14.14$\pm$0.09 & 13.84 & 12.69 & 13.36\\ 
 \ion{C}{i*} & 13.25$\pm$0.01 & 12.95 & 12.67 & 12.80\\
 \ion{C}{i**} & 13.03$\pm$0.01 & 12.89 & 12.49 & 12.76\\
 \ion{C}{i***} & 12.27$\pm$0.04 & 12.33 & 11.84 & 12.22\\
 \ion{C}{ii*} & 12.88$\pm$0.05 & 13.12 & 13.10 & 13.71\\
 \ion{Si}{ii} & 13.85$\pm$0.10 & 13.99 & 14.11 & 14.52\\
 \ion{S}{ii} & 13.67$\pm$0.07 & 13.67 & 13.79 & 14.20\\
 \ion{Fe}{ii} & 13.73$\pm$0.08 & 13.80 & 13.91 & 14.32\\
\hline
\end{tabular}
\begin{tablenotes}
\item $^\textit{a}$ KS15 is our best-fitting model for component 9 using the metagalactic background from \citet{Khaire2015a}. The physical parameters of this model are listed in Table \ref{tab:paras_7}. The HM12 model has the same parameters as the KS15 model, but uses the metagalactic background from \citet{HM2012} instead. The SF model constitutes radiation from \textit{in situ} star formation in addition to the Khaire-Srianand background. It also has lower cosmic ray ionization rate of 10$^{-15.7}$ s$^{-1}$, in order to account for the increased ionization due to more ultraviolet photons.
\end{tablenotes} 
\end{threeparttable}
\end{table*}

\subsection{Advantage of component-wise modelling}

Besides the individual H$_{2}$ components, we also compute a model assuming the entire DLA as a single cloud. We then add the column density predictions from the H$_{2}$ component models species-wise, and compare them with this average model. In the average model, we make incident an unattenuated radiation field comprising of the metagalactic background and the cosmic microwave background, on both faces of the constant pressure cloud. The total hydrogen density of the gas at each illuminated face of the cloud is $\sim$ 2 cm$^{-3}$, and increases to 9.4 cm$^{-3}$ in the molecular regions. The atomic hydrogen density $n_\mathrm{H}$, in these molecular regions is 9 cm$^{-3}$. In comparison, the \ion{C}{i} rate equations indicate the atomic hydrogen density to be in the range 22--67 cm$^{-3}$. We require cosmic ray ionization rate of 10$^{-15.17}$ s$^{-1}$ for our optimal model. The values for the cosmic ray ionization rate obtained for the H$_{2}$ components lie between 10$^{-15.70}$ and 10$^{-15.10}$ s$^{-1}$. The Doppler parameter for the different H$_{2}$ components lies between 0.7--8.0 km s$^{-1}$. For the average model, we use the value 7.5 km s$^{-1}$, which lies in this range. \par
\setlength{\parindent}{2ex}
We use the mean metallicity 0.3 $Z_\odot$ deduced for the DLA from the observed [S/H]. The elemental abundances of Si, P, Cr, Fe, Ni and Zn derived from the corresponding observed column densities too, are used in the model. Though the carbon abundance is not constrained observationally, we find through our model for the DLA that [C/H] = -1.41, by scaling the carbon abundance to match the column densities of the \ion{C}{i} and \ion{C}{ii*} fine structure levels. The mean [C/H] from the H$_2$ component models is -1.39, which closely agrees with the value we obtain through the average model. We use in the average model, dust-to-gas ratio of 0.45, which is only slightly higher than the observed value of 0.34$\pm$0.07. The dust grains constitute silicates and graphites ranging from 0.0025-0.125 $\umu$m in radius, and distributed as per the MRN size distribution. All the input parameters of the average {\tiny CLOUDY} model are listed in Table \ref{tab:paras_tot}. We compare the predicted column densities with both observed column densities for the DLA and the species-wise summed column densities from the H$_{2}$ component models. These values are presented in Table \ref{tab:model_tot}.\par
\setlength{\parindent}{2ex}
It is clear that the summed H$_{2}$ component models reproduce the total observed H$_{2}$ rotational levels and \ion{C}{i} fine structure levels better than the average model. However, as the summation model traces only the H$_{2}$ components, it naturally produces less \ion{H}{i}, \ion{S}{ii}, \ion{Si}{ii} and \ion{Fe}{ii} than expected for the entire DLA. The H$_{2}$ component models and the average DLA model predict total log[\textit{N}(\ion{Cl}{i})(cm$^{-2}$)] = 12.52 and 12.74 respectively. Though both values are higher than the observationally deduced upper limit of 12.44, it is the prediction from the H$_{2}$ component models that is closer to observation. Further, it may be possible to obtain a total column density prediction in agreement with the observed upper limit, by considering a slightly lower chlorine abundance in the component models. The total \textit{N}(\ion{Cl}{i}) predicted by the H$_{2}$ components can be treated as representative of the total \ion{Cl}{i} content of the DLA. \ion{Cl}{i} is known to occur mainly in the molecular regions, being formed through charge exchange reactions involving H$_{2}$ and \ion{Cl}{ii} \citep{Jura1974a}. As a result, the \ion{Cl}{i} contribution from other regions of the DLA is likely to be negligible in comparison. The advantage of component-wise modelling is evident as we observe all the column density predictions of the average model and summation model. The intricate structure of the DLA can be traced only by detailed modelling. This is reiterated below, as we study the physical properties constrained by the average model. \par
\begin{figure*}
\centering{\includegraphics[width=\textwidth, height=17 cm, keepaspectratio]{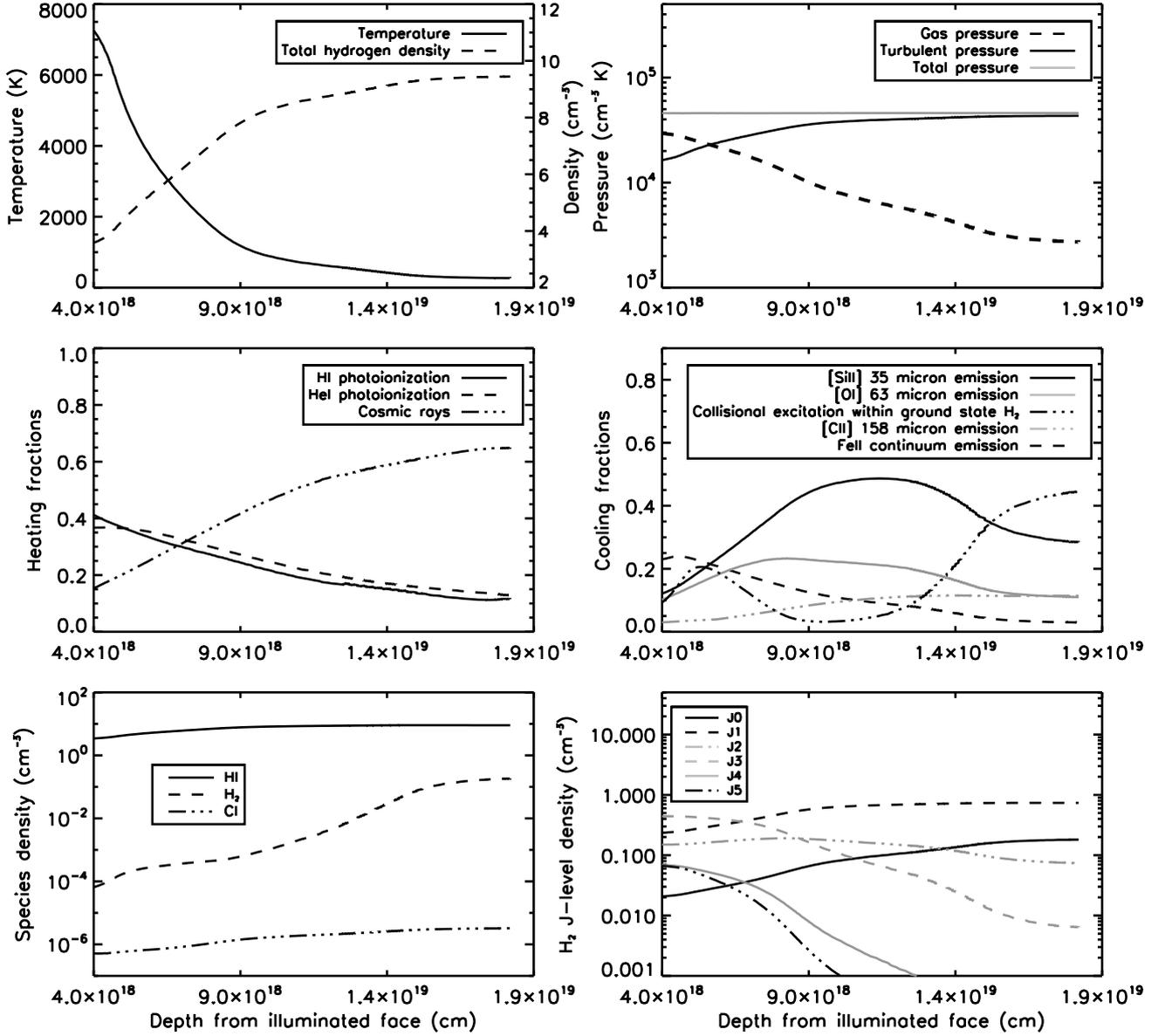}}
\caption{The physical conditions within the DLA when computed as a single cloud, are summarized in this figure. Each panel shows the variation of different physical properties with depth from one of the illuminated faces of the cloud. The plots here represent one half of the cloud, with the other half having a symmetrical profile. The properties include gas temperature and total hydrogen density (top row, first panel), total pressure and its major constituents (top row, second panel), heating and cooling fractions due to various physical processes (middle row), density of \ion{H}{i}, H$_2$ and \ion{C}{i} (bottom row, first panel), and the densities of various rotational levels of H$_{2}$ (bottom row, second panel).}
\label{fig:phycon_tot}
\end{figure*}

\setlength{\parindent}{2ex}
The electron temperature in the innermost region of the cloud is 269 K. In comparison, we calculate $T_{01}$ = $118^{+41}_{-24}$ K from the observed column densities of the H$_{2}$ (0) and $H_{2}$ (1) levels summed over the entire DLA. Thus, the average model of the DLA fails to trace the cooler regions that exist in the DLA, and that are reproduced by the H$_{2}$ component models. Gas pressure $P/k$ varies between 36,100--2,720 cm$^{-3}$ K, while going from the hotter to the colder phase. The gas pressure in the molecular region of the average model is much lower compared to the corresponding values in the component models. However, like most of the component models, the average model too, shows gas pressure to be the dominant contributor to the total pressure of the system at shallow depths into the cloud. Turbulent pressure becomes dominant in the shielded molecular region. We plot the variation of density, temperature and pressure constituents with depth in the top row of Fig. \ref{fig:phycon_tot}. All panels in this plot show the variation in physical properties with depth from one of the illuminated faces of the cloud, with the other half of the cloud having a symmetrical profile. From the average model, the extent of the cloud is found to be 11.8 pc. The H$_{2}$ component models span a total extent of 7.2 pc. As these models account for only the H$_{2}$ regions, we can treat this value as a lower limit to the size of the DLA. \par 
\setlength{\parindent}{2ex}	
As in the component models, the important heating processes are \ion{H}{i} and \ion{He}{i} photoionization, and heating due to cosmic rays. The total heating at various depths in the DLA lies in the range of $10^{-25}-10^{-23}$ erg cm$^{-2}$ s$^{-1}$. Cooling occurs through various processes. \ion{Fe}{ii} continuum emission is dominant at shallower depths, while collisional excitation within the ground state of H$_{2}$ and [\ion{C}{ii}] 158 $\umu$m emission become significant only deeper into the cloud. [\ion{Si}{ii}] and [\ion{O}{i}] fine structure emission play an important role in the intermediate regime. [\ion{Si}{ii}] emission continues to be a major cooling process in the inner molecular regions of the cloud. The middle row of plots in Fig. \ref{fig:phycon_tot} show the heating and cooling fractions associated with the physical processes discussed here.  \par
\setlength{\parindent}{2ex}
The first panel in the lowest row of Fig. \ref{fig:phycon_tot} shows the density of \ion{H}{i}, H$_2$ and \ion{C}{i} at different depths. The \ion{H}{i} density shows a mild increase as we move deeper into the cloud. At shallow depths, ionized hydrogen is the dominant form. H$_2$ becomes more abundant in the inner, shielded region of the cloud. Meanwhile, the \ion{C}{i} density increases only slightly with depth. The last panel of Fig. \ref{fig:phycon_tot} plots the density of each of the observed rotational levels of H$_{2}$ in different regions of the cloud, and shows the depth regimes that are responsible for the population of the different levels.

\begin{table*}
 \centering
  \caption{Physical parameters for the average {\tiny CLOUDY} model of the DLA}
  \label{tab:paras_tot}
  \begin{tabular}{@{}lll@{}}
  \hline
   Physical parameter     &   Model value & Observed value\\
 \hline
 Radiation field & KS15 background \textsuperscript{\textit{a}} & - \\
 Total hydrogen density at illuminated face & 2.1 cm$^{-3}$ (0.33 in log-scale) & -\\
 Metallicity (log-scale) & -0.52 & -0.52$\pm$0.06\\
 $[C/H]$ & -1.41 & - \\
 $[Mg/H]$ & -1.18 & - \\
 $[Si/H]$ & -0.69 & -0.63$\pm$0.06 \\
 $[P/H]$ & -0.17 & -0.11$\pm$0.06 \\
 $[Cr/H]$ & -0.68 & -0.62$\pm$0.06 \\
 $[Fe/H]$ & -0.92 & -0.87$\pm$0.05 \\
 $[Ni/H]$ & -0.95 & -0.89$\pm$0.06 \\
 $[Zn/H]$ & -0.35 & -0.32$\pm$0.07 \\
 Dust-to-gas ratio & 0.45 & 0.34$\pm$0.07 \\
 Size range of dust grains & 0.0025-0.125 $\umu$m & - \\
 Micro-turbulence & 7.5 km s$^{-1}$ & - \\
 Cosmic ray ionization rate & 10$^{-15.17}$ s$^{-1}$ & - \\
 \hline
 \multicolumn{3}{l}{\textsuperscript{a}\footnotesize{Khaire-Srianand background \citep{Khaire2015b, Khaire2015a}}}
\end{tabular}
\end{table*}

\begin{table*}
 \centering
  \caption{Observed and {\tiny CLOUDY} model column densities for the DLA}
  \label{tab:model_tot}
  \begin{tabular}{@{}lccc@{}}
  \hline
   Species     &     Observed log(\textit{N}) (cm$^{-2}$) & Average model log(\textit{N}) (cm$^{-2}$) & Summed H$_{2}$ component models log(\textit{N}) (cm$^{-2}$)\\
 \hline
 \ion{H}{i} & 20.35$\pm$0.05 & 20.38 & 20.18\\
 H$_2$ & 17.99$\pm$0.05 & 18.03 & 18.04 \\
 H$_2$ (0) & 17.46$\pm$0.09 & 17.26 & 17.38 \\
 H$_2$ (1) & 17.80$\pm$0.07 & 17.90 & 17.89 \\
 H$_2$ (2) & $\leq$ 16.75 & 16.97 & 16.79 \\
 H$_2$ (3) & 16.08$\pm$0.03 & 16.11 & 16.06 \\ 
 H$_2$ (4) & 14.68$\pm$0.04 & 14.39 & 14.61 \\ 
 H$_2$ (5) & $\leq$ 14.73 & 13.99 & 14.31 \\ 
 \ion{C}{i*} & 13.89$\pm$0.02 & 13.63 & 13.68 \\
 \ion{C}{i**} & 13.51$\pm$0.01 & 13.21 & 13.59 \\
 \ion{C}{i***} & 12.86$\pm$0.04 & 12.44 & 13.02 \\
 \ion{C}{ii*} & 13.75$\pm$0.05 & 13.51 & 13.69 \\
 \ion{Mg}{i} & $\leq$ 13.14 & 13.07 & - \\
 \ion{Al}{iii} & 13.14$\pm$0.04 & 13.05 & -\\
 \ion{Si}{ii} & 15.23$\pm$0.04 & 15.21 & 14.99 \\
 \ion{P}{ii} & 13.65$\pm$0.04 & 13.62 & - \\
 \ion{S}{ii} & 14.95$\pm$0.03 & 14.98 & 14.79 \\
 \ion{Cr}{ii} & 13.37$\pm$0.03 & 13.35 & - \\
 \ion{Fe}{ii} & 14.98$\pm$0.02 & 14.97 & 14.66 \\
 \ion{Ni}{ii} & 13.68$\pm$0.03 & 13.66 & - \\
 \ion{Zn}{ii} & 12.59$\pm$0.05 & 12.47 & - \\
\hline
\end{tabular}
\end{table*}

\section{Summary \& Conclusions}
\label{sec:conclude}
We have performed detailed spectroscopic analysis and numerical modelling of the DLA at $z_{abs}$ = 2.05 towards the quasar FBQS J2340-0053. Metal absorption features associated with this system arise from fourteen distinct velocity components spread over an interval spanning $\Delta v_{90}$ = 114 km s$^{-1}$. Of these, seven components harbour H$_{2}$ and \ion{C}{i}. The multicomponent absorption features enable us to study the physical environment of the H$_{2}$ components individually. Here, we present a brief summary of the results of our study.
\begin{itemize}[label=\textbullet]
\item{We measure the total \textit{N}(\ion{H}{i}), \textit{N}(H$_{2}$) and \textit{N}(HD) to be 20.35$\pm$0.05, 17.99$\pm$0.05 and 14.28$\pm$0.08 (log cm$^{-2}$) respectively. H$_{2}$ is detected in the lowest six rotational levels of the ground vibrational state.}
\item{The DLA has an average metallicity, Z = 0.3 $Z_{\sun}$ ([S/H] = -0.52$\pm$0.06) and dust-to-gas ratio, $\kappa$ = 0.34$\pm$0.07. The mean metallicity and dust-to-gas ratio from our models for the H$_{2}$ components are consistent with these values. }
\item{The observed \ion{C}{ii*} column density indicates that the DLA is exposed to an intensity of ultraviolet radiation 0.35 times the intensity in the Milky Way. Our numerical models show that the physical environment of the DLA can be recreated by subjecting the DLA to solely the Khaire-Srianand metagalactic background radiation, along with mean cosmic ray ionization rate of $\sim$ 10$^{-15.37}$ s$^{-1}$ for the H$_2$ components.} 
\item{Our models indicate that the DLA harbours grains that are smaller than the grains in the Galactic ISM. We incorporate both silicate and graphite grains distributed as per the MRN distribution, but having sizes half those of the grains in the ISM (in the range 0.0025-0.125 $\umu$m).} 
\item{From the numerical models, we constrain neutral hydrogen density $n_\mathrm{H}$, in the different molecular regions of the DLA to be in the range 30--120 cm$^{-3}$. Gas temperature lies between 140 and 360 K for the different H$_{2}$ components, with five of the seven components having electron temperature greater than 200 K in the inner shielded regions. Hence, these components do not trace the cold phase, but are rather associated with a warmer phase of H$_{2}$. The gas pressure in the molecular regions of the H$_{2}$ components lies in the range 7,000--23,000 cm$^{-3}$ K, towards the higher end of the pressure range seen in H$_{2}$-DLAs. Our models also enable us to study various contributing factors to the total pressure in the DLA. We conclude that micro-turbulence plays a crucial role in the molecular region.}
\item{Metal abundances vary across the DLA components. We constrain the abundances of sulphur, silicon, iron and carbon through the numerical models. The mean abundances of sulphur, silicon and iron across the seven H$_{2}$ components agree with the respective  mean observed abundances obtained for the entire DLA. Though the carbon abundance cannot be constrained observationally, we are able to predict the mean carbon abundance from our models. We predict that [C/H] = -1.39. The abundances for phosphorus, chromium, nickel and zinc predicted by the average DLA model are similar to the observationally constrained values.}
\item{We are able to determine the distribution of \ion{H}{i} across the H$_{2}$ and non-H$_{2}$ components through the {\tiny CLOUDY} models. The non-H$_{2}$ components give rise to log[\textit{N}(\ion{H}{i})(cm$^{-2}$)] = 19.86. This accounts for $\sim$ 32 percent of the total \ion{H}{i} in the DLA. These non-H$_{2}$ components have mean metallicity, Z = -0.54, which is slightly lower than the mean metallicity of -0.49 derived for the H$_2$ components, but agrees with the mean metallicity of 0.3 $Z_{\sun}$ derived for the entire DLA using the observed total column densities.}
\item{The extent of the DLA along the line-of-sight is obtained from the numerical models. The average model of the DLA yields a size of 11.8 pc, while the total extent of the H$_{2}$ components is 7.2 pc. This is in agreement with previous studies which show that H$_{2}$ absorption arises in regions $\leq$ 15 pc across.}
\item{We find that the H$_{2}$ component models trace the physical structure of the cloud better than the average model. Summing the species-wise column densities over all the H$_{2}$ components yields resultant column densities which are closer to the observed H$_{2}$ level population and \ion{C}{i} fine structure levels, as compared to the average model. The average model fails to trace the cold phase of H$_{2}$ seen in some of the components.}
\end{itemize}

\section*{Acknowledgements}
This work makes use of spectra observed with the Ultraviolet \& Visual Echelle Spectrograph on the Very Large Telescope at Cerro Paranal, Chile (Programme ID 082.A-0569), and with the High Resolution Echelle Spectrometer on the Keck I Telescope at Maunakea, Hawaii (PI: Prochaska). The Keck Telescope data presented in this work have been obtained from the Keck Observatory Database of Ionized Absorbers toward QSOs (KODIAQ), which was funded through NASA ADAP grant NNX10AE84G. The authors are thankful to Vikram Khaire for providing the metagalactic background. GS acknowledges support from the DST projects (D.O. No. SR/FTP/PS-133/2011 and SR/WOS-A/PM-9/2017). KR is grateful to IUCAA for hospitality during academic visits related to this work. The authors wish to thank the anonymous referee for insightful comments which have led to significant improvement in the quality of the manuscript.




\bibliographystyle{mnras}
\bibliography{dla} 



\appendix
\section{Details of H$_{2}$ component models}
\begin{figure*}
\centering{\includegraphics[width=\textwidth, height=17 cm, keepaspectratio]{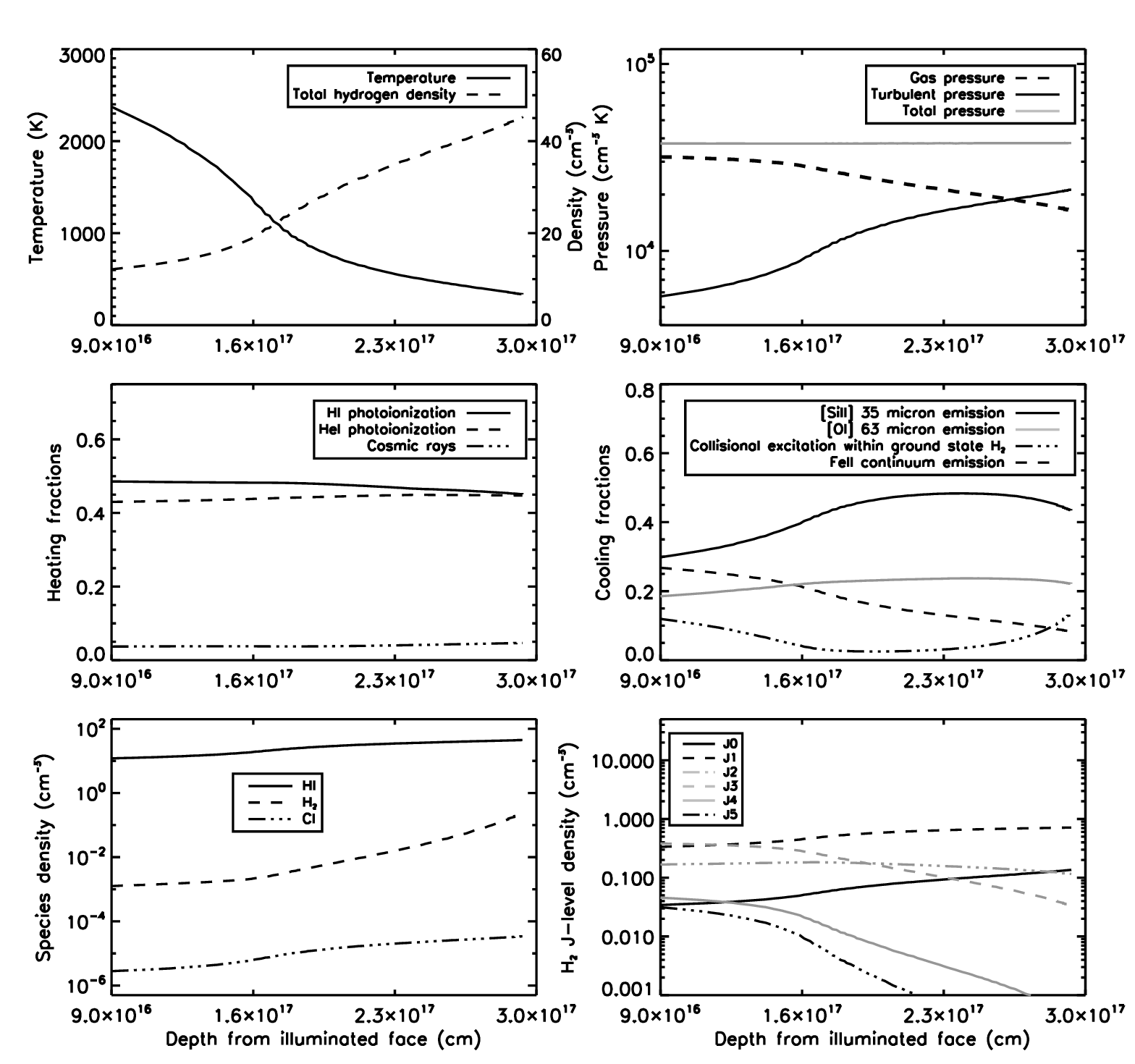}}
\caption{The physical conditions within H$_{2}$ component 4 are summarized in this figure. Each panel shows the variation of different physical properties with depth from one of the illuminated faces of the cloud. The plots here represent one half of the cloud, with the other half having a symmetrical profile. The properties include gas temperature and total hydrogen density (top row, first panel), total pressure and its major constituents (top row, second panel), heating and cooling fractions due to various physical processes (middle row), density of \ion{H}{i}, H$_2$ and \ion{C}{i} (bottom row, first panel), and the densities of various rotational levels of H$_{2}$ (bottom row, second panel).}
\label{fig:phycon_4}
\end{figure*}

\begin{figure*}
\centering{\includegraphics[width=\textwidth, height=17 cm, keepaspectratio]{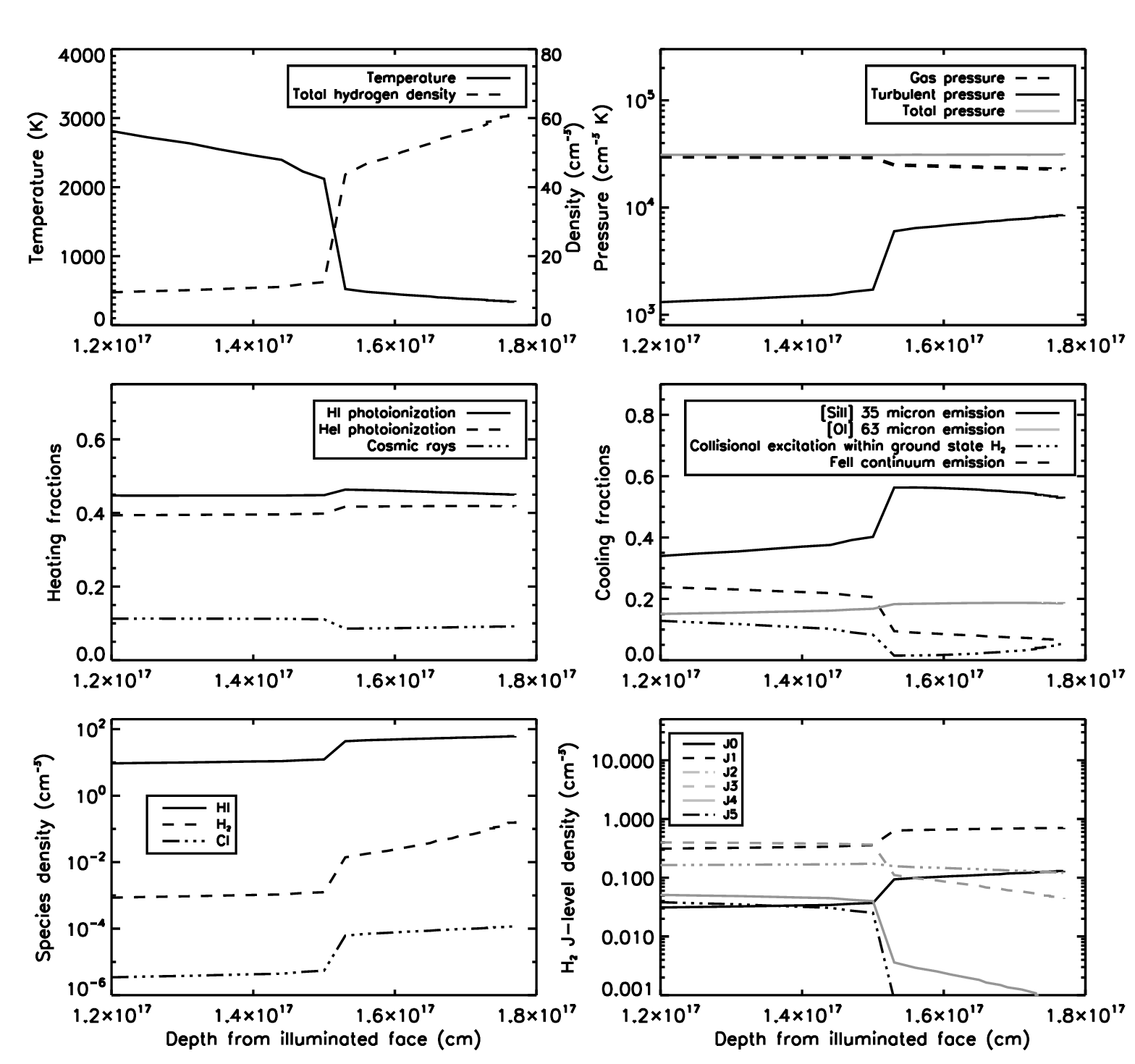}}
\caption{The physical conditions within H$_{2}$ component 5 are summarized in this figure. Each panel shows the variation of different physical properties with depth from one of the illuminated faces of the cloud. The plots here represent one half of the cloud, with the other half having a symmetrical profile. The properties include gas temperature and total hydrogen density (top row, first panel), total pressure and its major constituents (top row, second panel), heating and cooling fractions due to various physical processes (middle row), density of \ion{H}{i}, H$_2$ and \ion{C}{i} (bottom row, first panel), and the densities of various rotational levels of H$_{2}$ (bottom row, second panel).}
\label{fig:phycon_5}
\end{figure*}

\begin{figure*}
\centering{\includegraphics[width=\textwidth, height=17 cm, keepaspectratio]{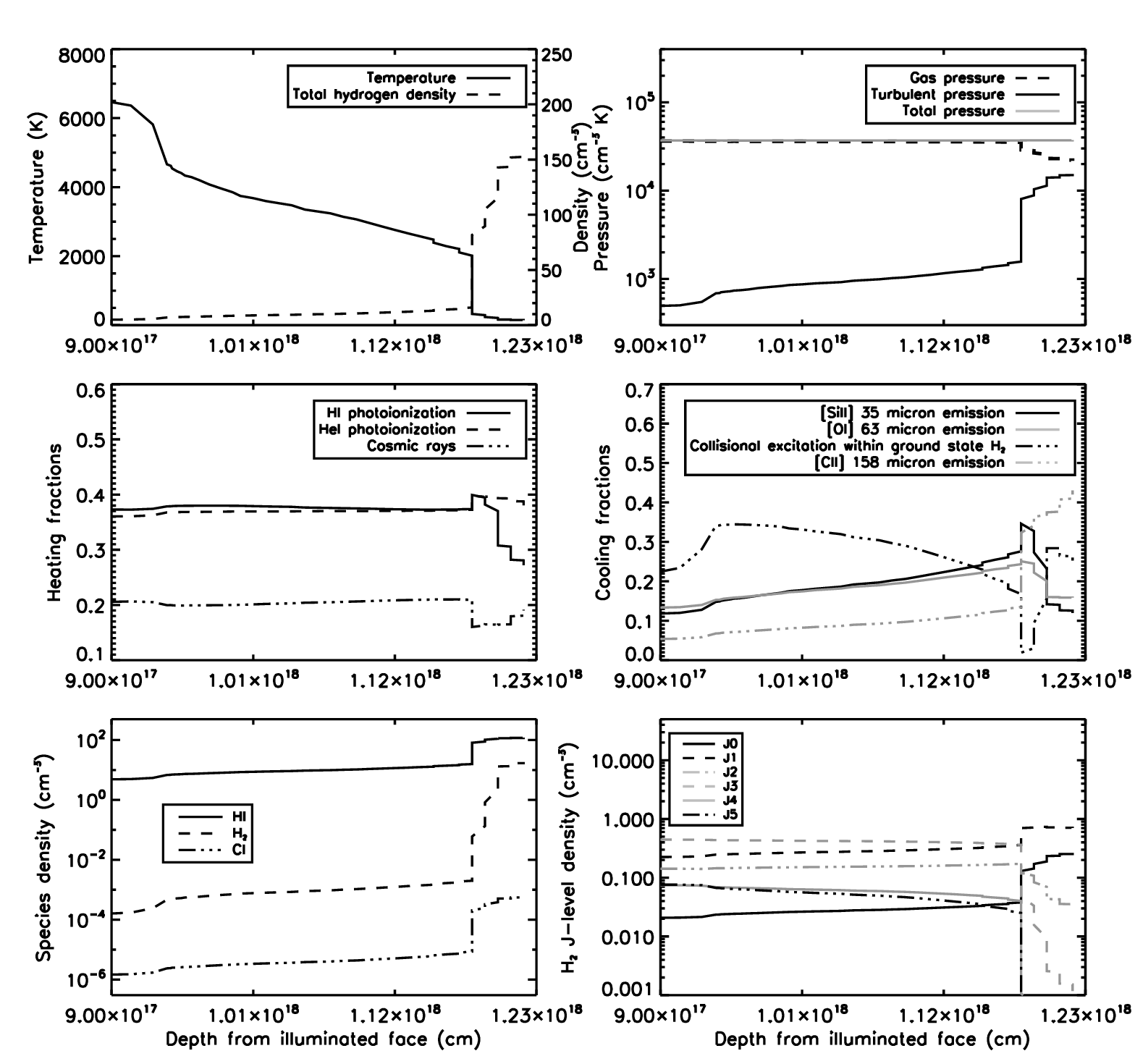}}
\caption{The physical conditions within H$_{2}$ component 7 are summarized in this figure. Each panel shows the variation of different physical properties with depth from one of the illuminated faces of the cloud. The plots here represent one half of the cloud, with the other half having a symmetrical profile. The properties include gas temperature and total hydrogen density (top row, first panel), total pressure and its major constituents (top row, second panel), heating and cooling fractions due to various physical processes (middle row), density of \ion{H}{i}, H$_2$ and \ion{C}{i} (bottom row, first panel), and the densities of various rotational levels of H$_{2}$ (bottom row, second panel).}
\label{fig:phycon_7}
\end{figure*}

\begin{figure*}
\centering{\includegraphics[width=\textwidth, height=17 cm, keepaspectratio]{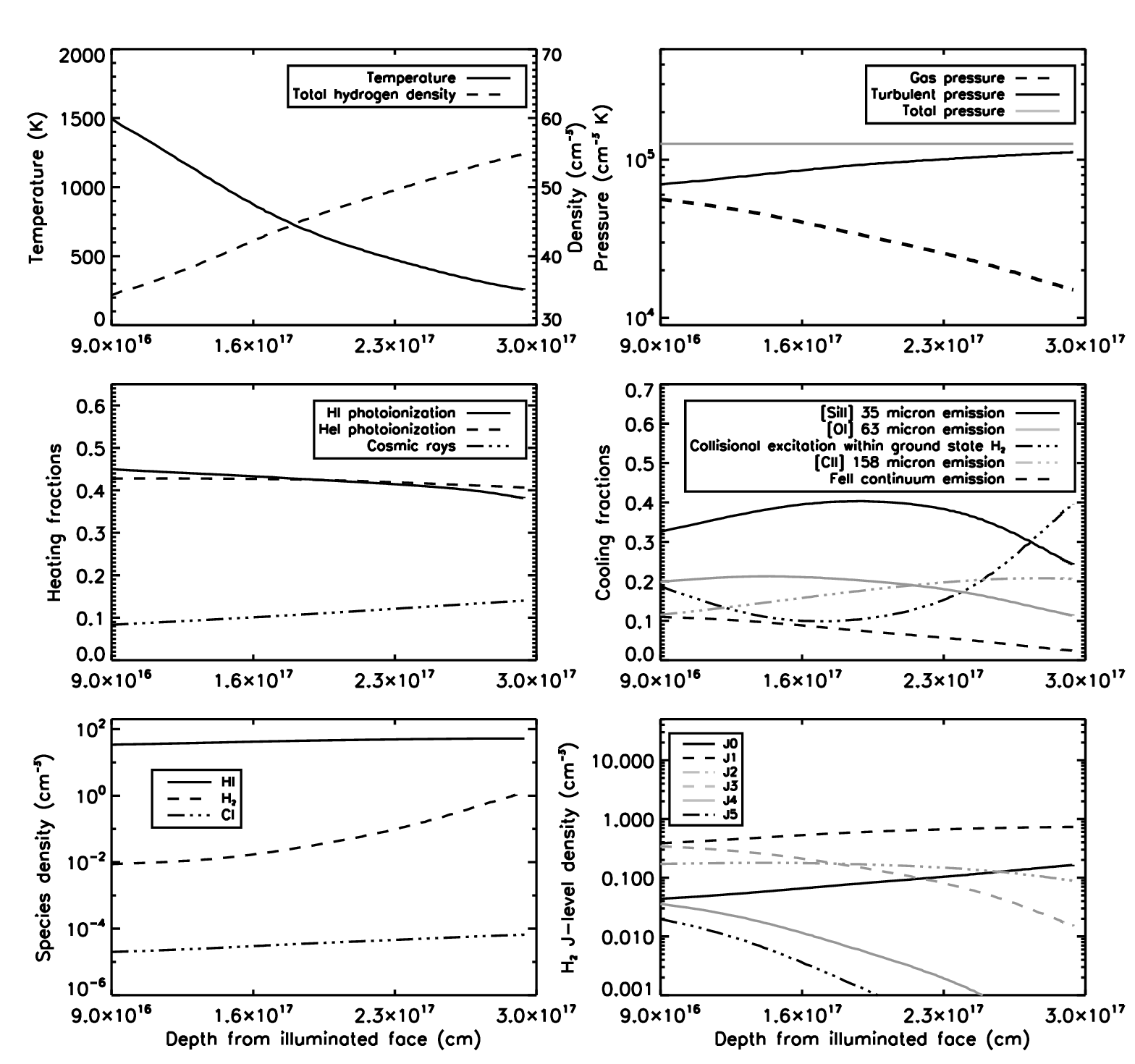}}
\caption{The physical conditions within H$_{2}$ component 9 are summarized in this figure. Each panel shows the variation of different physical properties with depth from one of the illuminated faces of the cloud. The plots here represent one half of the cloud, with the other half having a symmetrical profile. The properties include gas temperature and total hydrogen density (top row, first panel), total pressure and its major constituents (top row, second panel), heating and cooling fractions due to various physical processes (middle row), density of \ion{H}{i}, H$_2$ and \ion{C}{i} (bottom row, first panel), and the densities of various rotational levels of H$_{2}$ (bottom row, second panel).}
\label{fig:phycon_9}
\end{figure*}

\begin{figure*}
\centering{\includegraphics[width=\textwidth, height=17 cm, keepaspectratio]{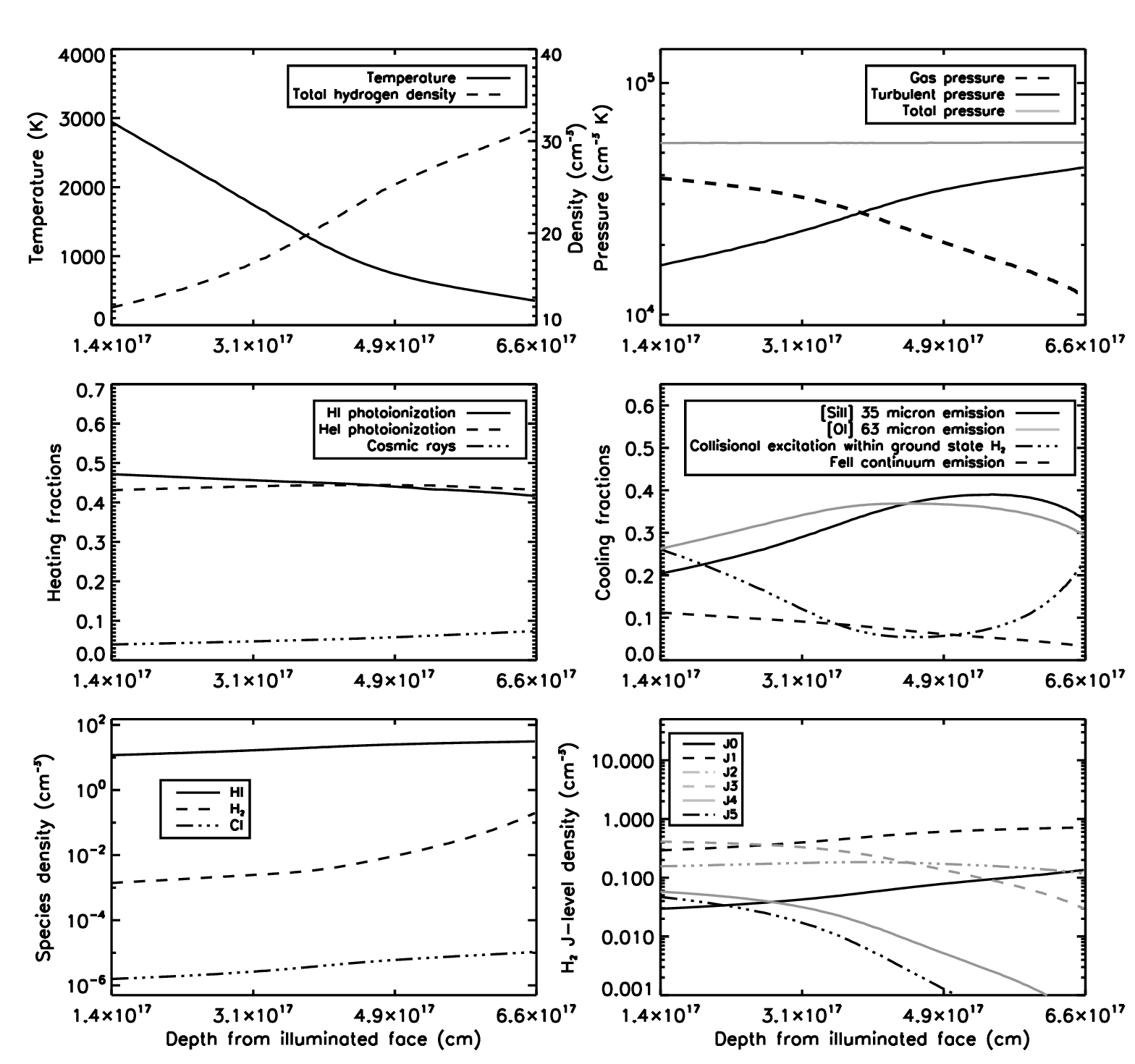}}
\caption{The physical conditions within the H$_{2}$ region comprising components 11 \& 12 are summarized in this figure. Each panel shows the variation of different physical properties with depth from one of the illuminated faces of the cloud. The plots here represent one half of the cloud, with the other half having a symmetrical profile. The properties include gas temperature and total hydrogen density (top row, first panel), total pressure and its major constituents (top row, second panel), heating and cooling fractions due to various physical processes (middle row), density of \ion{H}{i}, H$_2$ and \ion{C}{i} (bottom row, first panel), and the densities of various rotational levels of H$_{2}$ (bottom row, second panel).}
\label{fig:phycon_11}
\end{figure*}

\begin{figure*}
\centering{\includegraphics[width=\textwidth, height=17 cm, keepaspectratio]{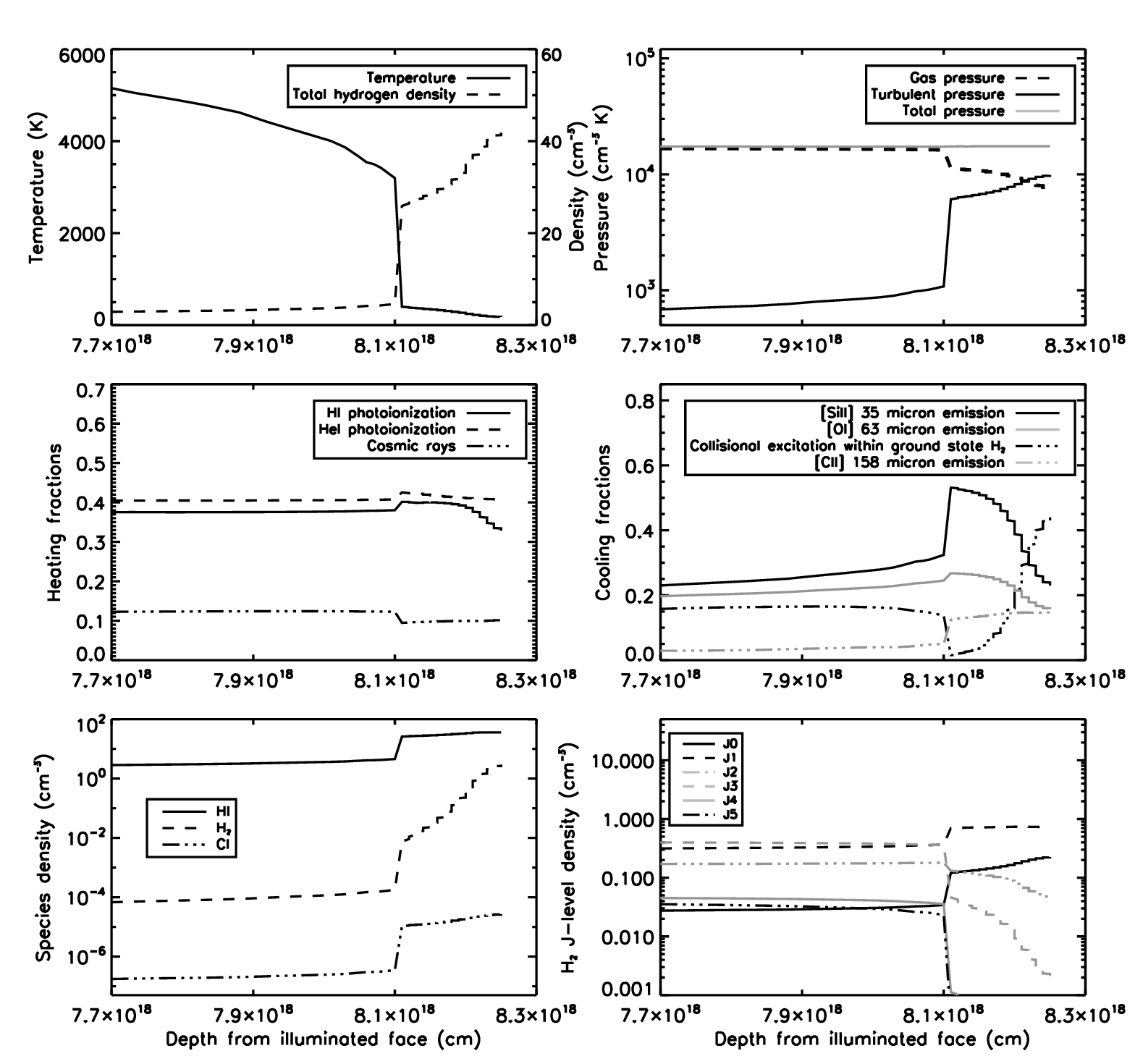}}
\caption{The physical conditions within H$_{2}$ component 13 are summarized in this figure. Each panel shows the variation of different physical properties with depth from one of the illuminated faces of the cloud. The plots here represent one half of the cloud, with the other half having a symmetrical profile. The properties include gas temperature and total hydrogen density (top row, first panel), total pressure and its major constituents (top row, second panel), heating and cooling fractions due to various physical processes (middle row), density of \ion{H}{i}, H$_2$ and \ion{C}{i} (bottom row, first panel), and the densities of various rotational levels of H$_{2}$ (bottom row, second panel).}
\label{fig:phycon_13}
\end{figure*}

In Section \ref{ssec:approach}, we have described our {\tiny CLOUDY} modelling approach for the H$_{2}$ components. H$_{2}$ absorption is seen in the seven metal components 4, 5, 7, 8, 9, 11 and 13. We have discussed the model for component 8 in detail in Section \ref{ssec:8_model}. Here, we present details of the remaining component models. A combined model of the H$_{2}$ component 11 and the non-H$_{2}$ component 12 is constructed, as the \ion{C}{i} features are closer to the metal component 12. Tables \ref{tab:paras_4}, \ref{tab:paras_5}, \ref{tab:paras_7}, \ref{tab:paras_9}, \ref{tab:paras_11} and \ref{tab:paras_13} list the variable parameters of the models for components 4, 5, 7, 9, (11+12) and 13 respectively. The observed and model column densities for these components are compared in Tables \ref{tab:model_4}, \ref{tab:model_5}, \ref{tab:model_7}, \ref{tab:model_9}, \ref{tab:model_11} and \ref{tab:model_13} respectively. Further, we also plot the temperature, density and pressure profiles, heating and cooling fractions, density of \ion{H}{i}, H$_2$ and \ion{C}{i}, and the densities of the rotational levels of H$_{2}$. These plots as a function of depth from one of the two illuminated faces of the cloud are made for each component, in a manner similar to Figs. \ref{fig:phycon_8} and \ref{fig:phycon_tot}. Figs. \ref{fig:phycon_4}, \ref{fig:phycon_5}, \ref{fig:phycon_7}, \ref{fig:phycon_9}, \ref{fig:phycon_11} and \ref{fig:phycon_13} correspond to components 4, 5, 7, 9, (11+12) and 13 respectively. 


\begin{table}
 \centering
  \caption{Physical parameters for {\tiny CLOUDY} model of component 4}
  \label{tab:paras_4}
  \begin{tabular}{@{}ll@{}}
 \hline
   Physical parameter     &     Model value\\
 \hline
 Radiation field & KS15 background$^{\textit{a}}$\\
 Total hydrogen density & 7.9 cm$^{-3}$ (0.90 in log-scale) \\
 (at illuminated face) & \\
 Metallicity (log-scale) & -0.46 \\
 $[C/H]$ & -1.50 \\
 $[Si/H]$ & -0.62 \\
 $[Fe/H]$ & -0.62 \\
 Dust-to-gas ratio & 0.40 (-0.06)$^{\textit{b}}$ \\
 Size range of dust grains & 0.0025-0.125 $\umu$m \\
 Micro-turbulence & 2.4 (2.5$\pm$0.1)$^{\textit{b}}$ km s$^{-1}$ \\
 Cosmic ray ionization rate & 10$^{-15.70}$ s$^{-1}$ \\
 \hline
\end{tabular}
\newline
\raggedright $^{\textit{a}}$ Khaire-Srianand background \citep{Khaire2015b, Khaire2015a}\\
\raggedright $^{\textit{b}}$ Numbers within the brackets indicate observed values
\end{table}

\begin{table}
 \centering
  \caption{Observed and {\tiny CLOUDY} model column densities for component 4}
  \label{tab:model_4}
  \begin{tabular}{@{}lcc@{}}
    \hline
   Species     &     Observed log(\textit{N}) (cm$^{-2}$) & Model log(\textit{N}) (cm$^{-2}$) \\
 \hline
 \ion{H}{i} & - & 19.10 \\
 H$_2$ & 15.96$\pm$0.03 & 16.00 \\
 H$_2$ (0) & 15.35$\pm$0.05 & 15.06 \\
 H$_2$ (1) & 15.74$\pm$0.05 & 15.83 \\
 H$_2$ (2) & 15.03$\pm$0.06 & 15.13 \\
 H$_2$ (3) & 14.30$\pm$0.07 & 14.86 \\ 
 H$_2$ (4) & 13.17$\pm$0.36 & 13.51 \\ 
 H$_2$ (5) & $\leq$ 13.67 & 13.17 \\ 
 \ion{C}{i*} & 12.42$\pm$0.09 & 12.52 \\
 \ion{C}{i**} & 12.10$\pm$0.08 & 12.36 \\
 \ion{C}{i***} & 11.92$\pm$0.14 & 11.73 \\
 \ion{C}{ii*} & 12.38$\pm$0.09 & 12.56 \\
 \ion{Si}{ii} & 14.23$\pm$0.02 & 14.00 \\
 \ion{S}{ii} & 13.97$\pm$0.02 & 13.77 \\
 \ion{Fe}{ii} & 14.27$\pm$0.02 & 13.99 \\
\hline
\end{tabular}
\end{table}

\begin{table}
 \centering
  \caption{Physical parameters for {\tiny CLOUDY} model of component 5}
  \label{tab:paras_5}
  \begin{tabular}{@{}ll@{}}
  \hline
   Physical parameter     &     Model value\\
 \hline
 Radiation field & KS15 background$^{\textit{a}}$\\
 Total hydrogen density & 4.6 cm$^{-3}$ (0.66 in log-scale) \\
 (at illuminated face) & \\
 Metallicity (log-scale) & -0.46 \\
 $[C/H]$ & -1.25  \\
 $[Si/H]$ & -0.48 \\
 $[Fe/H]$ & -0.68 \\
 Dust-to-gas ratio & 0.40 (0.38$\pm$0.08)$^{\textit{b}}$ \\
 Size range of dust grains & 0.0025-0.125 $\umu$m \\
 Micro-turbulence & 1.3 (1.5$\pm$0.2)$^{\textit{b}}$ km s$^{-1}$ \\
 Cosmic ray ionization rate & 10$^{-15.20}$ s$^{-1}$ \\
 \hline
\end{tabular}
\newline
\raggedright $^{\textit{a}}$ Khaire-Srianand background \citep{Khaire2015b, Khaire2015a}\\
\raggedright $^{\textit{b}}$ Numbers within the brackets indicate observed values
\end{table}

\begin{table}
 \centering
  \caption{Observed and {\tiny CLOUDY} model column densities for component 5}
  \label{tab:model_5}
  \begin{tabular}{@{}lcc@{}}
  \hline
   Species     &     Observed log(\textit{N}) (cm$^{-2}$) & Model log(\textit{N}) (cm$^{-2}$) \\
 \hline
 \ion{H}{i} & - & 18.70 \\
 H$_2$ & 15.49$\pm$0.05 & 15.45 \\
 H$_2$ (0) & 14.77$\pm$0.10 & 14.50 \\
 H$_2$ (1) & 15.15$\pm$0.08 & 15.27 \\
 H$_2$ (2) & 14.79$\pm$0.12 & 14.58 \\
 H$_2$ (3) & 14.62$\pm$0.08 & 14.38 \\ 
 H$_2$ (4) & 13.55$\pm$0.18 & 13.16 \\ 
 H$_2$ (5) & $\leq$ 13.90 & 12.96 \\ 
 \ion{C}{i*} & 12.13$\pm$0.07 & 12.42 \\
 \ion{C}{i**} & 12.40$\pm$0.04 & 12.32 \\
 \ion{C}{i***} & 12.29$\pm$0.04 & 11.74 \\
 \ion{C}{ii*} & 12.35$\pm$0.13 & 12.43 \\
 \ion{Si}{ii} & 13.87$\pm$0.15 & 13.74 \\
 \ion{S}{ii} & 13.37$\pm$0.23 & 13.36 \\
 \ion{Fe}{ii} & 13.59$\pm$0.17 & 13.53 \\
\hline
\end{tabular}
\end{table}

\begin{table}
 \centering
  \caption{Physical parameters for {\tiny CLOUDY} model of component 7}
  \label{tab:paras_7}
  \begin{tabular}{@{}ll@{}}
    \hline
   Physical parameter     &      Model value \\
 \hline
 Radiation field & KS15 background$^{\textit{a}}$ \\
 Total hydrogen density & 4.0 cm$^{-3}$ (0.60 in log-scale) \\
 (at illuminated face) & \\
 Metallicity (log-scale) & -0.60 \\
 $[C/H]$ & -1.14 \\
 $[Si/H]$ & -0.85 \\
 $[Fe/H]$ & -1.22 \\
 Dust-to-gas ratio & 0.35 (0.42$\pm$0.07)$^{\textit{b}}$ \\
 Size range of dust grains & 0.0025-0.125 $\umu$m \\
 Micro-turbulence & 1.1 (0.9$\pm$0.2)$^{\textit{b}}$ km s$^{-1}$ \\
 Cosmic ray ionization rate & 10$^{-15.10}$ s$^{-1}$ \\
 \hline
\end{tabular}
\newline
\raggedright $^{\textit{a}}$ Khaire-Srianand background \citep{Khaire2015b, Khaire2015a}\\
\raggedright $^{\textit{b}}$ Numbers within the brackets indicate observed values
\end{table}

\begin{table}
 \centering
  \caption{Observed and {\tiny CLOUDY} model column densities for component 7}
  \label{tab:model_7}
  \begin{tabular}{@{}lcc@{}}
 \hline
   Species     &     Observed log(\textit{N}) (cm$^{-2}$) & Model log(\textit{N}) (cm$^{-2}$) \\
 \hline
 \ion{H}{i} & - & 19.34 \\
 H$_2$ & 17.79$\pm$0.08 & 17.86 \\
 H$_2$ (0) & 17.38$\pm$0.11 & 17.25 \\
 H$_2$ (1) & 17.57$\pm$0.11 & 17.71 \\
 H$_2$ (2) & $\leq$ 15.90 & 16.46 \\
 H$_2$ (3) & 15.24$\pm$0.21 & 15.30 \\ 
 H$_2$ (4) & 13.77$\pm$0.15 & 13.70 \\ 
 H$_2$ (5) & 13.92$\pm$0.22 & 13.58 \\ 
 \ion{C}{i*} & 13.53$\pm$0.03 & 13.24 \\
 \ion{C}{i**} & 13.06$\pm$0.01 & 13.23 \\
 \ion{C}{i***} & 11.87$\pm$0.14 & 12.71 \\
 \ion{C}{ii*} & 13.24$\pm$0.13 & 13.18 \\
 \ion{Si}{ii} & 13.95$\pm$0.08 & 14.03 \\
 \ion{S}{ii} & 13.82$\pm$0.07 & 13.87 \\
 \ion{Fe}{ii} & 13.52$\pm$0.13 & 13.65 \\
\hline 
\end{tabular}
\end{table}

\begin{table}
 \centering
  \caption{Physical parameters for {\tiny CLOUDY} model of component 9}
  \label{tab:paras_9}
  \begin{tabular}{@{}ll@{}}
 \hline
   Physical parameter     &    Model value\\
 \hline
 Radiation field & KS15 background$^{\textit{a}}$\\
 Total hydrogen density & 26.9 cm$^{-3}$ (1.43 in log-scale) \\
 (at illuminated face) & \\
 Metallicity (log-scale) & -0.82 \\
 $[C/H]$ & -1.35 \\
 $[Si/H]$ & -0.89 \\
 $[Fe/H]$ & -1.08 \\
 Dust-to-gas ratio & 0.35 (0.25$\pm$0.11)$^{\textit{b}}$\\
 Size range of dust grains & 0.0025-0.125 $\umu$m \\
 Micro-turbulence & 5.0 (5.0$\pm$0.2)$^{\textit{b}}$ km s$^{-1}$ \\
 Cosmic ray ionization rate & 10$^{-15.35}$ s$^{-1}$ \\
 \hline
\end{tabular}
\newline
\raggedright $^{\textit{a}}$ Khaire-Srianand background \citep{Khaire2015b, Khaire2015a}\\
\raggedright $^{\textit{b}}$ Numbers within the brackets indicate observed values
\end{table}

\begin{table}
 \centering
  \caption{Observed and {\tiny CLOUDY} model column densities for component 9}
  \label{tab:model_9}
  \begin{tabular}{@{}lcc@{}}
  \hline
   Species     &     Observed log(\textit{N}) (cm$^{-2}$) & Model log(\textit{N}) (cm$^{-2}$) \\
 \hline
 \ion{H}{i} & - & 19.38 \\
 H$_2$ & 16.80$\pm$0.06 & 16.81 \\
 H$_2$ (0) & 15.76$\pm$0.06 & 15.94 \\
 H$_2$ (1) & 16.60$\pm$0.09 & 16.65 \\
 H$_2$ (2) & 16.07$\pm$0.06 & 15.88 \\
 H$_2$ (3) & 15.74$\pm$0.03 & 15.52 \\ 
 H$_2$ (4) & 14.37$\pm$0.04 & 14.17 \\ 
 H$_2$ (5) & 14.14$\pm$0.09 & 13.84 \\ 
 \ion{C}{i*} & 13.25$\pm$0.01 & 12.95 \\
 \ion{C}{i**} & 13.03$\pm$0.01 & 12.89 \\
 \ion{C}{i***} & 12.27$\pm$0.04 & 12.33 \\
 \ion{C}{ii*} & 12.88$\pm$0.05 & 13.12 \\
 \ion{Si}{ii} & 13.85$\pm$0.10 & 13.99 \\
 \ion{S}{ii} & 13.67$\pm$0.07 & 13.67 \\
 \ion{Fe}{ii} & 13.73$\pm$0.08 & 13.80 \\
\hline
\end{tabular}
\end{table}

\begin{table}
 \centering
  \caption{Physical parameters for the combined {\tiny CLOUDY} model of components 11 \& 12}
  \label{tab:paras_11}
  \begin{tabular}{@{}ll@{}}
    \hline
   Physical parameter     &      Model value\\
 \hline
 Radiation field & KS15 background$^{\textit{a}}$\\
 Total hydrogen density & 8.9 cm$^{-3}$ (0.95 in log-scale) \\
 (at illuminated face) & \\
 Metallicity (log-scale) & -0.40\\
 $[C/H]$ & -1.70 \\
 $[Si/H]$ & -0.80 \\
 $[Fe/H]$ & -1.10 \\
 Dust-to-gas ratio & 0.50 (0.40$\pm$0.09)$^{\textit{b}}$\\
 Size range of dust grains & 0.0025-0.125 $\umu$m \\
 Micro-turbulence & 4.1 (4.2$\pm$0.1)$^{\textit{b}}$ km s$^{-1}$ \\
 Cosmic ray ionization rate & 10$^{-15.70}$ s$^{-1}$ \\
 \hline
\end{tabular}
\newline
\raggedright $^{\textit{a}}$ Khaire-Srianand background \citep{Khaire2015b, Khaire2015a}\\
\raggedright $^{\textit{b}}$ Numbers within the brackets indicate observed values
\end{table}

\begin{table}
 \centering
  \caption{Observed and {\tiny CLOUDY} model column densities for the combined model of components 11 \& 12}
  \label{tab:model_11}
  \begin{tabular}{@{}lcc@{}}
   \hline
   Species     &     Observed log(\textit{N}) (cm$^{-2}$) & Model log(\textit{N}) (cm$^{-2}$) \\
 \hline
 \ion{H}{i} & - & 19.39 \\
 H$_2$ & 16.39$\pm$0.03 & 16.30 \\
 H$_2$ (0) & 15.57$\pm$0.03 & 15.34 \\
 H$_2$ (1) & 16.18$\pm$0.04 & 16.12 \\
 H$_2$ (2) & 15.66$\pm$0.06 & 15.45 \\
 H$_2$ (3) & 15.13$\pm$0.02 & 15.22 \\ 
 H$_2$ (4) & 13.77$\pm$0.12 & 13.96 \\ 
 H$_2$ (5) & $\leq$ 13.88 & 13.69 \\ 
 \ion{C}{i*} & 12.42$\pm$0.04 & 12.48 \\
 \ion{C}{i**} & 11.93$\pm$0.14 & 12.27 \\
 \ion{C}{i***} & 11.75$\pm$0.23 & 11.62 \\
 \ion{C}{ii*} & 12.59$\pm$0.24 & 12.58 \\
 \ion{Si}{ii} & 14.29$\pm$0.19 & 14.11 \\
 \ion{S}{ii} & 13.98$\pm$0.17 & 14.11 \\
 \ion{Fe}{ii} & 13.87$\pm$0.11 & 13.80 \\
 \hline
\end{tabular}
\end{table}

\begin{table}
 \centering
  \caption{Physical parameters for {\tiny CLOUDY} model of component 13}
  \label{tab:paras_13}
  \begin{tabular}{@{}ll@{}}
 \hline
   Physical parameter     &     Model value \\
 \hline
 Radiation field & KS15 background$^{\textit{a}}$\\
 Total hydrogen density & 1.7 cm$^{-3}$ (0.23 in log-scale) \\
 (at illuminated face) & \\
 Metallicity (log-scale) & -0.52\\
 $[C/H]$ & -1.52 \\
 $[Si/H]$ & -0.68 \\
 $[Fe/H]$ & -1.38 \\
 Dust-to-gas ratio & 0.50 (0.43$\pm$0.08)$^{\textit{b}}$\\
 Size range of dust grains & 0.0025-0.125 $\umu$m \\
 Micro-turbulence & 1.7 (1.9$\pm$0.2)$^{\textit{b}}$ km s$^{-1}$ \\
 Cosmic ray ionization rate & 10$^{-15.70}$ s$^{-1}$ \\
 \hline
\end{tabular}
\newline
\raggedright $^{\textit{a}}$ Khaire-Srianand background \citep{Khaire2015b, Khaire2015a}\\
\raggedright $^{\textit{b}}$ Numbers within the brackets indicate observed values
\end{table}

\begin{table}
 \centering
  \caption{Observed and {\tiny CLOUDY} model column densities for component 13}
  \label{tab:model_13}
 \begin{tabular}{@{}lcc@{}}
 \hline
  Species     &     Observed log(\textit{N}) (cm$^{-2}$) & Model log(\textit{N}) (cm$^{-2}$) \\
 \hline
 \ion{H}{i} & - & 19.62 \\
 H$_2$ & 17.29$\pm$0.09 & 17.25 \\
 H$_2$ (0) & 16.43$\pm$0.17 & 16.57 \\
 H$_2$ (1) & 17.20$\pm$0.11 & 17.11 \\
 H$_2$ (2) & 16.03$\pm$0.11 & 16.01 \\
 H$_2$ (3) & 14.69$\pm$0.06 & 14.98 \\ 
 H$_2$ (4) & 13.29$\pm$0.30 & 13.41 \\ 
 H$_2$ (5) & 13.55$\pm$0.30 & 13.23 \\ 
 \ion{C}{i*} & 12.75$\pm$0.18 & 12.66 \\
 \ion{C}{i**} & 12.06$\pm$0.08 & 12.34 \\
 \ion{C}{i***} & 11.88$\pm$0.11 & 11.58 \\
 \ion{C}{ii*} & 12.72$\pm$0.16 & 12.59 \\
 \ion{Si}{ii} & 14.32$\pm$0.15 & 14.47 \\
 \ion{S}{ii} & 13.96$\pm$0.15 & 14.19 \\
 \ion{Fe}{ii} & 13.49$\pm$0.21 & 13.76 \\
\hline
\end{tabular}
\end{table}

\bsp	
\label{lastpage}
\end{document}